\documentclass[useAMS,usenatbib,usegraphicx]{mn2e}

\newcommand{\lya}{Ly$\alpha$\ }
\newcommand{\hkpc}{h^{-1}{\rm kpc}}
\newcommand{\hmpc}{h^{-1}{\rm Mpc}}

\newcommand{\kms}{\;{\rm km}\,{\rm s}^{-1}}

\newcommand{\vw}{{v_{\rm wind}}}

\newcommand{\gad}{{\sc Gadget-2}}

\title[Simulations of IGM Enrichment]{Cosmological Simulations of Intergalactic Medium Enrichment from Galactic Outflows} 
\author[B. D. Oppenheimer \& R. Dav\'e]{Benjamin D. Oppenheimer$^1$,
Romeel Dav\'e$^1$ \\$^{1}$Astronomy Department, University of Arizona,
Tucson, AZ 85721}

\begin{document}

\pagerange{\pageref{firstpage}--\pageref{lastpage}} \pubyear{2006}

\maketitle

\label{firstpage}

\begin{abstract}

We investigate models of self-consistent chemical enrichment of
the intergalactic medium (IGM) from $z=6.0\rightarrow 1.5$, based
on hydrodynamic simulations of structure formation that explicitly
incorporate outflows from star forming galaxies.  Our main result is that
outflow parameterizations derived from observations of local starburst
galaxies, in particular momentum-driven wind scenarios, provide the
best agreement with observations of \hbox{C\,{\sc iv}} absorption at
$z\sim 2-5$.  Such models sufficiently enrich the high-$z$ IGM to produce
a global mass density of \hbox{C\,{\sc iv}} absorbers that is relatively
invariant from $z=5.5\rightarrow 1.5$, in agreement with observations.
This occurs despite continual IGM enrichment causing an increase in
volume-averaged metallicity by $\sim\times 5-10$ over this redshift range,
because energy input accompanying the enriching outflows causes a drop
in the global ionization fraction of \hbox{C\,{\sc iv}}.  Comparisons to
observed \hbox{C\,{\sc iv}} column density and linewidth distributions and
\hbox{C\,{\sc iv}}-based pixel optical depth ratios provide significant
constraints on wind models.  Our best-fitting outflow models show
mean IGM temperatures only slightly above our no-outflow case, metal
filling factors of just a few percent with volume-weighted metallicities around
$10^{-3}$ at $z\sim 3$, significant amounts of collisionally-ionized
\hbox{C\,{\sc iv}} absorption, and a metallicity-density relationship
that rises rapidly at low overdensities and flattens at higher ones.
In general, we find that outflow speeds must be high enough to enrich
the low-density IGM at early times but low enough not to overheat it,
and concurrently must significantly suppress early star formation while
still producing enough early metals.  It is therefore non-trivial that
locally-calibrated momentum-driven wind scenarios naturally yield the
desired strength and evolution of outflows, and suggest that such models
represent a significant step towards understanding the impact of galactic
outflows on galaxies and the IGM across cosmic time.

\end{abstract}

\begin{keywords}
intergalactic medium, galaxies: formation, galaxies: high-redshift, cosmology: theory, methods: numerical
\end{keywords}

\section{Introduction} \label{sec: intro}

The intergalactic medium (IGM) contains the majority of the Universe's
baryons \citep{dav01} and its largest structures, yet paradoxically
can typically only be painstakingly observed in absorption against a
background source.  The marriage of high-resolution Echelle
spectroscopy on 8-10m class telescopes with cosmological hydrodynamic
simulations in the last decade has led to the understanding that the
Lyman-$\alpha$ forest arises from highly-photoionized \hbox{H\,{\sc
i}} tracing fluctuations in the underlying IGM density in the
so-called Fluctuating Gunn-Peterson Approximation \citep{cro98}.  This
ionization state of the absorbing gas is governed by a balance between
photo-ionizational heating due to the metagalactic UV background and
adiabatic cooling from Hubble expansion, yielding a tight
density-temperature relation known as the IGM equation of state
\citep{hui97}.

A surprising discovery was that the diffuse IGM is enriched with metals,
seen as \hbox{C\,{\sc iv}}, \hbox{Si\,{\sc iv}}, and \hbox{O\,{\sc
vi}} absorption lines in quasar spectra.  Despite our
understanding of \hbox{H\,{\sc i}} absorption, the origin of these
metals far from sites of star formation remains puzzling.  Currently the
leading candidate for enriching the IGM is supernova-driven outflows,
because dynamical disruption of galaxies is too inefficient \citep[][see
also Figure~\ref{fig:rhoZ}]{agu01}.  Understanding such galactic feedback
processes is crucial for developing a complete theory of galaxy formation
and evolution \citep{dek86}, as observations such as the shallow faint-end
slope of the galaxy stellar mass function \citep{col01} versus the
halo mass function \citep{jen01}, the galaxy mass-metallicity relation
\citep{tre04,erb06}, and the overproduction of stars in the Universe
in models without feedback \citep{dav01,spr03b} indicate that outflows
significantly affect galaxy properties.

Observations of metal abundance in various ions that trace different
densities can quantify the metallicity-density relation in the IGM
\citep{dav98}, thereby constraining the nature of the enriching outflows.
Early results from Keck's High Resolution Echelle Spectrograph (HIRES)
constrained typical IGM metallicities as traced by \hbox{C\,{\sc iv}}
to be $-3\la$[C/H]$\la -2$ \citep{son96,dav98}.  Subsequent observations
by \citet{son01}, \citet{sch03} (hereafter S03), and \citet{son05}
further showed that \hbox{C\,{\sc iv}} absorption evolves very little
from $z\sim 5\rightarrow 2$, which is surprising considering that the
vast majority of stars in the Universe form at $z<5$.  These results
have been interpreted as implying early enrichment by primeval galaxies
and/or Population~III stars at $z>6$, where physical distances are
small and shallow potential wells allow winds to distribute metals over
large comoving volumes \citep{sca02}.  In contrast, \citet{ade03, ade05}
observe enhanced \hbox{C\,{\sc iv}} and \hbox{O\,{\sc vi}} absorption in
the vicinity of galaxies at $z\sim 2-3$, suggesting that enrichment is
ongoing at lower redshifts in the form of superwinds from Lyman-break
galaxies \citep{pet01,sha03}.  \citet{por05} counter this by showing
that such strong correlations can arise from metals ejected from dwarf
galaxies between $z\approx 6-12$, because outflows from highly biased
early galaxies should lie in highly overdense regions at lower redshifts.
Hence the sources and epoch of IGM enrichment remain controversial.

Recent observations of local starbursts have provided new insights into
the nature and impact of supernova-driven winds.  Observations of dwarf
starbursts and luminous infrared galaxies generally find outflows when
detectable \citep{mar05,rup05}, observed as blueshifted \hbox{Na\,{\sc
i}} absorption arising from cold clumps entrained in hot metal-rich
outflowing gas.  While earlier data showed little trend of outflow
properties with host galaxy properties \citep{mar99,hec00}, these more
recent data suggest that massive galaxies with higher star formation
rates (SFR's) drive faster and more energetic winds.  It is of particular
interest here that recent data exhibit trends that are broadly consistent
with theoretical expectations for momentum-driven (or radiation-driven)
winds \citep{mar05,mur05}, thereby providing for the first time some
intuition on the physical mechanisms that drive outflows.  In such a
model, radiation from young stars impinges on dust in the outflow,
which then couples to the gas and propels matter out of the galaxy
\citep{mur05}.  This is somewhat different than the canonical scenario
where the overpressure from local ISM heating causes a bubble that
eventually bursts out of the galaxy \citep[e.g.]{fuj04}.  An advantage of
momentum-driven winds is that, unlike heat, momentum cannot be radiated
away, and hence can plausibly drive winds over large distances.

For our purposes, such scenarios for outflows provide a physically
and observationally-motivated model connecting wind properties with
host galaxy properties, which we can exploit in order to understand
the global impact of outflows on the IGM across cosmic time.  Doing so
requires modeling large-scale outflows within the context of hierarchical
structure formation.  This is the main focus of this paper.  

In this paper we explore the impact of various feedback mechanisms on
IGM metallicity using cosmological hydrodynamic simulations, based on an
explicit implementation of superwind feedback pioneered by \citet{spr03b}
(hereafter SH03).  Our simulations also include metal-line cooling,
which significantly affects the temperature structure of metal-enriched
IGM \citep{agu05}.  Our goal here is to construct an enrichment model
that can reproduce both the metal-line observations of the IGM as well
as the star formation history of the universe.  We focus primarily on
\hbox{C\,{\sc iv}} absorption line observations between $z=1.5-6.0$
since this species is the cleanest tracer of IGM metallicity observable
over this redshift range \citep{son96,hel98}.

This paper is organized as follows. \S2 discusses the simulations
including the modifications we made to \gad, the various feedback
models we run, and the generation of simulated spectra.  \S3 gives an
overview of the global IGM physical properties in our suite of outflow
simulations, while \S4 discusses the physical characteristics of IGM
metals and \hbox{C\,{\sc iv}} absorbers.  \S5 compares our findings to
observations analyzed by Voigt profile fitting of lines and the Pixel
Optical Depth (POD) method.  In \S6 we show that our results are robust
against the effects of numerical resolution.  We present our
conclusions in \S7.

\section{Simulations} \label{sec: sims}

\subsection{Hydrodynamic Simulations of Structure Formation}\label{sec: gad}

We employ the N-body+hydrodynamic code \gad\ \citep{spr05}, which uses a
tree-particle-mesh data structure to efficiently compute gravitational
forces on a system of particles, along with an entropy-conservative
formulation of Smoothed Particle Hydrodynamics \citep[SPH;][]{spr02}
to model the pressure forces and shocks in the gaseous component.
The code is fully adaptive in space and time, enabling simulations
with large dynamic range crucial for studying galaxies together with
large-scale structure.

Additionally, \gad~models a number of physical processes important for
the formation of galaxies.  Star formation is modeled using a subgrid
recipe in which each gas particle above a critical density where
fragmentation becomes possible (calculated from the thermal Jeans mass
based on the local cooling rate) is treated as a set of cold clouds
embedded in a warm ionized medium, similar to the interstellar medium
of our own Galaxy.  The processes of evaporation and condensation are
followed analytically within each particle using the formalism of
\citet{mck77}.  Stars are allowed to form in the cold clouds at a rate
proportional to its density squared.  This produces a disk surface
density-star formation rate relationship that is in agreement with the
relation observed by \citet{ken98}, when a single free parameter, the
star formation timescale, is set to 2~Gyr \citep{spr03a}.  Feedback
energy from Type II supernovae is then added to the hot phase of the
ISM, using an instantaneous recycling approximation.  SH03 found that
the multi-phase model produces self-regulated star formation that does
not suffer from runaway star formation with increasing resolution,
as seen in previous simulations \citep{bal01,dav01}.  However, the
converged star formation rate was still found to be too high when
compared with observations, motivating SH03 to include galactic
outflows as we describe in \S\ref{sec: winds}.

Photo-ionization heating is incorporated based on a spatially-uniform
ultraviolet background taken from \citet{haa01}, described in detail
in \S\ref{sec: ionize}.  \gad~also includes radiative cooling.  In the
simulations run by SH03, the cooling rates were computed assuming
primordial composition with 76\% hydrogen by mass.  In \S\ref{sec:
cooling} we describe improvements to the cooling module and extend it
to include metal-line cooling.

\gad~tracks the metallicity of gas and stellar particles.  Gas
particles that are eligible to form stars continually enrich
themselves with metals based on an assumption of instantaneous
recycling and using a yield of 0.02 (i.e. solar metallicity).  This
yield is roughly what one would expect using supernova yields from
\citet{woo95} in a Chabrier IMF.  When a particle is converted into a
star (which happens in two stages, such that each star particle has
half the mass of its original gas particle), then the star particle
inherits the metallicity of its parent gas particle at the time of
conversion.

\subsection{Radiative Cooling With Metal Lines}\label{sec: cooling}

In the simulations of SH03, cooling is done using an implicit scheme
wherein the cooling timestep is the same as the hydrodynamical timestep
as set by the Courant condition \citep{spr02}.  While in most cases this
is a stable and accurate method, in dense regions surrounding galaxies the
cooling time can be considerably shorter than the sound crossing time, but
still not so rapid so that the particle reaches thermal equilibrium within
such a time.  Since the implicit method uses the cooling rate at the end
of the timestep to evolve the thermal energy over the entire timestep,
it can give inaccurate answers in regions where the slope of the cooling
curve is varying rapidly.  This happens at moderately high densities,
and at temperatures $\sim 10^{4.5}-10^{5.5}$~K~\citep[see][for some
examples of cooling curves]{kat96}.

In order to handle this regime better, we have modified \gad~to
implement a scheme to follow cooling on the cooling timescale.
As with \citet{spr03a}, we assume that the particle's density and
non-radiative rate of change of thermal energy remain fixed during the
dynamical timestep; these assumptions are of course not valid in detail,
but are made to facilitate ease of computation.  We also continue to
assume ionization equilibrium at all times, although the algorithm we
have implemented makes it easier to follow non-equilibrium evolution;
non-equilibrium effects are not expected to be important in the moderately
overdense IGM that will be the focus of this paper.

We compute the cooling time as
 \begin{equation}
  t_{\rm cool} = \epsilon_{\rm cool} {u\over du/dt},
 \end{equation} 
where $u$ is the thermal energy, $du/dt$ is the rate of change of thermal
energy, and $\epsilon_{\rm cool}$ is a tolerance factor that we set to
0.002.  This choice means that in a single cooling timestep a particle
cannot cool away more than 0.2\% of its thermal energy.  We further
limit the cooling timescale such that it cannot be less than 0.2\% the
dynamical timescale, so a particle can take up to 500 cooling timesteps
for each dynamical timestep.  The value of this tolerance parameter was
chosen based on numerous tests on individual particles in the moderate
overdensity, warm-hot temperature regime.

$du/dt$ for a given particle is obtained from a lookup table based
on its density and temperature.  The lookup table is computed for a
given strength of the photo-ionizing background, redshift (for Compton
cooling), and assuming primordial composition.  The strength of the
ionizing background is interpolated to the system redshift from the
\citet{haa01} model, and when its strength has changed by more than 1\%
the lookup tables are recomputed.  The lookup tables are also recomputed
whenever $\Delta z>1$ since the last lookup table computation, which
is important in the early universe when Compton cooling off microwave
background photons is strong.  The rate balance for primordial species
are calculated as described in \citet{kat96}.  If a particle is enriched,
we add cooling due to metal lines as described below.

We advance the particle's thermal energy explicitly based on the cooling
rate computed at the beginning of the cooling timestep, and then recompute
the cooling rate based on the new thermal energy.  If the cooling rate
has changed sign, then the particle has passed an equilibrium point, and
we return to the original state and reduce the timestep until either it
no longer changes sign, or the timestep falls below the minimum value.
If the cooling rate has not changed sign, we advance the thermal energy
using the average of the cooling rates calculated at the beginning and
end of the cooling timestep, thereby preserving second order accuracy.
We continue to advance the particle until it has been evolved over its
dynamical (or Courant) timestep.  

To quantify the difference between the old implicit method and our new
algorithm, consider particles with $n_H\approx 10^{-3}$ (i.e. overdensity
of 100 at $z=2.5$) and $T\approx 10^{4.5}$~K.  By running \gad\ with
the old and new versions of cooling, we found that such a particle has a
mean absolute difference in the final temperature of 1.4\% over a single
hydrodynamical timestep.  At $T=10^{5}$~K it is 0.5\%, and at higher
temperatures it rapidly becomes irrelevant.  These values are small
but non-trivial, and may accumulate over many timesteps.  At higher
densities the effects become more significant: For $n_H=10^{-2}$ and
$T=10^{5}$~K, the mean difference is 8\% (with values discrepant up to
$\sim\times 2$) and it is highly systematic in the sense that the new
method produces temperatures higher by about 7\%.  This depends on the
exact temperature, however, because whether the implicit method over
or underestimates the temperature depends on the sign of the cooling
curve slope at that temperature.  Although the new method increases the
total run time by typically 20-30\%, it seems worthwhile in order to
track the thermal history of particles more accurately in the moderately
shock-heated, moderately overdense regime, since as we shall show (see
Figure~\ref{fig:c4cont}) a substantial amount of \hbox{C\,{\sc iv}}
absorption arises here.

As \gad~tracks gas-phase metallicities, it is possible to use this
information directly to compute the additional contribution to cooling
rates from metal line cooling.  To do so, we employ the collisional
ionization equilibrium models of \citet{sut93} to generate a lookup
table of metal cooling versus temperature and metallicity, obtained
by subtracting their zero-metallicity models from their metal-enriched
cooling curves.  If enriched, a gas particle then experiences additional
cooling from its metals based on a bilinear interpolation within the
metal cooling table.  We also account for the impact of metal cooling
on the multi-phase subgrid ISM model, since the density at which the
fragmentation sets in and star formation begins depends on the cooling
rate.

\subsection{Superwind Feedback}\label{sec: winds}

SH03 found that even with the resolution-converged multi-phase ISM
model for star formation, the global star formation rate predicted
by simulations exceeded observations by $\sim\times 3$.  Hence they
additionally included an explicit model for superwind feedback in order
to reduce the reservoir of gas available for star formation.  In \gad,
particles that are capable of star formation are given a probability of
entering into a superwind based on their current star formation rate.
If a particle enters a superwind, then it is kicked with a velocity given
by $\vw$, in a direction given by ${\bf v}\times{\bf a}$ (which would
yield a polar outflow in the case of a thin disk).  Furthermore, the
wind particle is not allowed to interact hydrodynamically until it has
escaped from the star forming region such that its SPH density is less
than 10\% of the critical density for multi-phase collapse; SH03 find
that the results are not very sensitive to the choice of this value,
so long as the winds escape the dense star forming region.  This is
intended to mimic a free-flowing chimney of gas extending
outside the star-forming region as observed in local starbursts.

In the SH03 prescription there are two main free parameters: The wind
speed $\vw$ and the mass loading factor $\eta$, which is the rate of
material being ejected from the galaxy relative to its star formation
rate.  Following observations by \citet{mar99} and \citet{hec00} and
IGM enrichment considerations from \citet{agu01}, these values were both
taken to be constant in the runs done by SH03, at values of 484~km/s and
2, respectively.  We will call this the {\it constant wind} (cw) model.
This model resulted in a $z=0$ stellar mass density in broad agreement
with observations.

The constant wind prescription, while simple and effective, has some
deficiencies.  Although the nature of the winds from the smallest
protogalaxies are currently unobtainable by observations, using the
same large wind velocities for these galaxies would heat the IGM too
much by $z=3$ to agree with \hbox{C\,{\sc iv}} observations
\citep{agu05}, though metal-line cooling performed self-consistently
may alter that conclusion.  Additionally, we find that this model
has poor resolution convergence in terms of global metal enrichment--
a higher-resolution simulation that resolves small galaxies earlier
will distribute a great deal more metals throughout the IGM at early
times as compared to a lower-resolution run.

Observations of outflows from starburst galaxies have improved
considerably in recent years.  \citet{mar05} found that the terminal wind
velocity scales roughly linearly with circular velocity, with top winds
speeds around three times the galaxy's circular velocity.  \citet{rup05}
studied a large sample of luminous infrared galaxies to find that,
at least when combined with smaller systems from \citet{mar05}, those
trends continue to quite large systems.  It is worth noting that these
observations generally target cold clouds entrained in the hot wind as
traced by \hbox{Na\,{\sc i}} absorption, not the hot wind itself that
carries most of metals.  Still, \citet{hec03} argues that the wind speeds
and mass loading factors are likely to be more accurately inferred from
this cold component owing to the observational difficulty of detecting
X-ray emission from hot gas.

A feasible physical scenario for the wind driving mechanism is deduced by
noting that the observed scaling are well explained by a momentum-driven
wind model \citep{mar05}, such as that outlined in \citet{mur05}.  In such
a scenario, it is the radiation pressure of the starburst that drives the
outflow, possibly by transferring momentum to an absorptive component
(such as dust) that then couples to the bulk outflowing material.
The presence of large amounts of dust surrounding the classic starburst
galaxy M82 \citep{eng06} lends circumstantial support to this type of
scenario.  The momentum-driven wind model provides us with a physically
motivated and observationally constrained way to tie outflow properties
to the star forming properties of the host galaxy.  Furthermore, it
provides the impetus for our approach of tying local observations of
starbursts with outflows across cosmic time.

The idea of using winds observed in local starbursts as a template for
winds at all epochs in all galaxies may seem like quite a leap of
faith.  Locally, starbursts are relatively rare objects, so it is
unclear whether galactic winds are ubiquitous and tied solely to the
galaxy's star formation rate.  However, as \citet{hec00} showed there
is a threshold of star formation surface density of around
0.1~$M_\odot {\rm yr}^{-1} {\rm kpc}^{-2}$ above which winds are
typically seen, and unlike with local star forming disks, virtually
{\it all} star forming galaxies at high redshift satisfy this
criterion because they are more compact and more vigorously forming
stars \citep{hec00,erb06}.  Hence down to $z\sim 1.5$ at least, it is
plausible that essentially all galaxies are forming stars that drive
galactic superwinds.  Indeed, direct observations of Lyman break
galaxies at $z\sim 2-3$ by \citet{pet01} and \citet{sha03} show
outflows of hundreds of km/s.  Of course, there is no guarantee that
outflows from these high-redshift systems follow similar relations as
local starbursts, but as we shall see this Occam's razor assumption
turns out to be remarkably successful.

Guided by local observations, we choose to focus mainly on a class of
models based on momentum-driven winds.  As \citet{mur05} describes, in
such a model the wind speed scales as the galaxy velocity dispersion (see
their eqn.~17), as observed by \citet{mar05}.  Since in momentum-driven
winds the amount of input momentum per unit star formation is constant (their
eqn.~12), this implies that the mass loading factor must be inversely
proportional to the velocity dispersion (their eqn.~13).  We therefore
implement the following relations:
 \begin{eqnarray}
  \vw &=& 3\sigma \sqrt{f_L-1}, \label{eqn: windspeed} \\
  \eta &=& {\sigma_0\over \sigma} \label{eqn: massload},
 \end{eqnarray} 
where $f_L$ is the luminosity factor, which is the luminosity of the
galaxy in units of the critical (or sometimes called Eddington) luminosity
of the galaxy, and $\sigma_0$ provides a normalization for the mass
loading factor.  Here we assume that the final radius that the wind is
driven to is approximately $100\times$ the initial radius, i.e. $r/R_0\sim
100$ in the \citet{mur05} formalism; the dependence on this
term is weak.  Since we are unable to reliably calculate galaxy stellar
velocity dispersions directly in our simulations owing to a lack of
resolution, we approximate $\sigma$ using virial theorem as $\sigma =
\sqrt{-{1\over 2}\Phi}$, where $\Phi$ is the gravitational potential at
the location of the particle being placed into the wind.

We enlist observations and some theoretical considerations to determine
the free parameters $\sigma_0$ and $f_L$.  \citet{mur05} argue that
for a Salpeter IMF and a typical starburst SED, $\sigma_0=300$~km/s.
As we will show, the mass loading factor controls star formation at
early times, so $\sigma_0$ can also be set by requiring a match to the
observed global star formation rate.  It turns out that for our assumed
cosmology, $\sigma_0=300$~km/s broadly matches this constraint as well,
so we will use this value throughout for our momentum-driven wind models.

\citet{mar05} suggests $f_L\approx 2$, while \citet{rup05} find a range
of values $f_L\approx 1.05-2$.  In a momentum-driven wind scenario,
the relevant luminosity is set by the Lyman continuum emission from
the stars that is absorbed by the dust particles that propel the wind.
In such a radiation pressure scenario, stars with lower metallicity that
produce greater Lyman continuum emission (such as those in the early
universe) would be expected to drive stronger winds.  Stellar models by
\citet{sch03b} suggest an approximate functional form for far-UV emission
as a function of metallicity (his equation~1), which we use to obtain
the following relation:
 \begin{equation}
  f_L = f_{L,\odot} \times 10^{-0.0029*(\log{Z}+9)^{2.5} + 0.417694}.
\label{eqn: zfact}
 \end{equation}
For $Z=0.02$ (solar metallicity), the value of the exponent is
zero, making $f_L=f_{L,\odot}$, while e.g. for a metallicity of
$10^{-3}Z_\odot$, $f_L=1.7f_{L,\odot}$.  The metallicity used to determine $f_L$
is that of the gas particle entering the wind.  We typically use
$f_{L,\odot}=2$ from \citet{mar05}, since the galaxies observed in that
sample generally have around solar metallicity (or perhaps a tenth-solar,
which makes little difference).  We will also consider models where we
do not include this low-metallicity boost, and yet other models where
we allow the $f_{L,\odot}$ to randomly vary between $1.05-2$ in accord
with \citet{rup05}.

The form of the mass loading factor $\eta$ is also uncertain.  While
the momentum-driven wind model of \citet{mur05} predicts $\eta\propto
1/\sigma$, observations seem to find little if any trend of $\eta$ with
$\sigma$ \citep{mar99,rup05}.  There does appear to be a large scatter,
so some trends may be hidden in the scatter, and it is worth noting
that constraining the mass loading factor is even more uncertain than
determining outflow speeds.  Hence we use equation~\ref{eqn: massload}
as our fiducial relations, but we will also consider a model where $\vw$
varies according to equation~\ref{eqn: windspeed} and $\eta$ is constant.

Finally, we must circumvent another technical difficulty with
implementing momentum-driven winds within a cosmological simulation.
The winds are generally driven in a manner that continually increases
its velocity out to some decoupling radius as in equation~(15) of
\citet{mur05}; this radius can be quite large, perhaps $\sim 150\hkpc$
for typical winds.  However, our implementation only gives an
instantaneous kick to a gas particle entering a wind, and does not
continue to accelerate it further.  In order to account for this
discrepancy, we optionally give the winds an additional kick
corresponding to the local escape velocity (given by $2\sigma$), so
that the escaping wind will have approximately the desired velocity as
it leaves its galaxy's halo.  This may be an overcorrection, but if
the wind is actively driven to well outside the halo scale radius,
this should be a fairly good approximation.

\subsection{Runs and Outflow Models}\label{sec:runs}

In summary, we consider the following outflow models:
\begin{itemize}
\item {\bf ``nw" model:} No winds.
\item {\bf ``cw" model:} Constant winds: $\vw=484$~km/s, $\eta=2$.
This is the model used in SH03 and many subsequent papers 
using those simulations.
\item {\bf ``zw" model:} $\vw$ from eqn.~\ref{eqn: windspeed}, a constant
$\eta=2$, and $f_L$ from eqn. 4.  Note that this is not formally a
momentum-driven wind model.
\item {\bf ``mw" model:} Momentum-driven winds: $\vw$ and $\eta$ from
eqns. 2 \& 3, but without a $2\sigma$ kick and using $f_L=f_{L,\odot}=2$ (no
metallicity dependence).
\item {\bf ``mzw" model:} Momentum-driven winds as above, with a
varying $f_L$ as given in eqn. 4, $f_{L,\odot}=2$ and with a $2\sigma$ kick.
\item {\bf ``vzw" model:} Like mzw, but where 
$f_{L,\odot}$ is allowed to randomly vary between $1.05-2$, as observed by
\citet{rup05}.
\end{itemize}

\begin{table}
\caption{Simulation parameters}
\begin{tabular}{lcccccc}
\hline
Name$^{a}$ &
$L^{b}$ &
$\epsilon^{c}$ &
$m_{\rm SPH}^{d}$ &
$m_{\rm dark}^{d}$ &
$M_{\rm *,min}^{d,e}$ &
$z_{end}$ \\
\hline
w8n256 & $8$ & $0.625$ & $0.484$ & $3.15$ & $15.5$ & $3.0$\\
w16n256 & $16$ & $1.25$ & $3.87$ & $25.2$ & $124$  & $1.5$\\
w32n256 & $32$ & $2.5$ & $31.0$ & $201$ & $991$    & $1.0$\\
\hline
\end{tabular}
$^a$Additionally, a suffix is added to denote a particular
wind model as described in \S\ref{sec:runs}.\\
$^b$Box length of cubic volume, in comoving $\hmpc$.\\
$^c$Equivalent Plummer gravitational softening length, in comoving
 $\hkpc$.\\
$^d$All masses quoted in units of $10^6M_\odot$.\\
$^e$Minimum resolved galaxy stellar mass.\\
\label{table:sims}
\end{table}

Table~\ref{table:sims} lists parameters for our three simulation volumes
having 8, 16, and 32 $h^{-1}$Mpc (comoving) box lengths.  Each volume
is run for all of the above wind models.  The initial conditions used
are identical for all the wind models, and are generated when the
universe was still well within the linear regime using an \citet{eis99}
transfer function with $\Omega_m=0.3$, $\Omega_\Lambda=0.7$, $h=0.7$, $n=1$,
$\sigma_8=0.9$, and $\Omega_b=0.04$.  Each run has $256^3$ gas and dark
matter particles, with gas particle mass resolutions spanning $5\times
10^5 M_\odot$ to $3\times 10^7 M_\odot$.  Our smallest volume still
doesn't quite resolve the Jeans mass in the high-$z$ IGM \citep{sch00},
but because we will mostly examine \hbox{C\,{\sc iv}} absorption which
arises in moderate overdensity regions, our resolution constraints
are less stringent, as we discuss in \S\ref{sec: resolution}.  We will
mainly focus on the 16~$h^{-1}$Mpc box simulations, because as we show
in our resolution convergence study in \S\ref{sec: resolution}, this is
the largest volume for which we can robustly predict \hbox{C\,{\sc iv}}
observables.

\subsection{Spectral Generation} \label{sec: extract}

From these simulations, we extract the optical depths along random
lines of sight for 25 ions representing 12 species, along with density,
temperature, and metallicity for each atomic species (which differ
because of thermal broadening), all as a function of redshift.  We use
CLOUDY to calculate ionization fractions assuming an optically thin
slab of gas with the given density, temperature, and impinging ionizing
radiation field (see \S\ref{sec: ionize}).  The 25 ions selected have
been previously observed in optical/UV absorption line spectra, even
though many ions are too weak to show up in our simulated spectra, as
well as theoretical expectations for IGM densities and temperatures
\citep{hel98}.  The three ions that are considered in this work are
\hbox{H\,{\sc i}}, \hbox{C\,{\sc iv}}, and \hbox{C\,{\sc iii}}, while
the remainder are included only to simulate chance contamination.
The main contaminants are \hbox{N\,{\sc v}}, \hbox{O\,{\sc vi}},
\hbox{Si\,{\sc ii}}, \hbox{Si\,{\sc iii}}, and \hbox{Si\,{\sc iv}}.
When using the pixel optical depth (POD) method \citep{ell99, agu02,
ara04}, we must generate the contaminants as accurately as possible,
and we use alpha-enhanced abundances ([N/C]=-0.7, [O/C]=0.5, [Si/C]=0.4
where [C/Fe]=0.0), even though this increases the total amount of metals
by about 150\% because half of all metals are oxygen.  For much of our
analysis however, we generate uncontaminated \hbox{C\,{\sc iv}} spectra
and compare to data where contaminants have been removed.  Our goal
in this paper is to explore how simulated observations evolve over our
redshift range, so we leave for the future a more painstakingly detailed
comparison \citep[as done for \lya\ in e.g.][]{dav01b}.

Our software, called {\tt specexbin}, extracts spectra at angles such
that the lines of sight can wrap around the periodic simulation box
and continuously sample different structure.  We generally run 30
lines of sight at 30 separate angles between 10 and 82 degrees (angles
too close to 0 and 90 will keep sampling the same structure) for a
simulation beginning at $z=6$ and ending at the redshift of the final
output (usually $z=1.5$).  We choose 30 lines of sight because we want
a sample size comparable to a survey of quasar spectra achievable
currently or in the near-term future, so that we can calculate
relevant error bars in our plots.  A single line of sight between
$z=6.0\rightarrow 1.5$ will traverse a 16 $\hmpc$ box straight across
approximately 150 times.  We extract from simulation snapshots at
least every $0.25 z$, while varying the ionization background as
described in the next subsection and smoothly accounting for Hubble
expansion.

{\tt specexbin} first calculates the physical properties (gas density,
temperature, metallicity, and velocity) as a function of position along
the line of sight by averaging the contribution of every SPH particle
whose smoothing kernel overlaps the current pixel.  To generate a spectrum
in velocity space (redshift space), one needs to include the effects
of Hubble expansion, bulk motions of the gas, and thermal broadening.
For each ionic species, the ionization fraction is determined at each
position by using a lookup table where the inputs are density and
temperature, assuming ionization equilibrium.  The ionization fraction
is then re-binned into velocity space with the inclusion of thermal
broadening specific to each atomic species.  The oscillator strength of
the line converts this value to an optical depth.  The metallicity for
each atomic species in velocity space is saved along with the optical
depths assuming solar metallicity so the we can apply any desired
metallicity distribution.  We typically take the metallicity directly
from our simulations, but this approach retains the option of applying
an external $\rho-Z$ relation.

We then compute fluxes from the optical depths outputted by {\tt
specexbin}.  We generate 0.05 \AA-resolution spectra, then convolve
with an instrumental profile typical of an Echelle spectrograph such as
HIRES or the Very Large Telescope's Ultraviolet and Visual Echelle
Spectrograph (VLT/UVES), in particular $R\equiv\lambda/d\lambda=43,000$,
and finally add Gaussian noise ($S/N=100$ per pixel).  As alluded to
above, we make two type of spectra: \hbox{C\,{\sc iv}}-only spectra with
only the 1548 \AA~\hbox{C\,{\sc iv}} component, and complete spectra
with all 25 ions included.  We use \hbox{C\,{\sc iv}}-only spectra to
measure \hbox{C\,{\sc iv}} column densities and b-parameters, and to
calculate $\Omega$(\hbox{C\,{\sc iv}}) (see \S\ref{sec: col}).  We use
the complete spectra when we apply the pixel optical depth (POD) method
(see \S\ref{sec: pod}), because contamination can affect the \hbox{C\,{\sc
iv}} flux decrement.  We make multiple spectra for each line of sight,
placing the quasar redshifts such that all portions of the \hbox{C\,{\sc
iv}} forest from $z=6.0\rightarrow 1.5$ are ``observed'' at wavelengths
uncontaminated by the Ly$\alpha$ forest (e.g. $z_{QSO}=$6.0, 5.0, 4.1,
3.4, etc.).

\begin{figure*}
\includegraphics[angle=-90,scale=.66]{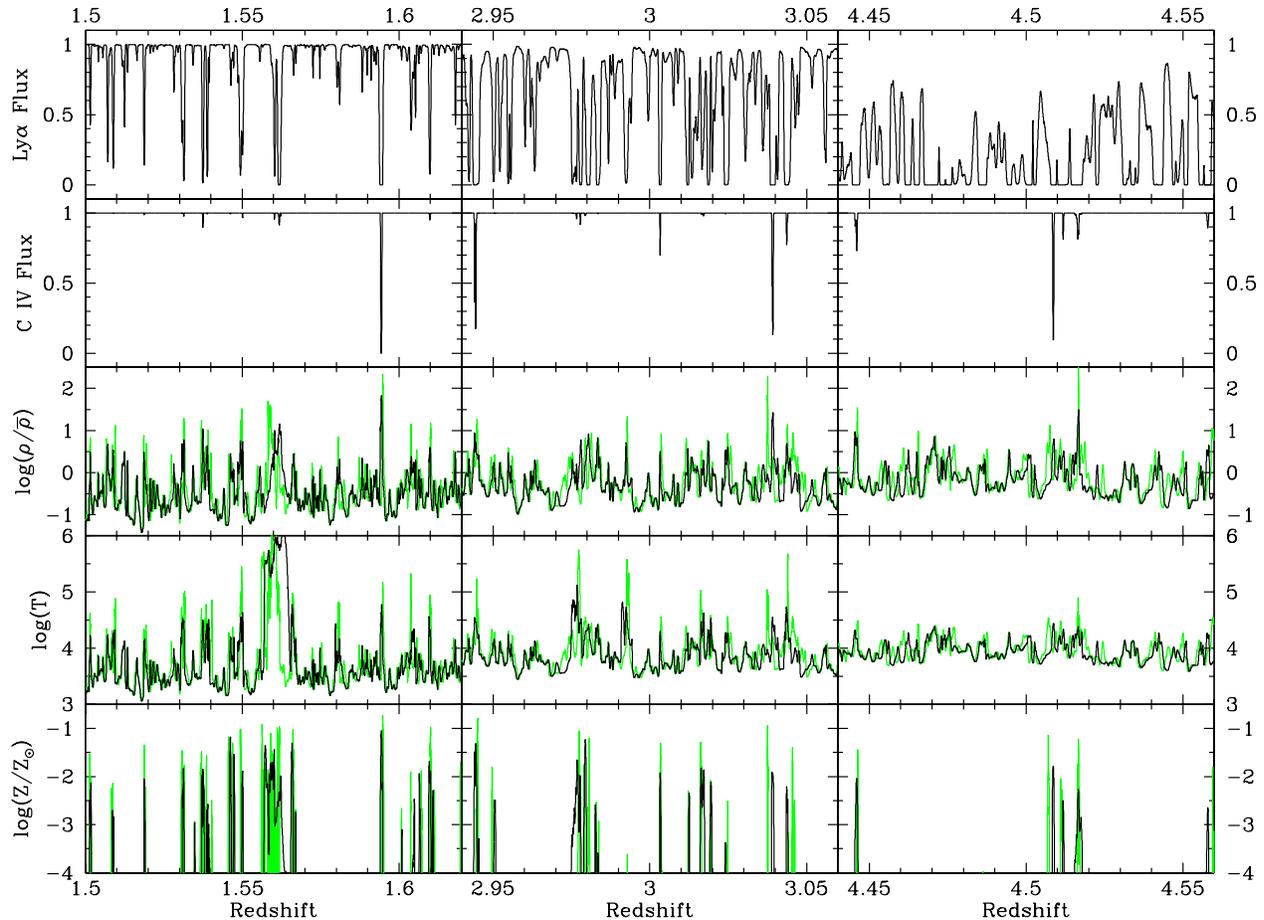}
\caption[]{Example simulated spectra from the w16n256vzw
simulation at three redshifts, $z\approx$ 1.5, 3.0, \& 4.5, showing
Ly$\alpha$ absorption, \hbox{C\,{\sc iv}} absorption, and the underlying
overdensity, temperature and metallicity distributions.  Black lines
show the properties with peculiar velocities and thermal broadening
applied, while green lines show the underlying physical distribution.  }
\label{fig:los_w16n256vzw}
\end{figure*}

Figure~\ref{fig:los_w16n256vzw} demonstrates how Ly$\alpha$ flux,
\hbox{C\,{\sc iv}} flux, density, temperature, and metallicity
appear at three different redshifts in one of our lines of sight.
Some of the trends apparent in this model spectrum relate to the main
conclusions of this paper.  First, the evolution in \hbox{C\,{\sc iv}}
absorption is significantly less than that seen for Ly$\alpha$ due
to the complicated interplay between enrichment, photoionization, and
IGM heating in expanding large-scale structure.  At higher redshifts,
\hbox{C\,{\sc iv}} absorption faithfully traces metal enrichment, but
at lower redshifts this correlation becomes significantly weaker.

Except where noted, we will group individual Voigt-profile fit
\hbox{C\,{\sc iv}} absorbers (i.e. ``components'') into ``systems" if
they lie within 100~$\kms$ of another.  This is done in order to
facilitate a more robust comparison among data of varying quality and
to avoid systematics arising from different line fitting techniques.
For example, in Figure~\ref{fig:los_w16n256vzw}, the \hbox{C\,{\sc
iv}} line at $z=1.594$ is a single component while the lines around
$z=4.1$ form a multi-component system.

\subsection{UV Background}\label{sec: ionize}

We apply a spatially-uniform photoionizing background is taken from
\cite{haa01} to the matter distribution, both during the simulation
runs and during the spectral extraction.  Unless otherwise noted, the
background used is the one that is comprised of quasars and 10\% of UV
photons escaping from star-forming galaxies (referred to as QG), which
turns on at $z\approx 9$.  The slope of the QG background is softer than
the \cite{haa96} quasar-only background short-wards of 1050 \AA~($\alpha =
-1.8$ vs. $\alpha=-1.55$).

The amplitude of the QG background can be constrained by approximately
matching the Ly$\alpha$ flux decrement, $D_{Ly\alpha}$, to observed values
over this redshift range.  To do so, we divide the background amplitude by
1.6 at all redshifts to match the observed mean Ly$\alpha$ flux decrement,
$D_{Ly\alpha}$.  We do not do this during the simulation run, but only in
post-processing as we extract the spectra, but as \citet{cro98} showed,
because photoionization is subdominant in gas dynamics such a post facto
correction yields virtually identical results as having done the entire
run with the lower background.

\begin{figure}
\includegraphics[angle=-90,scale=.32]{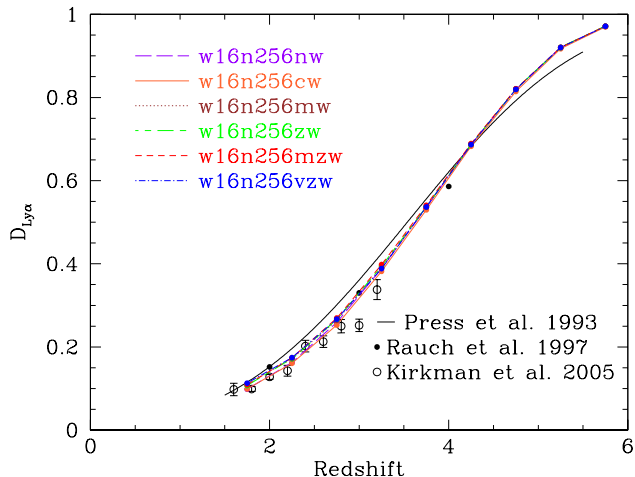} 
\caption[]{Lyman-alpha flux decrement ($D_{Ly\alpha}$) averaged over
30 lines of sight generated from each outflow simulation, using a
HM01 background amplitude divided by 1.6 across all redshifts.
This simple correction results in a simulated $D_{Ly\alpha}$ evolution
generally in accord with observations.  Our outflow
models have little effect on $D_{Ly\alpha}$.}
\label{fig:lyadec}
\end{figure}

In Figure~\ref{fig:lyadec} we compare the mean flux decrement in our
spectra to measurements by \citet{pre93}, \citet{rau97}, and
\citet{kir05}, which have been corrected for metal line contamination.
The \citet{pre93} functional fit and Rauch measurements are meant to
measure the total $D_{Ly\alpha}$ as they apply corrections for their
continuum fitting at high redshift, hence we do not continuum-fit our
spectra for this comparison.  Continuum fitting would have a
non-trivial effect on $z\ga 3$ where the Ly$\alpha$ transmission
almost never reaches 100\% in our spectra.  The difference declines at
lower redshift; for instance, \citet{dav97} find that continuum
fitting to the tops of Ly$\alpha$ peaks results in 5.7\% flux loss at
$z=3$ and 1.2\% at $z=2$.  The data points from \citet{kir05}, which
have been continuum fitted, show deviations in the highest redshift
bins as expected.  Their data agree within reason with our
measurements, except at $z=3.0$ where they find a 22\% lower
$D_{Ly\alpha}$.  Other than this measurement, we find that the HM01
ionizing background simply divided by 1.6 provides a good fit to the
available data.  The fact that we don't need a redshift-dependent
correction indicates that the HM01 background in conjunction with
density growth in our assumed $\Lambda$CDM universe yields the correct
evolution for $D_{Ly\alpha}$.

Interestingly, our outflow models have typically a negligible effect
on the Ly$\alpha$ flux decrements, with the values mostly agreeing to
2\% among the various simulations.  This indicates that winds are not
affecting the density and temperature structure over much of the
Universe's volume, or put another way, the filling factor of winds is
fairly small (as we will show more quantitatively in \S\ref{sec:
globalIGM}).  This is consistent with the results of \citet{the02} and
\citet{ber06}; the latter finding that the only quantitative effect of
galactic winds on the Lyman-$\alpha$ forest is an increase in the the
number of sub-\AA~saturated regions resulting from dense shells plowed
up by the winds.

\section{Global Physical Properties} \label{sec: global}

\subsection{Star Formation Rate Density}

A major impetus for including superwind feedback is to suppress star
formation and solve the overcooling problem.  As SH03 showed, the constant
wind model broadly matches the observed star formation history of the
universe, the so-called Madau plot \citep{mad96}.  Here we examine Madau
plots for our various wind models to ensure that they fall within the
observed range as well.

\begin{figure}
\includegraphics[angle=-90,scale=.32]{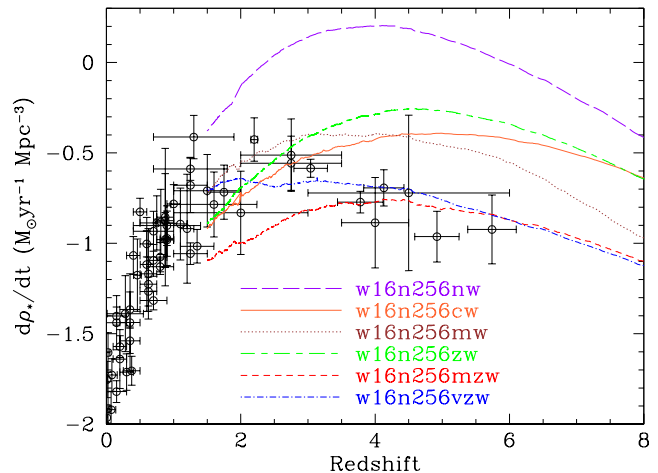}
\caption[]{The total integrated star formation history compared to
observations compiled by \citet{hop04}.  Wind models suppress star formation
from the no-wind case (long-dashed) to bring then into broad agreement
with data.  The mass loading factor controls early star formation,
while wind speeds become more important later on.  Outflow models with
large $\eta$ at early times suppress star formation, resulting in 
a peak in global star formation at later times.
}
\label{fig:sfr}
\end{figure}

Figure~\ref{fig:sfr} shows the total star formation rate density as
a function of redshift averaged over our three simulation volumes,
for all our wind models.  Data points shown are from a compilation by
\citet{hop04}.  We include all star formation in each volume, without
selecting out any particular galaxy population.  In general, all our
wind models roughly fall within the observed range down to their lowest
redshift ($z=1.5$ or $z=2$), with possible discrepancies versus $z\ga 3$
data that are more poorly constrained.  With no wind feedback, far too
many stars are produced as found by SH03, while our various wind models
suppress star formation at $z\sim 3$ by $\sim\times3-5$ relative to
no winds.  Hence all our wind models broadly pass the Madau plot test.

More detailed examination reveals that there are some differences
between wind models that yield insights into the impact of varying
feedback parameters.  For instance, in models where we have a variable
$\eta$ (mzw,mw,vzw) that produces high mass loading factors at early
times, we end up with a global star formation history that is peaked
towards later epochs ($z_{\rm peak}\approx 2-4$) than in the constant
$\eta=2$ cases (cw,zw) which show $z_{\rm peak}\approx 5$.  This
arises because the high early mass loading factors in small early
galaxies keep gas puffy and warm, suppressing early star formation
despite the fact that the wind velocities are low.  As a case in
point, note the vzw and mzw models show virtually identical star
formation histories at high redshift despite having significantly
different (mean) values of $\vw$, showing that it is indeed $\eta$
that primarily governs early star formation.  The wind speed is not
irrelevant, though, as that represents the only difference between vzw
and mzw, and they show noticeable differences at later epochs.  There,
the universe becomes less dense, winds travel farther, and cooling
rates are lower, so the wind speed $\vw$ increasingly governs the
global star formation rate.

At face value, variable $\eta$ models appear to be in better agreement
with observations that show $z_{\rm peak}\sim 2-3$, particularly the
vzw model which shows a peak at $z\sim 2$.  However, since we are not
selecting galaxy populations in the simulation analogous to observed
samples from which the data are calculated, such comparisons are at best
preliminary.  For now, we simply note that the feedback recipe can have
significant impact on the global star formation history, including the
peak of the star formation rate density in the Universe, and all wind
models we consider here fall broadly within the allowed range.

\subsection{Outflow Properties}

As outflows represent the main new feature of our simulations, it is
worth examining their properties in more detail.  Here we highlight
the differences between wind models in their typical speeds, mass
loading factors, and energy deposition rates as a function of redshift.
These provide useful background information for understanding the behavior
of \hbox{C\,{\sc iv}} absorption in our various outflows scenarios.

\begin{figure*}
\includegraphics[angle=-90, scale=0.64]{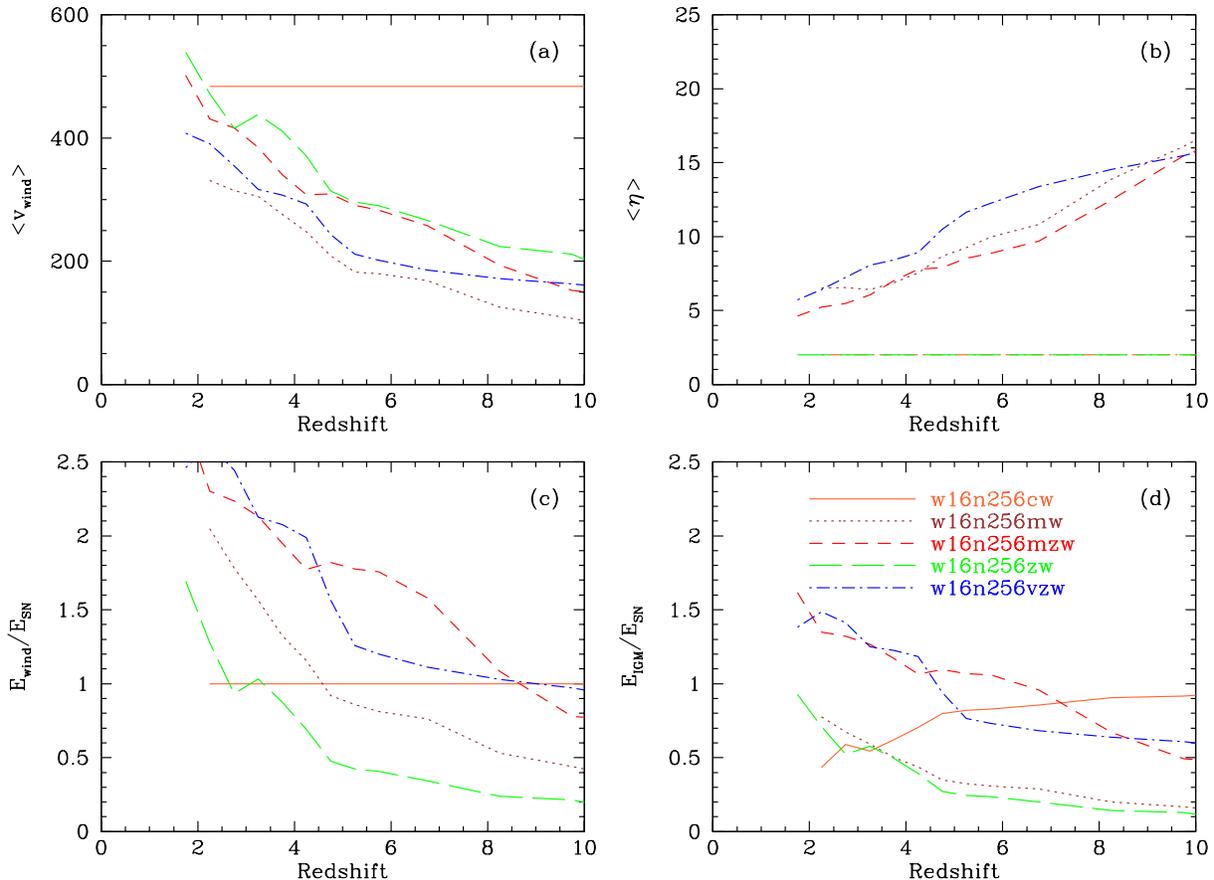}
\caption[]{Wind property averages as a function of redshift for
various outflow models.  Panels (a) and (b) show the average wind
velocity and mass loading factor, respectively.  Panel (c) shows the
energy inputted into the wind particles, and panel (d) shows the
energy injected into the IGM, after we subtract the energy needed to
leave the potential well of the galaxy.  The energies are normalized
by the energy injection from supernovae.  The wind speeds and mass
loading factors at $z\sim 2-3$ in momentum-driven wind models are in
agreement with observations by \citet{pet01,erb06}.  The energy input
can significantly exceed that from supernovae in our momentum-driven
wind models, but this is physically plausible since radiation energy
is the primary driver of momentum-driven winds.}
\label{fig:winds}
\end{figure*}

Figure~\ref{fig:winds} (upper left) shows the mean wind speed as a
function of redshift for our various wind models.  The zw and mzw models
show similar wind speeds because they both employ equation~\ref{eqn:
windspeed}, while the mw model shows lowered wind speeds primarily
because it lacks the $2\sigma$ kick.  The mean vzw wind speed is more
in line with the mw model at $z\la 6$, but vzw produces a wider range
of wind velocities due to its random assignment of $f_{L,\odot}$,
and hence ends up being quite different than mw in many other
ways.  The boost from low-metallicity stars (not present in mw)
is only important at very early times, because stars enrich themselves
fairly quickly \citep{dav06}, and once the metallicity exceeds around 1\%
of solar the boost becomes fairly small.  The mean wind speed when we
employ equation~\ref{eqn: windspeed} becomes comparable to the constant
wind case (horizontal line) at $z\sim 2$.  It is worth noting that the
wind speeds at $z\sim 3$ in all our models are comparable to the range
of outflow velocities observed in Lyman break galaxies \citep{pet01}.

The upper right panel of Figure~\ref{fig:winds} shows that allowing mass
loading factor dependence according to equation~\ref{eqn: windspeed}
results in much more material being ejected from early galaxies.
Of course, because the wind speeds are small and material is infalling
gravitationally, the winds do not reach very far before being re-accreted.
However, they do keep the gas somewhat hotter and more diffusely
distributed around galaxies, so the high mass loading factors are
effective at suppressing early star formation.  Note that \citet{dav06}
found that such suppression is necessary in order to match observed $z\sim
6$ galaxy luminosity functions.  At low redshifts, \citet{erb06} estimated
a required typical mass loading of around 4 in order to understand the
trends observed in the galaxy mass-metallicity relation at $z\approx 2$;
our momentum-driven winds broadly agree with this.

In the bottom two panels of Figure~\ref{fig:winds}, we show the kinetic
energy injected by the winds (left), and the amount of energy that reaches
the IGM once subtracting off the energy required to leave the potential
well and enter the IGM (right).  We normalize these quantities to the
average supernova energy for a Salpeter IMF, namely $4\times 10^{48}$~ergs
per $M_{\odot}$ of star formation (the fiducial value used by SH03).
The cw model was constructed by SH03 to return 100\% of the supernova
energy into kinetic feedback (although thermal feedback was included
in addition), while the energy input from the momentum-driven wind
models increases as galaxies grow.  Eventually, the energy inputted by
winds exceeds the supernova energy; this is physically possible because
these winds are driven by UV radiation from massive stars over their
entire lifetimes.  An important feature of momentum-driven winds is
that deeper potential wells in larger galaxies do not inhibit feedback
of energy into the IGM, in accord with observations by \citet{rup05}.
The reverse is true for the cw model, where the energy injection into the
IGM is quite high at high-$z$ and declines to lower redshift, which as
we shall see leads to excessive heating of the IGM \citep[see \S\ref{sec:
globalIGM} and][]{agu05}.  Indeed, \citet{fer00} showed that metal
enrichment via supernova-driven winds alone is energetically insufficient
to enrich the IGM to the observed levels, and another unknown mechanism
was required; momentum-driven winds provide such a mechanism.

\begin{figure*}
\includegraphics[scale=.80]{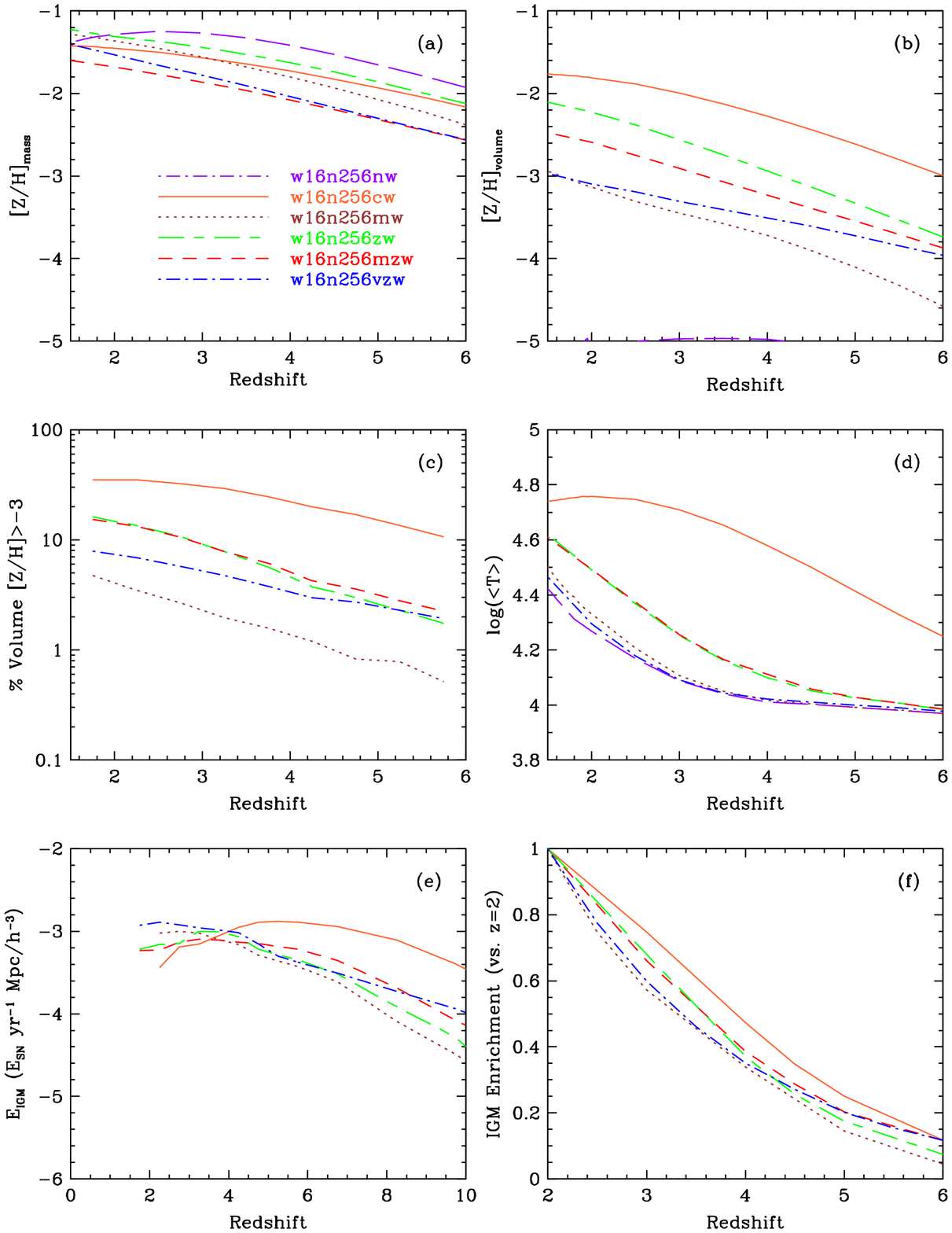}
\vskip -0.5in
\caption[]{IGM properties averaged over the simulation box for various
outflow models as a
function of redshift.  Panel (a) displays the global mass-weighted
gas-phase metallicity, a quantity proportional to stellar mass except
that the metals locked up in stars are not included.  Panels (b) shows
the volume-weighted metallicity, which is governed more by outflow
speeds than by metal production, as the relative trends between models
follow that seen in Figure~\ref{fig:winds}a.  Panel (c) shows the
fractional volume of the simulation enriched to over [Z/H]$>-3$, which
again is broadly governed by outflow speeds.  Note that the no-wind
cases is unable to enrich more than 0.1\% of the volume at any epoch.
Panel (d) shows the volume-weighted temperature, showing that constant
winds dramatically enhances IGM temperature, while momentum-driven
wind models heat the IGM substantially less.
Panel (e) shows the amount of energy reaching the IGM
in terms of supernova energy per year, per $\hmpc$; this is
essentially a cumulative version of Figure~\ref{fig:winds}d.  Panel
(f) shows the amount of metals in the diffuse IGM
($\rho/\bar\rho<100$) relative to the $z=2$ IGM metallicity; an
increase of $\sim\times10$ from $z=6\rightarrow 2$ is typical.}
\label{fig:globalIGM}
\end{figure*}

\subsection{IGM Enrichment and Heating} \label{sec: globalIGM}

Winds add metals and energy simultaneously to the intergalactic medium,
and both quantities affect the characteristics of metal absorption in
the diffuse IGM.  In this section we examine how outflow parameters
affect the evolution of metallicity and temperature in the IGM.

Figure~\ref{fig:globalIGM}, upper left, plots the mass-averaged
metallicity in gas in our various runs.  These curves track the cumulative
amount of star formation in each run, with the slight difference that
metals locked up in stars are not included in this plot.  Consistent with
expectations from Figure~\ref{fig:sfr}, the no-wind case has the largest
amount of metal mass, while mzw and vzw have the smallest global metal
production.

Comparing that plot to the volume-averaged metallicity in the upper
right panel shows some interesting differences.  Note that when
quantifying IGM metallicity using absorption lines, it most directly
traces the volume-averaged metallicity.  In our outflow models, the
amount of metals that enters into the IGM depends on a complicated
interplay between the mass of metals (or stars) formed, the wind
speeds, and the mass-loading factor.  The large wind speeds in the
constant wind case, together with relatively vigorous star formation,
produce the most widely distributed metals, with a mean metallicity at
$z\sim 3$ of $\approx 1$\%~solar.  The difference between zw and mzw,
which have similar wind speeds, arises because the low mass loading
factor in zw yields more star formation and hence more metals.  The
lower wind speeds in mw and vzw produce relatively low volume-averaged
metallicities.  Canonical values inferred from \hbox{C\,{\sc iv}}
observations at $z\sim 3$ suggest [C/H]$\approx -2.5$
\citep[e.g.][]{son96,dav98}, which is broadly consistent with all the
wind models.  We will engage in more careful comparisons to
observations in \S\ref{sec: observe}.

The no-wind model does not distribute metals into the IGM hardly at
all ([C/H]$_{volume}$ never exceeds -5.0), leaving most of its volume
pristine, and showing that dynamical stripping of enriched gas owing
to interactions cannot enrich the diffuse IGM.  Our results are in
qualitative agreement with high-resolution Eulerian simulations by
\citet{gne97} with no winds, having spatial resolution comparable to
our runs but in much smaller volumes.  They found a volume-averaged
metallicity of $\approx 3\times 10^{-4}$ at $z=4$, which is an order
of magnitude higher than what we find but is still much too low to
enrich the diffuse IGM to the observed levels.

The two middle panels show the filling factor with metallicity greater
than one-thousandth solar (left) and the volume-averaged temperature
(right).  The trends in these panels are related.  In order to expel
metals to large distances and fill volume, high wind speeds are required
that results in high IGM temperatures.  Hence the trends are similar among
our wind models in the two plots: Constant winds produce high filling
factors and large temperatures, zw and mzw are virtually identical,
while the mw and vzw models fill less of the volume and hardly heat the
IGM much above the primarily photoionization-established temperatures
seen in the no-wind case.  Interestingly, the spread of velocities in the
vzw model both enrich the IGM while injecting relatively little energy,
which as we will show turns out to be favored by observations.

The volume filling factor increases with time in all models because
(1) metals have escaped their potential wells and are coasting away
from their parent galaxies, and (2) galaxies are forming in less
biased regions at later times.  The volume filled in the cw model
begins to asymptote as hierarchical buildup occurs and metals ejected
by small galaxies falls back into larger halos.  The volume will
eventually asymptote for the other models, but at lower redshift since
the smaller enriched volumes around galaxies will not overlap until
later, and because the peak star formation is shifted to later epochs.

As a side note, \citet{sch00} demonstrated that at $z\sim3$, the
temperature of the IGM as traced by Ly$\alpha$ absorption line widths
is $\approx 1.5-2\times 10^4$ K.  This temperature is above that
expected for pure photoionization \citep[e.g.][or alternatively our
no-wind case]{her96}, which they interpret as arising from latent heat
owing to \hbox{He\,{\sc ii}} reionization.  Our mzw and zw models
broadly agree with the \citet{sch00} measurement, but in our case the
excess heat is due to outflow energy deposition.  Hence outflows can
produce elevated temperatures without requiring \hbox{He\,{\sc ii}}
reionization \citep{sok03}.  We leave more detailed studies of
Ly$\alpha$ line widths for future work.

The lower left panel of Figure~\ref{fig:globalIGM} shows the energy
injected into the IGM per year per $\hmpc^3$, in terms of amount of
supernova energy produced.  This energy injected is proportional to
$\eta\times$SFR$\times v^2_{\rm IGM}$ where $v_{\rm IGM}$ is the velocity
after leaving the potential well of the galaxy.  The large amounts of
early feedback energy in the cw model explains how its IGM becomes so
hot by $z=6$.  The momentum wind models peak in their energy output at
$z=2-4$, with weaker winds peaking later.  We will show later that this
heat input results in significant variations in the global ionization
fraction of \hbox{C\,{\sc iv}}, which is a key ingredient in understanding
\hbox{C\,{\sc iv}} evolution.

The lower right panel shows the amount of metals in the IGM where
$\rho/\bar\rho<100$ relative to that at $z=2$.  This shows that in all
models, the amount of IGM metals at $z=6$ is less than one-eighth of
that at $z=2$.  In other words, in all of our models the vast majority of
metals are injected into the IGM during $z=6\rightarrow 2$, rather than
at $z\ga 6$.  Note that we do not include any ``pre-enrichment" from
exotic star formation at early times; our models are instead intended
to test whether normal star formation in ordinary galaxies can enrich
the IGM with outflow models included.

Overall, these results highlight the importance of the oft-overlooked
connection between metal and energy input into the IGM.  The mass-loading
factor $\eta$ roughly governs the amount of total metal production
at early times, while the outflow velocity $\vw$ mostly determines the
physical extent to which the diffuse IGM ($\rho/\bar\rho<10$) is enriched
and heated.  As we shall see, the complicated interplay between metal
production rate, wind speed, mass loading factor, and radiative cooling
makes observations of IGM metallicity a highly sensitive probe of
the physics of large-scale outflows.

\section{Physical Properties of Metals and \hbox{C\,{\sc iv}} Absorbers}
\label{sec: phys}

\subsection{Evolution of Metals in the IGM} \label{sec: evo}

\begin{figure*}
\includegraphics[scale=1.00]{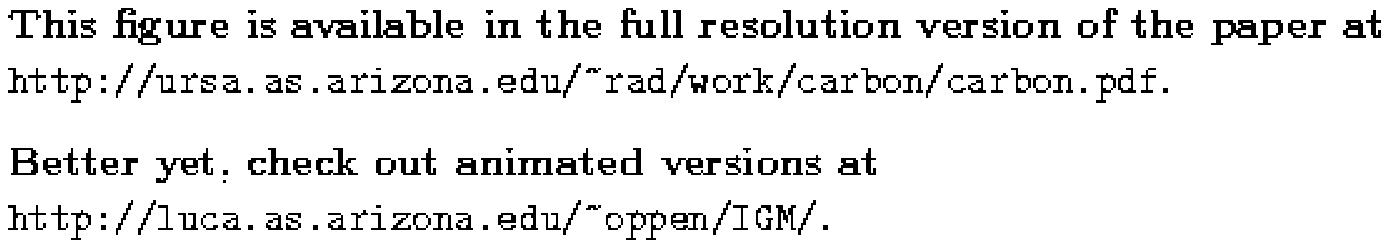}
\caption[]{A 100 $\kms$ slice of the w16n256vzw simulation centered on
a growing galaxy group at $z=$ 4.5, 3.0, and 1.5.  In addition to the
well-understood growth of large-scale structure (overdensity), the
middle panels show that metal enrichment is tied with the early
heating of the IGM, especially at high-$z$ before virialization
dominates at low redshift.  \hbox{C\,{\sc iv}} absorption traces nearly all
metals at high-$z$, and then declines to trace only the most overdense
structures at low-$z$ as the overall ionization correction rises.
Close inspection of the absorption structures indicates an evolution
from smooth and diffuse to clumpy knots tracing the higher
overdensities at low-$z$.}
\label{fig:visuals}
\end{figure*}

Figure~\ref{fig:visuals} shows density, temperature, metallicity, and
\hbox{C\,{\sc iv}} absorption at $z=$4.5, 3.0, and 1.5, in $15\times15$
$\hmpc$ slices that are 100 $\kms$ wide, from the w16n256vzw model.
We have deliberately centered the velocity slice on an overdense region
(both in density and \hbox{C\,{\sc iv}} absorption) to show the large
variety of structures formed.  This figure illustrates some of the trends
that will be quantified in upcoming sections\footnote{For movies of
this evolution, see {\tt http://luca.as.arizona.edu/\~{}oppen/IGM/}.}.

The growth of large-scale structure, dominated by gravitational
instability, is evident in the gas density snapshots (top panels).
Outflows increase the metal filling factor while growing the metallicity
level in previously enriched regions.  Bubbles of shocked gas grow
around star forming systems and trace the filamentary structures that
house galaxies.  In the high-redshift snapshot, much of the heating
is due to winds, as the hot bubbles trace precisely where the metals
appear.  Later, as a proto-group forms in the center of this region,
the temperature also tracks the virialization of gas in the growing
potential well.  Interestingly, cold mode accretion \citep{ker05} is
evident along the dense filaments feeding galaxies; hence despite the
heat input, outflows do little to stop rapid accretion at high redshifts,
because they tend to flow into lower density regions.

The \hbox{C\,{\sc iv}} absorption is shown in front of a backlit
screen to highlight the morphology of absorption, such as if our
Universe was infinite and static \citep{olb26}.  In reality, quasars
provide the backlight, and any quasar only probes a single pixel.
At $z=4.5$, nearly all metals show \hbox{C\,{\sc iv}} absorption when
the ionization fractions are highest in the IGM.  $\tau$(\hbox{C\,{\sc
iv}})$\sim0.05-0.10$ traces IGM gas with $\rho/\bar\rho=3-10$ and
$\tau$(\hbox{C\,{\sc iv}})$>0.10$ traces $\rho/\bar\rho=10-30$ between
$z=4.5-6.0$, a time when the vast majority of these overdensities remain
unenriched by our prescribed winds.  By $z=3.0$, metals have enriched
the filaments, which show up as strong absorbers connecting the growing
galaxy groups.  By $z=1.5$, much of the diffuse IGM carbon is ionized
to higher states, and \hbox{C\,{\sc iv}} preferentially traces higher
overdensity structure.

\begin{figure*}
\includegraphics[angle=-90,scale=.64]{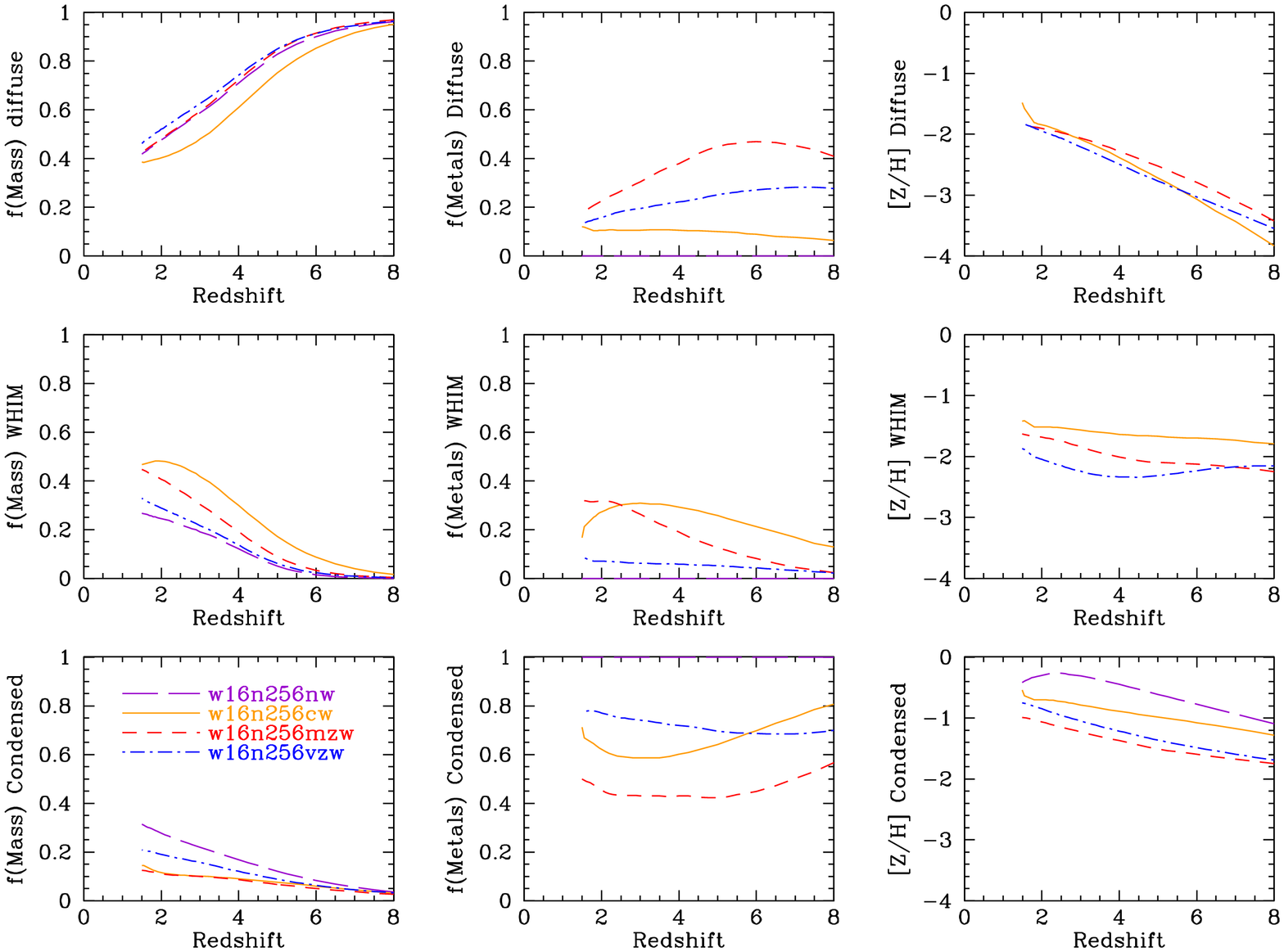}
\caption[]{The fractional mass, fractional metals, and mass-weighted 
metallicity of the IGM (from left to right) subdivided into diffuse IGM
($\rho/\bar\rho<100$, $T<30,000$ K), warm-hot IGM (WHIM,
$\rho/\bar\rho<100$, $T\ge30,000$ K), and condensed IGM
($\rho/\bar\rho\ge100$) (from top to bottom).  The condensed component
includes stellar mass and metals, although the [Z/H] measurement is
just for the gas.}
\label{fig:phaseIGM}
\end{figure*}

Figure~\ref{fig:phaseIGM} shows the mass and metal evolution in the
diffuse IGM ($\rho/\bar\rho<100$, $T<30,000$ K) in the top panels, the
warm-hot IGM (WHIM, $\rho/\bar\rho<100$, $T\ge30,000$ K) in the middle,
and the condensed IGM ($\rho/\bar\rho\ge100$, includes stars) on the
bottom for the cw, mzw, vzw, and nw models.  The mass fraction in diffuse
gas falls in all models from $z=6\rightarrow 2$, as mass transitions
mostly to the WHIM phase resulting from both shock heating on large-scale
structure and energy feedback.  Not surprisingly, the WHIM fraction
follows the mean temperature evolution in Figure~\ref{fig:globalIGM}.
The no wind model provides a baseline for the amount of WHIM formed purely
from growth of structure.  At $z=2$, the cw model has a WHIM fraction of
50\%, double the nw case, indicating half the WHIM results from feedback,
and even more at higher redshift.  The WHIM formed by energy feedback
is moderate in the mzw case, and slight in the vzw model, thus leading
to our conclusion that the fraction of WHIM formed by feedback depends
most on wind speed.  Meanwhile, the condensed phase understandably grows
the fastest in the nw case, while the vzw model clearly distinguishes
itself from other two feedback models shown here by producing more
condensed matter resulting from low wind velocities unable to escape
their parent haloes.

Both wind parameters, $\vw$ and $\eta$, have a significant effect on
the fraction of metals in various phases, especially compared to the
nw case where virtually no metals leave the condensed phase.  The cw
model has a significant amount of metals in the condensed phase due to
its lower mass loading factor, but the metals that do escape into the
IGM usually are shocked to WHIM temperatures.  In momentum-driven winds
(vzw and mzw), although $\eta$ is larger which suppresses star formation,
the lower wind speeds allow more metals to return to the condensed phase,
more so in the vzw case.

The mean gas metallicities (minus the stars in the condensed phase) are
plotted in the right set of Figure~\ref{fig:phaseIGM}.  The condensed gas
has a slowly-evolving metallicity of 5-20\% solar in all wind models at
these epochs at $z<4$.  This enrichment level and evolution is similar
to that seen for damped Ly$\alpha$ systems \citep{pro03}, which are
expected to arise in condensed gas.  In contrast, diffuse gas shows a
steadily increasing metallicity in all models, as star formation and
winds combine to drive metals into the diffuse IGM.  Interestingly, all
wind models show similar diffuse phase metallicities and mass fractions,
a result of the self-regulating nature of feedback (i.e. more feedback
curtails star formation and metallicity enrichment).  The WHIM at high
redshift is primarily the result of feedback and is more enriched than
the diffuse IGM.  As virialization forms more WHIM at lower redshift,
the WHIM metallicity becomes more similar to that of the diffuse IGM.

These figures illustrate that winds polluting the IGM also affect the
temperature structure of the IGM, making \hbox{C\,{\sc iv}} absorption
an evolving tracer of metallicity.  Next we quantify these trends from
the perspective of \hbox{C\,{\sc iv}} absorbers.

\subsection{\hbox{C\,{\sc iv}} Absorption in Phase Space} \label{sec: phase}

Currently the only observational probe we have of diffuse high-redshift
IGM gas are one-dimensional quasar absorption line spectra skewering
a complex matter distribution.  In this section we present physical
properties of the underlying gas for \hbox{C\,{\sc iv}} absorption in
our simulated spectra, in terms of the cosmic phase space (overdensity
\& temperature) of absorbing gas.  We focus on the cw, mw, mzw, and vzw
models as a representative range of wind scenarios: cw displays abundant
early energy and metal injection, mw has comparatively little metals
injection and almost no increased temperature compared to the no-wind
case (see Figure~\ref{fig:globalIGM}), while mzw and vzw represent two
intermediate cases that, as we shall see, are our favored models, with
vzw being slightly preferred.  Hence for illustrative plots we shall
utilize the vzw model, with the understanding that the qualitative trends
are similar in other models unless otherwise noted.

\begin{figure*}
\includegraphics[angle=-90, scale=0.65]{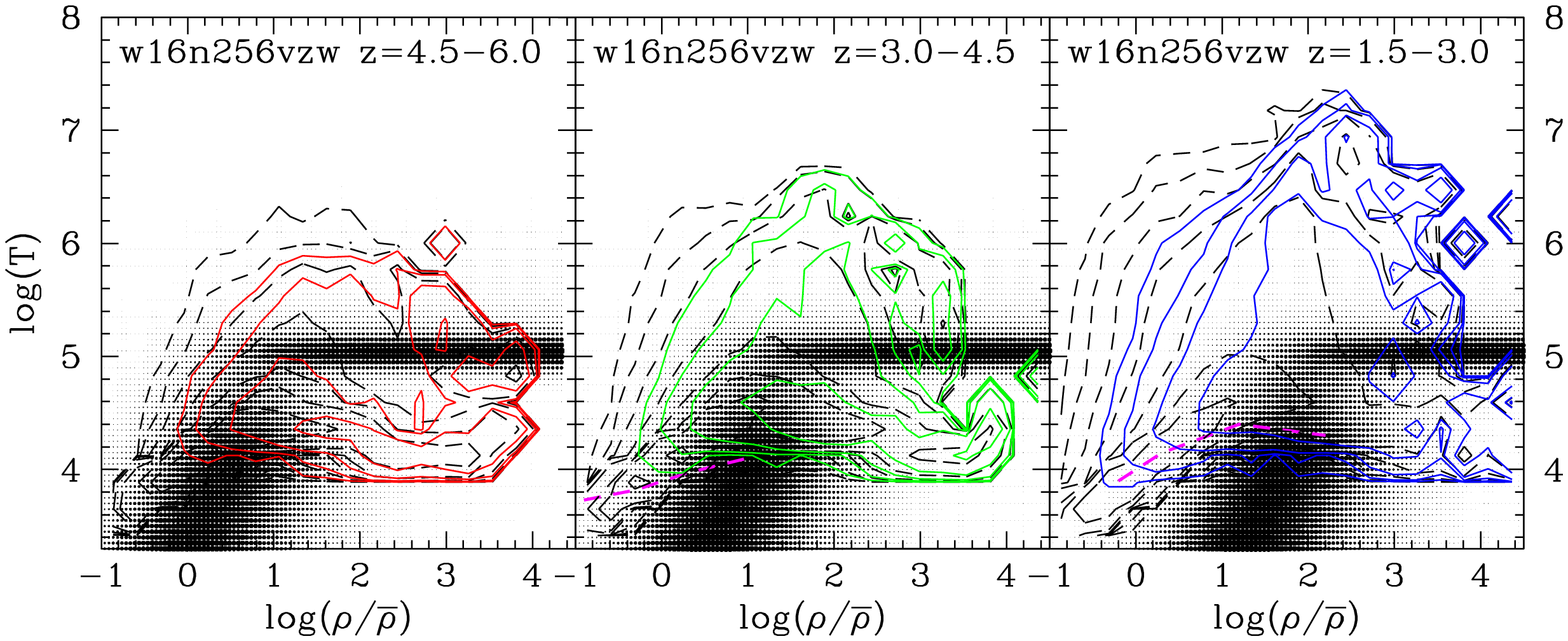}
\caption[]{Solid (colored) contours show metal mass, while dashed contours
show total mass plotted in $\rho-T$ phase space.  Shading corresponds
to the ionization fraction in \hbox{C\,{\sc iv}}.  These contours
have 0.5 dex steps and are normalized across the three redshift bins.
The shading goes from $0\rightarrow 40$\% \hbox{C\,{\sc iv}} ionization
fraction, where the highest fraction occurs in collisionally ionized
regions ($T\sim10^5$ at $\rho/\bar\rho\ga 10$).  The magenta dashed
line in the lower two redshift panels is the $\rho-T$ relation used by
\citet{sch03} to derive their metallicity-density-redshift relationship.
Metals occupy a large region of phase space leading to a variety of
ionization conditions that evolve with redshift.  }
\label{fig:metcont}
\end{figure*}

Figure~\ref{fig:metcont} shows the ionization fraction as a function of
density and temperature, in three different redshift intervals for the
w16n256vzw model.  The ionization fraction shows two distinct regions: A
horizontal band corresponding to collisional ionization around $10^5$~K,
and a diagonal band at lower temperatures and densities corresponding to
photoionization.  Dashed contours overlaid on top of these shaded regions
trace the total mass distribution from the vzw model.  \hbox{C\,{\sc iv}}
absorption would be visible where these contours overlap the shaded
regions, if metals are present.  Metallicity (solid) contours that overlap
shaded regions represent areas that are in fact traced by \hbox{C\,{\sc
iv}} for the vzw model.

One interesting result from this figure is that \hbox{C\,{\sc iv}} is
{\it not} an optimal tracer of photo-ionized IGM gas, particularly at
lower redshifts.  \hbox{C\,{\sc v}} (40.3\AA) would be a better
tracer, but such an observation requires X-ray spectroscopy far beyond
the current capabilities.  Conversely, there is significant amounts of
gas heated to around $10^5$ K in this and the other models, in which
\hbox{C\,{\sc iv}} should be prominent, even at high redshifts.  The
number of actual observed systems as a function of density is governed
by an additional factor, namely the physical cross-section of the
absorbing gas; this will be greater at lower densities, which is why
weak \hbox{C\,{\sc iv}} absorption is still more common despite having a
relatively low \hbox{C\,{\sc iv}} ionization fraction.

Comparing the solid and dashed contours also shows that in the diffuse
IGM, which holds the majority of baryons and the vast majority of volume
at these epochs, contains few metals.  Hence the filling factor of
metals is small (as seen in Figure~\ref{fig:globalIGM}), especially
at high-$z$.  The magenta dashed line in the two lower redshift
bins is the density-temperature relationship used by S03 to derive
their mass-metallicity relation, obtained by assuming that the metals
occupy the diffuse IGM and can be described by a 1-dimensional $\rho-T$
relationship corresponding to the IGM equation of state.  Their $\rho-T$
relation is flatter because they assume substantial latent heat from
\hbox{H\,{\sc i}} reionization at $z\sim 6$; we did not include this
because latest results from the Wilkinson Microwave Anisotropy Probe
show that the Universe was predominantly reionized out to $z\sim 10$
\citep{pag06}, and our \citet{haa01} model assumes reionization at
$z\approx 9$.  The temperatures they use result in a global ionization
correction that does not change much over their redshift range, leading to
their conclusion that their lack of evolution in the \hbox{C\,{\sc iv}}
systems implies little redshift evolution in the true IGM metallicity.
From this plot, it is clear that our self-consistent enrichment models
predict \hbox{C\,{\sc iv}} absorbers lying in a wider range of 
densities and temperature, and are not consistent with the assumption
that \hbox{C\,{\sc iv}} almost always arises in a photoionized phase.
We quantify this in the next section.

\begin{figure*}
\includegraphics[scale=0.80]{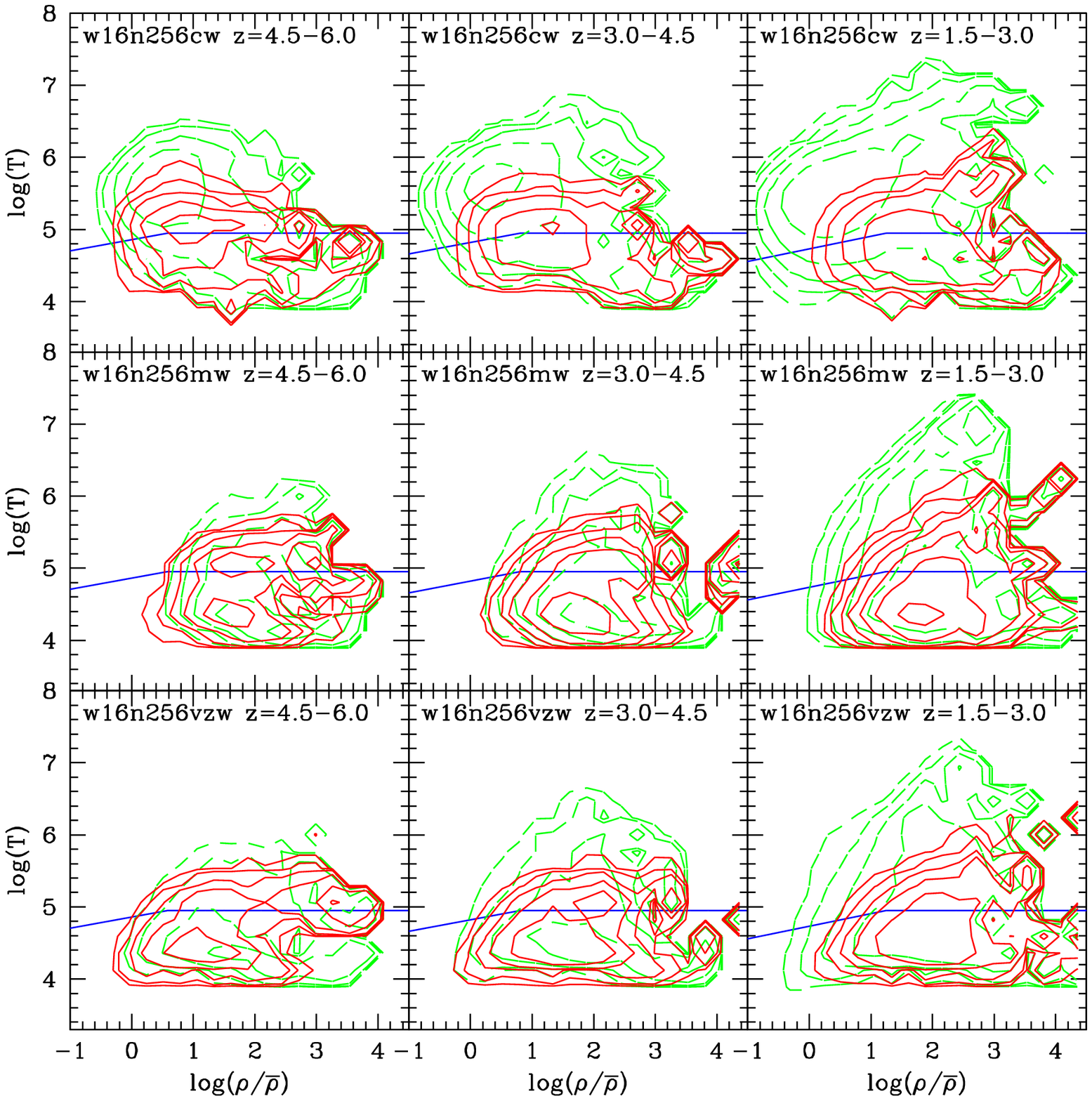}
\vskip -2in
\caption[]{Dashed contours corresponding to metallicity mass (similar
to those in Figure~\ref{fig:metcont}) and solid contours corresponding
to the summed \hbox{C\,{\sc iv}} optical depth (derived from lines of sight)
in $\rho-T$ phase space for three different redshift from the
w16n256cw, w16n256mw, and w16n256vzw models.  The blue line divides
regions dominated by collisional ionization (top) and photo-ionization
(bottom).  The cw model predominantly enriches the low density IGM
while causing significant shock-heating, whereas the mw model enriches
mostly higher overdensities and heating occurs mostly due to
virialization at low redshift.  The vzw model manages to enrich a
variety of overdensities while causing little more shock heating than
the mw model.  }
\label{fig:c4cont}
\end{figure*}

Figure~\ref{fig:c4cont} compares the metal mass contours from the
previous plot (shown as dashed contours) with the distribution of
the actual \hbox{C\,{\sc iv}} absorption (solid), in the $(\rho,T)$
plane, for the cw, mw, and vzw models (top, middle, and bottom panels,
respectively).  We obtain the \hbox{C\,{\sc iv}} absorption density from
the lines of sight in physical space, i.e. without peculiar velocities
and thermal broadening that can significantly alter the relationship
between absorption and physical properties.  The two sets of contours
illustrate the difference in phase space between the true metallicity
and the metallicity traced by \hbox{C\,{\sc iv}}.

The three wind models shown tell very different stories about IGM
metallicity, observed and actual.  The constant wind model has high
wind speeds and greater star formation at high redshift, resulting in
earlier enrichment of lower overdensities and the shock-heating of the
IGM.  The \hbox{C\,{\sc iv}} ionization fractions grow increasingly
smaller at later times as the hot metals are ionized to higher states
and cannot cool efficiently at such low densities.  In contrast, the
low wind speeds of the mw model copiously enrich high overdensities
while minimizing shock-heating, and lead to a higher overall
\hbox{C\,{\sc iv}} ionization fraction, allowing \hbox{C\,{\sc iv}} to
more closely trace the metals.  The intermediate vzw model with its
variable outflow velocity has both higher wind speeds capable of
injecting metals into the diffuse IGM and lower wind speeds
replenishing metals in the galactic halo gas.  The result is IGM
enrichment at a wider range of densities than mw with less feedback
heating than cw, which as we shall see is a better match to
observations.  All forms of feedback result in metals occupying a
large range of temperatures by $z=3$ resulting in many metal lines
being collisionally ionized, a prediction also made by \citet{the02}
in their simulations with feedback.

Generally, at high redshifts \hbox{C\,{\sc iv}} is a nice tracer of
diffuse metals, but it misses the low-temperature high-density regime
of the metal distribution (i.e. condensed phase metals) where much of
the metals are residing.  At $z<3$, \hbox{C\,{\sc iv}} traces diffuse
metals less well, because an increasing amount of metals arises in hot
gas that is too highly collisionally ionized for \hbox{C\,{\sc iv}}.
It also fails to trace very diffuse gas that is too highly
photoionized for \hbox{C\,{\sc iv}}.  Hence in what could be termed
the ``Age of \hbox{C\,{\sc iv}}'', this ion between $z=3-4.5$ is able
to trace metals over the largest range of overdensities corresponding
to the largest variety of structures.  As a side note, the relative
invariance in \hbox{C\,{\sc iv}} absorption across these redshifts can
be seen by the similar areas covered by the \hbox{C\,{\sc iv}}
contours, despite the fact that the total metallicity (the area of the
metallicity contour) is increasing with redshift; we will quantify
this effect later.

In the low redshift bin, \hbox{C\,{\sc iv}} traces typical overdensities
of $\rho/\bar\rho\sim 100$ despite the peak in metallicity moving
downwards to $\rho/\bar\rho\sim 30$.  This indicates that \hbox{C\,{\sc
iv}} absorption, particularly strong absorption, is arising primarily
in galactic halo gas.  Unfortunately for our simulations, if the gas
in these halos has a structure similar to our Milky Way with cold
clumps (i.e. high-velocity clouds) and a complex multi-phase structure
\citep{mo96,mal04}, then we cannot hope to resolve the detailed density
and temperature structure in our simulations.  Indeed, observations by
\citet{sim06} of quasar HS1700+6416 between $z=1.8\rightarrow 2.7$ suggest
sub-kpc length scales for many of the strongest absorbers.  However,
lensed quasar image pairs do not show significant deviations until above
a few hundred pc in \hbox{C\,{\sc iv}} line profiles indicating highly
ionized gas traces structure on at least kpc scales \citep{rau01}.
While these observations probe the IGM between $z=1.6\rightarrow 3.6$,
smaller separations are probed at higher redshifts, so it is difficult
to determine if the characteristic size of the absorber does in fact
decrease at low redshift as our models predict.

Figure~\ref{fig:c4cont} further shows that \hbox{C\,{\sc iv}} arises
from both photo-ionized and collisionally ionized absorbers; we show a
line above which collisional ionization dominates.  Interestingly, the
fraction of collisionally ionized \hbox{C\,{\sc iv}} is highest at the
earliest times despite a overall cooler IGM.  The diffuse IGM densities
are too high and the background is too low to ionize most metals to
\hbox{C\,{\sc iv}}, while at the same time the collisionally-ionized band
intersects regions with metals heated almost entirely by early feedback.
It is because of this that early observations of \hbox{C\,{\sc iv}}
absorption provide the greatest discrimination between feedback models,
as we shall show more explicitly later on.

The fractions of \hbox{C\,{\sc iv}} collisionally ionized between
$z=5.5-6.0$ are 68\% and 43\% for the cw and vzw models respectively.
These fractions fall precipitously to 29\% and 15\% by $z=4.0$, mainly
because the photo-ionized \hbox{C\,{\sc iv}} component rapidly increases
due to the changing ionization conditions.  The amount of collisionally
ionized \hbox{C\,{\sc iv}} never really drops, and grows rapidly below
$z=2.5$ in the vzw model as metal-enriched gas collapses back into halos
(especially overdensities $\rho/\bar\rho>300$) leading to a $\sim$50\%
collisionally ionized fraction at $z\sim2$.

In summary, \hbox{C\,{\sc iv}} absorption in cosmic phase space clearly
demonstrates that there is no one simple \hbox{C\,{\sc iv}} absorber type
relevant to all redshifts and large-scale structures we are investigating.
Instead, the photo-ionized and collisionally ionized absorption probe
two different regions of the metal-enriched IGM that evolve independently
over redshift.  As the collisionally ionized component in particular is
quite sensitive to the outflow model, \hbox{C\,{\sc iv}} can in principle
provide interesting constraints on galactic feedback, especially at
early epochs.

\subsection{Metallicity-Density Relationship} \label{sec: Zrhorelat}

Since metals are generated in galaxies that lie at highly biased peaks
of the early matter distribution \citep[e.g.][]{dav06}, it is expected
that there will be a positive correlation between metallicity and
local overdensity.  Indeed, such a relation was determined based on
\hbox{C\,{\sc iv}} pixel optical depth statistics by S03.  They
inferred that the median metallicity followed the relation [C/H] =
$-3.47+0.08\;(z-3)+0.65\;(\log(\rho/\bar\rho)-0.5)$ between
$z=1.8-4.1$, i.e. that IGM metallicity shows a moderate density
gradient with little redshift evolution.  Such a relationship provides
a constraint on models of metal injection, though as we shall see this
S03 relationship is sensitive to assumptions about the thermal state
of the gas which are not supported in our more successful wind models.

\begin{figure*}
\includegraphics[angle=-90,scale=.64]{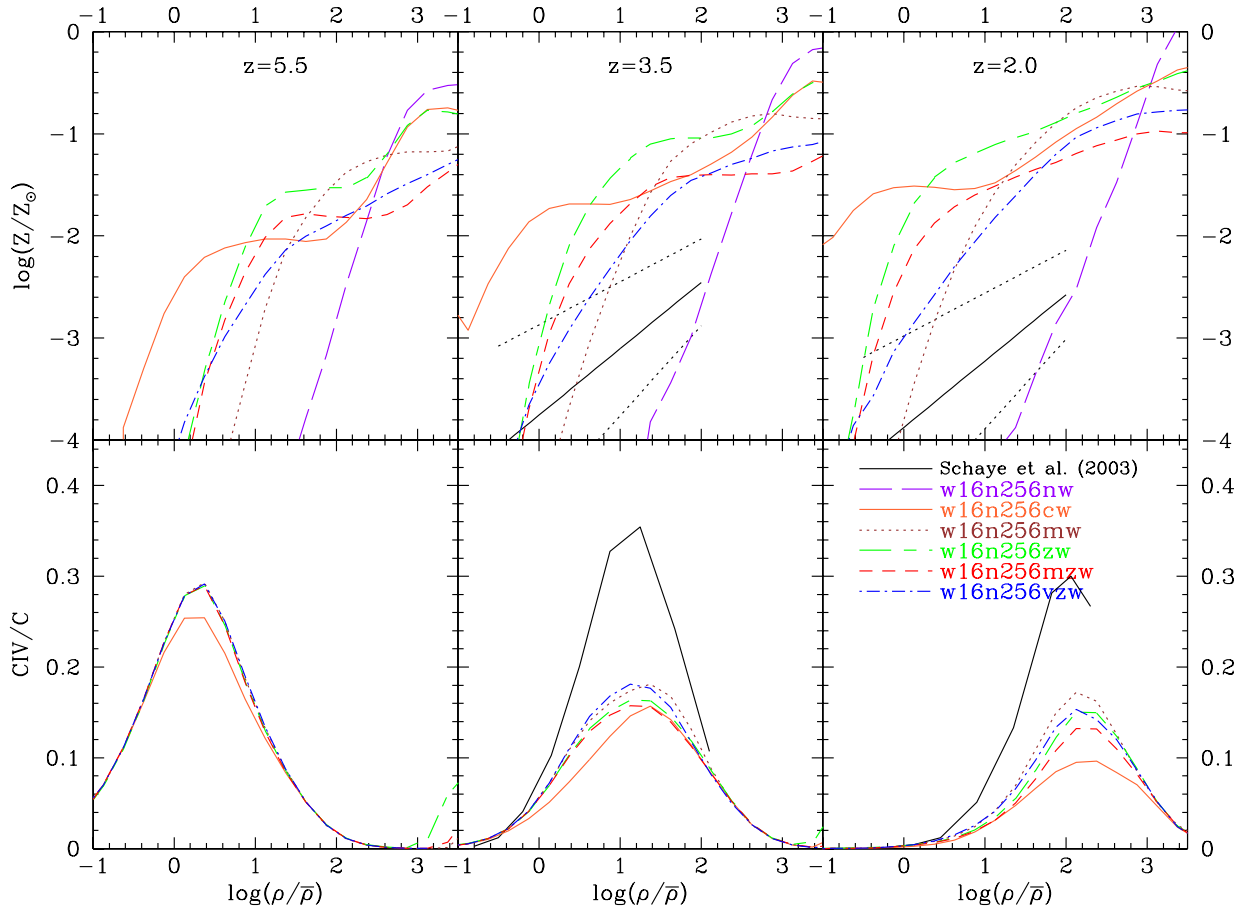}
\caption[]{Top panels show the mean metallicity as a function of
  overdensity for all our wind models in three redshift bins.  For
  comparison, the \citet{sch03} relation with its log-normal
  distribution is included at the two lower redshifts.  All models
  show steady growth in metallicity at all overdensities as outflows
  reach lower overdensities while the metals are continuously
  replenished around galaxies.  The bottom panels show the mean
  \hbox{C\,{\sc iv}}/C ratio for all IGM gas independent of
  metallicity, demonstrating how the overall ionization fraction
  decreases and \hbox{C\,{\sc iv}} preferentially traces more
  overdense gas. \hbox{C\,{\sc iv}} absorption will be observed where
  the ionization ratio and the metallicity are both significant.  The
  differences between the models result from the effect of feedback on
  the temperature structures of the IGM.  The density-temperature
  relation assumed by S03 yields a steeper ionization fraction as a
  function of density (solid line), resulting in an inferred
  metallicity-density relation that is shallower than our
  simulations.}
\label{fig:rhoZ}
\end{figure*}

The top panels of Figure~\ref{fig:rhoZ} show the metallicity-density
relationship for all our wind models at three redshifts.  These are not
log-linear relationships as have often been assumed in fits (e.g. SO3).
Instead, the metallicity-density relations are generally fairly flat out
to some moderate overdensity, and then drop increasingly rapidly to
lower overdensities.  Furthermore, metallicities are continually growing
at all overdensities in all our simulations as outflows extend to lower
overdensities and the condensed regions are constantly replenished with
new metals from galactic superwinds.

The cw model shows some different behaviors compared to the
momentum-driven wind cases.  The large early wind speeds in cw push
metals so far out that the $<Z/Z_{\odot}>$ shows almost no gradient
between $\rho/\bar\rho=1-10$, while various forms of the momentum wind
models clearly show a gradient.  For comparison, \citet{cen05} find a
``metallicity trough'' between $\rho/\bar\rho=0.1-1$ in their TIGER
simulations, because their winds propagate from the most massive galaxies
anisotropically, enriching the underdense IGM to higher metallicities than
the filaments.  The initial velocities in their model are 1469 $\kms$,
which quickly slow down to a few hundred $\kms$ after accumulating
material.  Our cw model is probably most similar to theirs, but with
lower outflow velocities that are not quite sufficient to produce a
dramatic inversion in the metallicity-density relation.

The momentum-driven wind models have a stronger metallicity-density
gradient, because smaller galaxies living at lower overdensities enrich
smaller volumes with smaller velocities.  The vzw model has the smoothest
gradient resulting from the assumed spread in velocities enriching
a variety of overdensities.  The zw model, despite its significantly
lower mass loading factor than mzw, enriches all overdensities more
efficiently, because it has a higher overall star formation rate.
Wind speed is the biggest determinant of metallicity in the diffuse IGM
(especially at $\rho/\bar\rho<10$) while mass loading factor determines
the metallicity of the condensed phases of the IGM ($\rho/\bar\rho>1000$);
with larger values of $\eta$ removing more gas from galaxies, curtailing
star formation, and decreasing enrichment.  These difference emphasize the
interplay of feedback, star formation, and energy deposition in the IGM.

Our metallicity-density relationship is generally higher than that
inferred by S03 at most overdensities The primary reason for this
discrepancy is the IGM temperature structure in our simulations.  As
we showed in \S\ref{sec: phys}, our simulated IGM is continually
growing hotter at lower redshift making the \hbox{C\,{\sc iv}}
ionization fraction smaller where the metals are, and hence requiring
higher true metallicities to match a given $\Omega$(\hbox{C\,{\sc
iv}}) absorption.  To demonstrate this, the bottom panels of
Figure~\ref{fig:rhoZ} plot the overall \hbox{C\,{\sc iv}}/C ratio
(i.e. the inverse of the ionization correction) for all IGM gas
independent of metallicity as a function of overdensity for the
various models, along with that derived from the density-temperature
relation assumed by S03 (the dashed line in Figure~\ref{fig:metcont}).
\hbox{C\,{\sc iv}} should be primarily observed at overdensities where
both the metallicity and this ratio are large.

The S03 relation shows a much steeper ionization fraction growth with
density in the mildly overdense regime than any of our simulations.
This is because they assume pure photoionization, whereas our models have
significant contributions from collisionally ionized \hbox{C\,{\sc iv}}.
This is even true of our no-wind model, owing purely to the growth of
structure around galaxies.  This explains why S03 obtains a shallower
metallicity-density relation in this regime from \hbox{C\,{\sc iv}}
observations, while (as we will show in \S\ref{sec: observe}) we
match similar data with our steeper metallicity-density gradient.
Interestingly, \citet{sca06} suggested that a top-hat distribution of
metals around galaxies is a better fit to the observed \hbox{C\,{\sc iv}}
auto-correlation function, as compared to the S03 fit.  Our simulations
produce a distribution that at least comes closer to this.

Figure~\ref{fig:rhoZ} is also helpful in visualizing some of the
trends described in \S\ref{sec: phase}.  First, the \hbox{C\,{\sc
iv}}/C ratio peak shifts from overdensities of about 1 to over 100
from $z=5.5\rightarrow 2.0$, a startling increase due to the decreasing
physical densities and (more weakly) the increasing ionizing background.
The overdensities of the observed \hbox{C\,{\sc iv}} do not shift
as much because lower overdensities have less optical depth, plus the
diffuse IGM has few or no metals at high-$z$ where the ratio is highest.
Although the photo-ionized and collisionally ionized regions are distinct
in $\rho-T$ phase space, the ratio merges into a single smooth peak when
the temperature axis is collapsed (to visualize this look at the shading
relative the the IGM dashed mass contours in Figure~\ref{fig:metcont}).
Also, the overall decrease in the \hbox{C\,{\sc iv}}/C ratio toward
low-$z$ demonstrates the decline in the ability to use \hbox{C\,{\sc iv}}
as a tracer of metals.

The differences between outflow models isolate the effect of feedback
on the temperature structure of the entire IGM.  At $z=5.5$, feedback
has not affected much the diffuse and underdense gas as can be seen by
the similarity in the models; only the cw model is able to inject energy
significantly to lower the ratio.  During the Age of \hbox{C\,{\sc iv}}
at $z=3.5$, the ionization ratio is high at metal-rich overdensities
high enough to create significant optical depth in all feedback
models.  The temperature structure causes the most divergence in the
ratio by $z=2$ because of the transition to collisionally ionized
\hbox{C\,{\sc iv}}, especially in the cw model.  Our models clearly
contrast with the S03 linear density-temperature relationship, which
intersects higher ionization fractions in the photo-ionized regime
(see Figure~\ref{fig:metcont}), leading to their estimates of lower
metallicity.

In short, \hbox{C\,{\sc iv}} ionization conditions evolve with density
and redshift, with an increasing contribution from collisionally-ionized
absorbers at lower redshifts.  Our metallicities at a given density are
higher than inferred previously by S03 because the broad distribution of
metals in the $\rho-T$ plane results in a higher overall \hbox{C\,{\sc
iv}} ionization correction.

\subsection{Evolution of Median \hbox{C\,{\sc iv}} Absorber in Phase Space} \label{sec:avec4lines}

\begin{figure*}
\includegraphics[angle=-90,scale=.66]{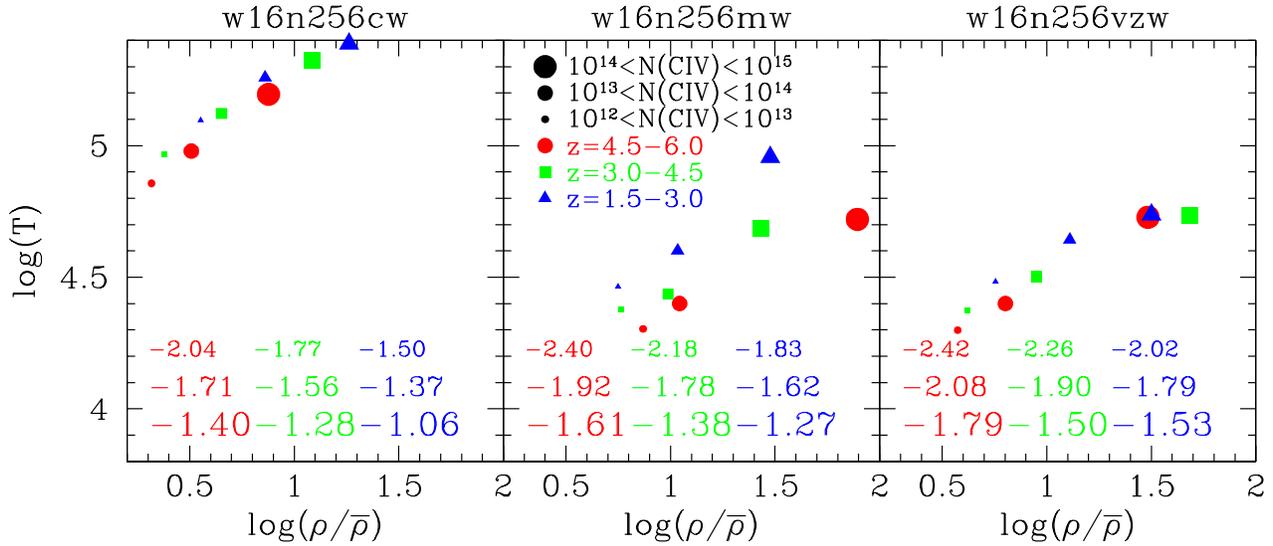}
\caption[]{Median values of \hbox{C\,{\sc iv}} absorbers
subdivided into column density bins (point size) and redshift (color and
shape) along with the median values of metallicity ([Z/H], written
below with larger text for higher column density).  Left, middle
and right panels show the w16n256cw,
w16n256mw, and w16n256vzw models.  The medians plotted are derived
from spectra that include peculiar velocities and thermal broadening.
The cw model is dominated by collisionally-ionized absorbers at all
redshifts, while mw is photo-ionization dominated.  The vzw model
probes the widest range of densities from weak to strong systems,
showing its ability to enrich a wide range of environments, while
still maintaining mostly photoionization-dominated temperatures.}
\label{fig:avec4lines}
\end{figure*}

Figure~\ref{fig:avec4lines} presents the evolution of absorbers in terms
of the median phase space properties of \hbox{C\,{\sc iv}} lines.  We show
three bins of column density (log($N$(\hbox{C\,{\sc iv}})) = 12-13 (weak),
13-14 (intermediate), 14-15 (strong)) over three different redshift
ranges ($z=6\rightarrow 4.5$, $4.5\rightarrow 3$, $3\rightarrow 2$),
for three of our wind models (cw, mw, vzw).  The values are obtained
by summing over parcels of gas in redshift space (i.e. including
peculiar velocities and thermal broadening), and is thereby likely
to underestimate the density and metallicity of the parcel(s) of gas
contributing most to the absorption within a given line.  This is more
true at lower redshift as \hbox{C\,{\sc iv}} absorbers become compressed
in comoving space while the comoving pathlength per redshift increases.
Nevertheless, these outflow models are clearly distinguished, allowing
us to draw some simple conclusions about the physics of the absorbers.

In the cw model, the median absorber exists at a temperature that is
dominated by collisional ionization ($\approx 10^5$ K), which as we
shall show produces poor agreement with observations.  The typical
overdensity increases with time at all column densities, as diffuse
gas becomes increasingly ionized to \hbox{C\,{\sc v}} and beyond.  In
contrast, the mw model's absorbers are almost always photo-ionized
except at the strongest lines at high-$z$, and show relatively little
evolution in overdensity over our redshift range, because the low wind
speeds do not eject metals much outside the relatively dense regions
that are well-traced by photo-ionized \hbox{C\,{\sc iv}}.  Meanwhile,
the vzw model manages to shock heat more metals creating a larger
collisionally ionized component, while concurrently allowing
significant metal enrichment of denser halo gas by virtue of some
outflow velocities being low.  This creates a wider range of
overdensities and temperatures traced by different column densities at
any given redshift as compared with either the cw or mw models, with
weak absorbers tracing $\rho/\bar\rho\approx 4-5$, and strong
absorbers tracing $\rho/\bar\rho\approx 30-100$.

In brief, the behavior of \hbox{C\,{\sc iv}} absorption in cosmic
phase space shows noticeably different behavior for various wind
models, suggesting that \hbox{C\,{\sc iv}} observations can provide a
sensitive probe of galactic outflows.  In the next section, we
demonstrate this explicitly by comparing our outflow models to
observations.

\section{Testing Outflow Models Against Observations}  \label{sec: observe}

\subsection{\hbox{C\,{\sc iv}} Mass Density} \label{sec: omega}

So far we have seen that outflows can have a significant impact on the
physical properties of the IGM.  We now compare \hbox{C\,{\sc iv}}
absorption line properties from our various outflow models in order
to determine which ones, if any, successfully reproduce available
observations.  We begin by examining the most global statistic that can
be derived from \hbox{C\,{\sc iv}} absorption, namely the total mass
density in \hbox{C\,{\sc iv}} absorbers, $\Omega$(\hbox{C\,{\sc iv}}).

$\Omega$(\hbox{C\,{\sc iv}}) provides a benchmark observational test
for models of IGM enrichment.  The observed lack of evolution of
$\Omega$(\hbox{C\,{\sc iv}}) from $z\approx 6\rightarrow 2$ has been
cited as evidence for very early ($z\ga 6$) enrichment of the IGM
\citep{son01,sca02,por05,son05} from a putative generation of early
massive stars and/or primeval galaxies.  Such a scenario is attractive
at face value because winds can easily escape the small potential wells
of early galaxies.  But it should be remembered that the metals must
be expelled to densities approaching the cosmic mean in order not to
have been gravitationally re-accreted onto galaxies by lower redshifts
\citep[e.g.][]{por05}.  This means winds must not only escape galaxies,
but must also overcome Hubble flow which is quite rapid at these early
times, approaching 1000~km/s.  A population of early galaxies that is
energetically able to expel winds at $\sim 1000$~km/s may be difficult
to reconcile with limits on pre-reionization massive star formation from
the latest WMAP Thompson optical depth results \citep{pag06}.

None of our models directly test this early IGM enrichment scenario.
Instead, our outflow models attempt to reproduce the observed
\hbox{C\,{\sc iv}} evolution self-consistently from standard stellar
populations combined with locally-calibrated outflow models.  We find
that our models are successful, thereby removing the need for widespread
early IGM enrichment.  We further find that observations non-trivially
constrain our outflow models.

\citet{son01} determined $\Omega$(\hbox{C\,{\sc iv}}) by integrating the
total column density of systems between $10^{12} \leq N$(\hbox{C\,{\sc
iv}})$ < 10^{15}$ cm$^{-2}$ and dividing by the pathlength using
\begin{equation}
\Omega({\rm CIV})(z) = {H_0 m_{CIV} \over c \rho_{crit}} {\Sigma N_{tot}({\rm CIV},z) \over \Delta X(z)}
\label{eqn:omega_civ}
\end{equation}
where 
\begin{equation}
\Delta X(z) = \int {(1+z)^2 dz \over \sqrt{\Omega_M (1+z)^3+\Omega_{\Lambda}}}.
\end{equation}
These limits are chosen because larger column density systems are likely
to arise from galactic halo gas, while the lower column densities
suffer from observational incompleteness.  When needed, we convert
observations to our simulations' $\Lambda$CDM cosmology ($\Omega_M=0.3$,
$\Omega_{\Lambda}=0.7$).  Measurements of $\Omega$(\hbox{C\,{\sc iv}})
have highlighted some interesting trends that already provide a key test
for our various feedback models.

\begin{figure} 
\includegraphics[scale=.40]{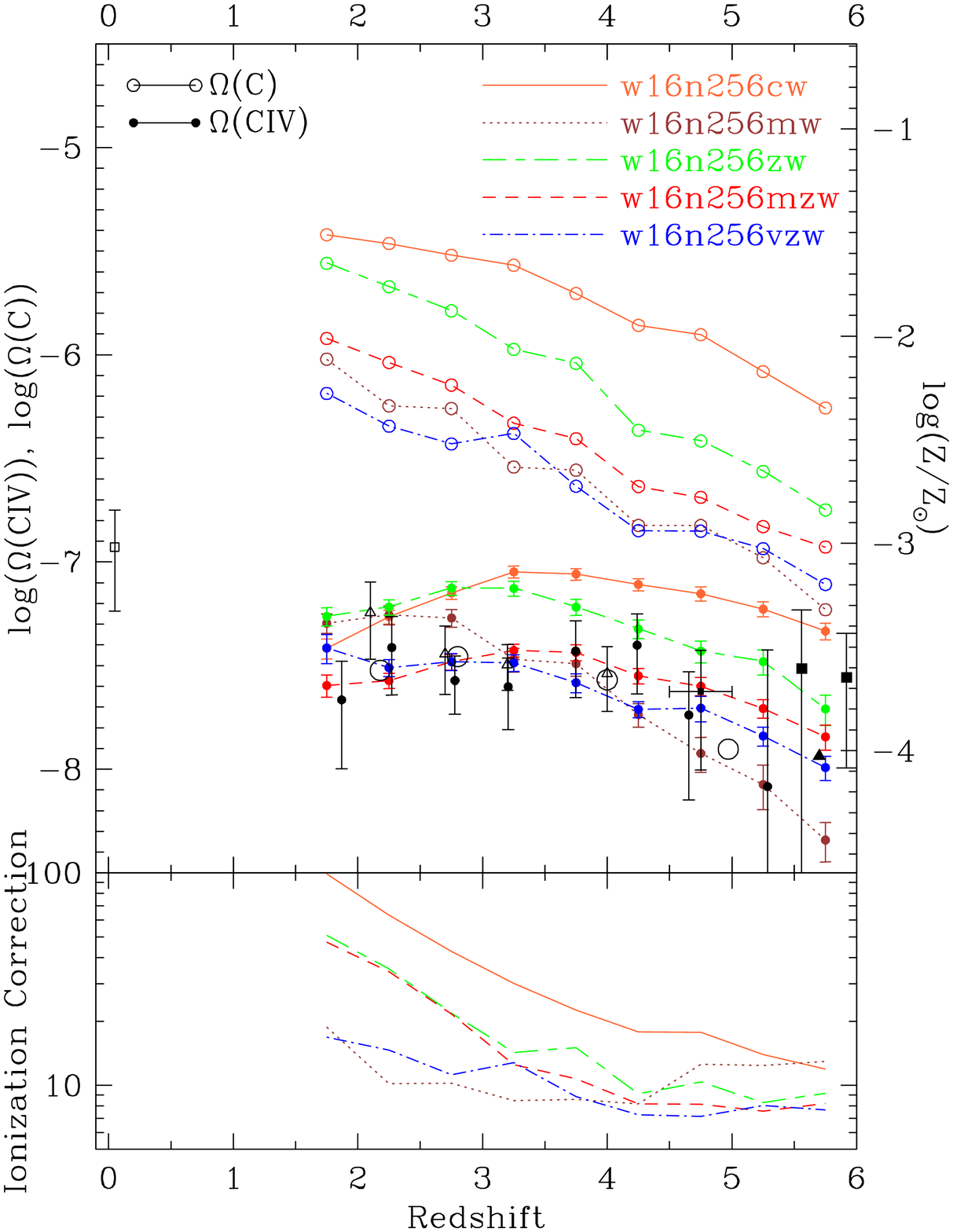}

\caption[]{Evolution of $\Omega$(\hbox{C\,{\sc iv}}) (closed circles)
and $\Omega$(C) (open circles), the mass density in \hbox{C\,{\sc iv}}
and metals, respectively.  Our five feedback models are compared to
observations from \citet{son01} (black circles), \citet{pet03} (small
filled black squares), \citet{bok03} (open triangles), \citet{fry03}
(open square), \citet{son05} (large open dots), \cite{sim06b} (large
black squares), and \cite{rya06} (black triangle is a lower limit).
All models show an increasing $\Omega$(C) over this redshift range,
but most show a relatively constant $\Omega$(\hbox{C\,{\sc iv}}).  The
ratio is shown as a ``global ionization correction" in the bottom
panel.  The cw model is the most metal rich, and has increasingly
large ionization corrections toward low redshift as the IGM is
significantly heated and metals are pushed predominantly into low
density regions.  The weak momentum wind model (mw) produces little
evolution in the ionization correction because it does not heat the
IGM significantly, and hence is the only model that shows a rising
$\Omega$(\hbox{C\,{\sc iv}}) over the entire redshift range.  The zw
model overproduces $\Omega$(\hbox{C\,{\sc iv}}) because it produces
too many metals overall.  The mzw and vzw models provide good fits to
the observations, with the greatest discrimination between the two
occurring at the highest redshifts.  If further observations can
confirm the lack of a downturn at $z>5,5$, another mechanism for early
enrichment may be required.}
\label{fig:omegac4} 
\end{figure}

Figure~\ref{fig:omegac4} shows the evolution of $\Omega$(\hbox{C\,{\sc
iv}}), as well as $\Omega$(C), calculated similarly to
equation~\ref{eqn:omega_civ}, except using the total metallicity along
our simulated lines of sight.  Observations of $\Omega$(\hbox{C\,{\sc
iv}}) by \citet{son01}, \citet{pet03}, \citet{bok03}, \citet{son05},
\citet{sim06b}, and \citet{rya06} are shown.  We also show a
measurement at $z\sim 0$ by \citet{fry03} for reference, although we
will not compare to this data here.

The first striking result from this plot is that while the true IGM
metallicity continually grows from $z=6\rightarrow 1.5$, most wind
models show $\Omega$(\hbox{C\,{\sc iv}}) to be relatively constant,
and perhaps even decreasing at the lowest redshifts.  This is
especially true for our stronger wind models (cw, mzw, zw), indicating
that outflows play a key role in governing \hbox{C\,{\sc iv}}
evolution.  The key point is that in our models,
$\Omega$(\hbox{C\,{\sc iv}}) {\it evolution is not a true indicator of
metallicity evolution.}

Two counteracting effects conspire to make $\Omega$(\hbox{C\,{\sc
iv}}) appear nearly constant.  All feedback models grow IGM
metallicity at a similar rate, but in the stronger wind models this is
offset by a decreasing \hbox{C\,{\sc iv}} ionization fraction owing
mainly to an increasing IGM temperature from feedback energy.  In the
case where the feedback energy is minimal such as in the mw model, the
globally-averaged ionization correction (shown in the bottom panel)
remains more constant.  Nevertheless, even this model's ionization
correction increases beyond $z=3$ as physical densities of the IGM
decrease and the ionization background reaches its maximum strength.
However, for feedback models that heat the IGM significantly
(cw,zw,mzw), the ionization correction from \hbox{C\,{\sc iv}} to
[C/H] increases significantly with time, resulting a fairly flat
$\Omega$(\hbox{C\,{\sc iv}}) from $z=5\rightarrow 2$ despite nearly an
order of magnitude increase in $\Omega$(C).

Intermediate wind-speed models, mzw and vzw, provide the best fit to
observations of $\Omega$(\hbox{C\,{\sc iv}}) over this redshift range.
Recall that these are both momentum-driven wind models where
$f_{L,\odot}=2$ and $f_{L,\odot}=1.05-2$, respectively.  Indeed, the
vzw model reproduces the most recent SuperPOD analysis by
\citet[][open circles]{son05} almost exactly, though without a more
careful comparison to data this should not be taken too seriously.
Interestingly, vzw and mzw provide good fits to $\Omega$(\hbox{C\,{\sc
iv}}) evolution for somewhat different reasons.  In the mzw model
substantial IGM enrichment is accompanied by significant heat input,
so that ionization corrections evolves rapidly to maintain a roughly
constant $\Omega$(\hbox{C\,{\sc iv}}).  In the vzw model, there is
less IGM enrichment owing to lower wind velocities, but also less heat
input.  Indeed, mzw and vzw seem to bracket the allowed range of
heat/energy input combinations: cw and zw produce too many metals in
the IGM, while mw provides too little heat to obtain an ionization
correction that grow sufficiently from $z=6\rightarrow 1.5$.  Hence
even from a relatively blunt tool such as $\Omega$(\hbox{C\,{\sc
iv}}), it is possible to place interesting constraints on outflow
models.

To further constrain models using  $\Omega$(\hbox{C\,{\sc iv}}) requires
probing to lower and higher redshifts.  Lower redshift \hbox{C\,{\sc
iv}} observations should in principle be able to break this model
degeneracy, but may be difficult to interpret because \hbox{C\,{\sc iv}}
traces increasingly denser regions that may be subject to multi-phase
collapse, and the photoionizing background is more poorly constrained.
Additionally, there is the observational challenge of \hbox{C\,{\sc
iv}} moving into the ultraviolet.  Nevertheless, the determination of
$\Omega$(\hbox{C\,{\sc iv}}) in the local universe by \citet{fry03}
would appear to favor a model with lower ionization corrections, since
the IGM metallicity should level off because of the rapidly dropping
global star formation rate at $z\la 1.5$.  We leave comparisons to
$z<1.5$ data for future work.

Conversely, moving to higher redshifts may provide more robust
constraints.  In particular, new observations by \citet{sim06b} and
\citet{rya06} at $z>5.5$ continue to show no apparent drop in
$\Omega$(\hbox{C\,{\sc iv}}) while our models predict a gradual
decline.  These measurements are dominated by a few strong systems (as
these are the only ones detectable with current technology), so it is
possible that may be associated with recent outflows that do not
occupy a significant volume rather than indicative of widespread
enrichment.  Upcoming near infrared spectra of high-$z$ quasars
discovered by the Sloan Digital Sky Survey should confirm just how
widespread the \hbox{C\,{\sc iv}} is, and whether our models require a
further enrichment mechanism at high-$z$ or perhaps an increase in
yields from the earliest stars.  The lack of strong \hbox{C\,{\sc iv}}
observed above $z>5.90$ by \citet{sim06b} along with the observation
of a sudden increase in \hbox{O\,{\sc i}} \citet{bec06} at $z>6$ may
mean a sudden change in the ionization conditions of metals at this
epoch corresponding to the last stages of reionization \citet{fan06}
(R. Simcoe, private communication).

Finally, we examine a potential source of systematic uncertainty in
comparing models to observations of $\Omega$(\hbox{C\,{\sc iv}}),
namely the effects of Helium reionization.  The epoch and duration of
\hbox{He\,{\sc ii}} reionization is controversial, with some
observations suggesting a rapid transition
\citep{son96,sch00,the02a,zhe04} at $z\sim 3$, while others favoring a
more gradual evolution before $z=3$ \citep{dav98,sok03}.  Although the
ionization potential of \hbox{C\,{\sc iv}} is around 3.5~Ryd while
\hbox{He\,{\sc ii}} is optically thick above 4~Ryd, there may still be
some residual effect on \hbox{C\,{\sc iv}} absorption.  To test this,
we generate spectra using a soft QG ionizing background where we have
reduced the photoionizing flux above 4~Ry by $\times 10$, and derive
the resulting $\Omega$(\hbox{C\,{\sc iv}}) evolution.

\begin{figure}
\includegraphics[angle=-90,scale=.32]{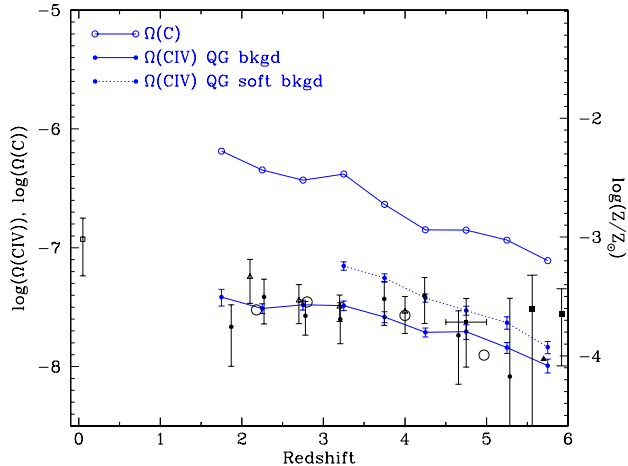}
\caption[]{The comparison of the evolution of $\Omega$(\hbox{C\,{\sc iv}})
with the standard \citet{haa01} ionization background (solid) and a softer
ionization background where we have reduced the ionization at $>4$ Ryd
by 90\% to simulate pre-\hbox{He\,{\sc ii}} reionization conditions above
$z=3$ (dotted).  The same data points described in Figure~\ref{fig:omegac4} are
displayed.  The soft background increases the observed \hbox{C\,{\sc iv}} by
as much as a factor of two.  This difference favors a gradual
\hbox{He\,{\sc ii}} reionization, as a sudden transition would produce an
observable drop in \hbox{C\,{\sc iv}}.  }
\label{fig:omegac4_he2}
\end{figure}

Figure~\ref{fig:omegac4_he2} shows the results for the vzw case; the
trends are similar for other outflow models.  \hbox{C\,{\sc iv}}
traces less overdense material especially between
$\rho/\bar\rho=1-10$, because of the lack of photons capable of
ionizing \hbox{C\,{\sc iv}} into \hbox{C\,{\sc v}}; thus decreasing
the overall \hbox{C\,{\sc iv}} ionization correction.  Hence if
\hbox{He\,{\sc ii}} reionization is a sudden process, there should a
sudden decrease in $\Omega$(\hbox{C\,{\sc iv}}) by a factor of around
two at the epoch.  Such a drop is not obvious in the data, although
error bars are large and could conceivably hide it.  At face value, it
instead appears that \hbox{He\,{\sc ii}} reionization is a gradual
process, which could also somewhat affect the evolution of
$\Omega$(\hbox{C\,{\sc iv}}) possibly making this measurement even
more constant over the redshift range.  In fact, \citet{bol06} find
evidence that the fluctuations in the \hbox{He\,{\sc
ii}}/\hbox{H\,{\sc i}} ratio require a patchy background at energies
$>4$~Ryd between $z=2-3$.  We plan to examine the effects of this in
more detail in the future.

To summarize, our models indicate that the invariance in
$\Omega$(\hbox{C\,{\sc iv}}) at $2\la z\la 5$ does not require the
majority of IGM enrichment to occur at $z>5$.  Our favored feedback
simulations prefer an interpretation where metals injection follows global
stellar mass production, but the accompanying energy injection conspires
to make $\Omega$(\hbox{C\,{\sc iv}}) roughly constant over the observed
redshift range.  We cannot rule out an early enrichment scenario that
invokes a speculative early generation of stars to distribute metals
in a manner that matches $\Omega$(\hbox{C\,{\sc iv}}) data, but our
self-consistent outflow models provide a more straightforward explanation.

\subsection{\hbox{C\,{\sc iv}} Line Parameters}

\subsubsection{\hbox{C\,{\sc iv}} Column Density Distributions} \label{sec: col}
While we showed that $\Omega$(\hbox{C\,{\sc iv}}) can provide interesting
constraints on outflow models, in fact there is more information contained
in the statistical properties of \hbox{C\,{\sc iv}} absorption systems
from which $\Omega$(\hbox{C\,{\sc iv}}) is derived.  In this
section we compare our models to observations of the column density
distribution (CDD), which is the number of lines per unit redshift
interval per log column density bin.  The CDD is observed to fairly
accurately follow a steep power law \citep{ell00,son01}.  The slope,
amplitude, and any deviations from this power law provide key diagnostics
for testing models of IGM enrichment.

\begin{figure*}
\includegraphics[angle=-90,scale=.64]{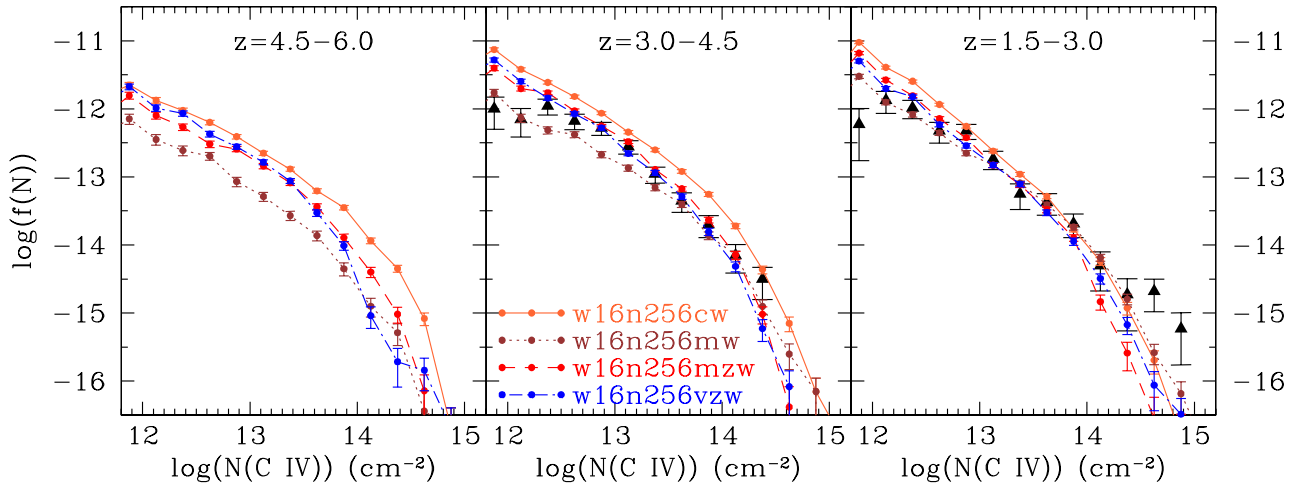}
\caption[]{\hbox{C\,{\sc iv}} Column density distributions for four of
our outflow models in three redshift bins, compared to observations by
BSR03 (black triangles).  The cw model clearly overpredicts the CDD,
while the mw predicts too few weak systems showing that its winds are
too weak to enrich the low-density IGM.  The mzw and vzw model
provide promising fits to the data; the overprediction at weak column
densities is likely a result of lower $S/N$ in the BSR03 data.  }
\label{fig:c4cdd}
\end{figure*}

We use the AutoVP package \citep{dav97} to fit 30 simulated spectra
per run generated continuously from $z=6\rightarrow 1.5$.  For now we
only consider the 1548 \AA~\hbox{C\,{\sc iv}} component to avoid
confusion; in real spectra, the sparseness of \hbox{C\,{\sc iv}}
absorption generally results in little ambiguity in line
identification and characterization.  This sparseness also provides a
robust determination of the quasar continuum, so we do not
continuum-fit our \hbox{C\,{\sc iv}} spectra.  \citet{the00}
demonstrated that AutoVP provided nearly identical fits to lines as
the standard fitting package VPFIT \citep{car87} for non-saturated
lines (as is virtually always the case for \hbox{C\,{\sc iv}}), and
the quality of the fits in the figure appears good even in complex
systems.  One technical difference in fitting simulated spectra is
that here we fit only the stronger component of the doublet, while
observations generally require the presence of both doublet components
to unambiguously identify \hbox{C\,{\sc iv}}.  Because of this,
observations are more likely to miss weak lines.  This difference is
negligible when we integrate the total metal mass, but can become
significant when we discuss the small-$N$(\hbox{C\,{\sc iv}}) end of
the CDD.

To compare our models with observations, we take line lists from 9
high-resolution spectra taken with HIRES published by \citet{bok03}
(hereafter BSR03).  The BSR03 data have typical signal-to-noise ratios
of $S/N\sim 100$ and span $z=1.6\rightarrow 4.5$.  We subdivide the
BSR03 dataset into systems with $z>3$ and $z<3$, which covers a total of
9.19 and 7.73 in \hbox{C\,{\sc iv}} redshift path length, respectively.
We have also compared with line lists from 19 VLT/UVES spectra kindly
provided to us by E. Scannapieco \citep{sca06}, which covers a path length
of 25.12 from $z=1.5-3.0$ with lower signal-to-noise than BSR03.  However,
the results are similar to BSR03, and the lower S/N and smaller redshift
range motivates us to focus on the BSR03 data for clarity.  In the future
we hope to obtain the spectra themselves in order to do more careful
comparisons including detailed noise characteristics and resolution
effects \citep[e.g.][]{dav01b}, but for now these data will suffice in
order to ascertain the basic trends versus our various wind models.

Figure~\ref{fig:c4cdd} plots CDDs for systems of \hbox{C\,{\sc iv}}
absorbers in three redshift bins, $z=1.5-3.0,3.0-4.5,4.5-6.0$.
We will refer to these as the low, intermediate, and high redshift
bins, respectively, throughout this paper (though we recognize that
$z=1.5-3$ is not often regarded as ``low" redshift).  For simplicity we
omit the zw and nw models from discussion, because the relevant trends
are well-illustrated by the other four models (cw, mzw, vzw, and mw).
Also shown are the BSR03 data (black triangles) in the relevant redshift
ranges.

As can be seen, all the simulations generally produce a power law between
$10^{12.5}\la N$(\hbox{C\,{\sc iv}})$\la 10^{14.5}$ whose shape is in
agreement with observations.  In fact, at early times the simulations show
more of a deviation from a power law in the sense of fewer weak lines,
because the enrichment has not spread yet to very low densities; such
a trend is vaguely noticeable in the data as well.  At the very lowest
column densities the effects of incompleteness become significant.  For
concreteness, here are our completeness limits from simulated spectra with
$S/N = 100$: 75\% for $10^{12.00}<N$(\hbox{C\,{\sc iv}})$<10^{12.25}$,
88\% for $10^{12.25}<N$(\hbox{C\,{\sc iv}})$<10^{12.50}$, and 97\% for
$10^{12.50}<N$(\hbox{C\,{\sc iv}})$<10^{12.75}$.  BSR03's completeness
limits are probably similar but somewhat lower given their requirement
of identifying the weaker doublet line.  As we are looking to understand
general physical trends rather than engage in a detailed comparison versus
observations, we do not attempt to mimic the data in any greater level
of detail.  We simply note that if a simulated CDD exceeds the data at the
small $N$(\hbox{C\,{\sc iv}}) end, one cannot draw conclusions from this
without more carefully analyzing the incompleteness level in the data.

While qualitatively similar, these four wind models do show
statistically significant differences, indeed quite large ones.  The
difference in CDD amplitude (and its evolution) tells basically the
same story as the evolution of $\Omega$(\hbox{C\,{\sc iv}}): cw
produces too much \hbox{C\,{\sc iv}}, while mw produces quite little,
particularly early on.  But the decomposition into column density bins
provides additional information.  For instance, the mw model does not
enrich the low-density IGM much owing to its small wind speeds, and
hence cannot produce enough weak lines at $z=3-4.5$ to agree with
data.  We note that the incompleteness in the observations is likely
{\it greater} than that in the simulated spectra, so this effect works
in the wrong direction to reconcile the mw model CDD.  So even though
mw looked promising at early times in the $\Omega$(\hbox{C\,{\sc iv}})
comparison, it actually fails significantly when examining CDDs.  In
contrast, the mzw and vzw models provides very good agreement with
observations, with the only discrepancies at $N$(\hbox{C\,{\sc
iv}})$\la 10^{12.5}$ which may be explained by larger incompleteness
in the observations.

Moving to the low redshift bin, model CDDs draw closer for two reasons.
First, \hbox{C\,{\sc iv}} traces denser gas where metals have been
injected earlier on, and second because the metallicity in the denser
regions starts to saturate, reaching an equilibrium between galaxies
producing new metals and outflows driving those metals to lower densities.
Again, there are still non-trivial differences, as cw still overproduces
\hbox{C\,{\sc iv}} absorption at all column densities, and mw still
produces fewer weak lines.  At face value the mw model seems to match the
weak lines best, but recall that additional incompleteness in the data is
quite significant here (particularly since the BSR03 observations with
HIRES are less sensitive in the blue), so the turnover in the data
at $N$(\hbox{C\,{\sc iv}})$<10^{13}$ may be at least partly artificial.
There is a hint that all models produce too few of the very strongest
systems with $N$(\hbox{C\,{\sc iv}})$>10^{14.5}$, which may be due
to an increasing population of \hbox{C\,{\sc iv}} absorbing gas in
small, cold clumps that our simulations cannot resolve.  For instance,
\citet{sim06} derive small thicknesses ($<1$ kpc) and high overdensities
($\rho/\bar\rho>100$) for many of their strongest \hbox{C\,{\sc iv}}
components, which is below our spatial resolution limit.

Overall, the \hbox{C\,{\sc iv}} CDD provides a sensitive test of the
manner in which outflow models push metals into the IGM.  Comparisons to
observed \hbox{C\,{\sc iv}} column density distributions favor the vzw and
mzw models, which shows little redshift evolution at all but the smallest
column densities, and whose total \hbox{C\,{\sc iv}} absorption is close
to that observed.  More generally, it demonstrates that outflows must
have wind speeds large enough to enrich the very diffuse IGM (giving
rise to weak \hbox{C\,{\sc iv}} absorption) at early times, but must
not be so strong as to overproduce \hbox{C\,{\sc iv}} absorption overall.

\subsubsection{\hbox{C\,{\sc iv}} Linewidths} \label{sec: bparam}

While column density distributions probe the spatial distribution
\hbox{C\,{\sc iv}} in the IGM, line widths are more directly related to
the temperature and dynamics of the absorbing gas.  For \hbox{H\,{\sc
i}}, the line widths at $z\ga 2$ are dominated by Hubble flow broadening
\citep[e.g.][]{her96}, i.e. the cosmic expansion velocity spread across an
unbound absorbing structure.  This becomes less true at lower redshifts,
where \hbox{H\,{\sc i}} line widths obtain an increasing contribution
from thermal broadening \citep{dav01b}, mainly because they arise in
higher overdensity structures that do not have substantial Hubble
flow across them \citep{dav99}.  In the case of \hbox{C\,{\sc iv}},
even at high redshifts they arise in fairly overdense regions (see
Figure~\ref{fig:c4cont}), so we can expect that their line widths would
have a significant contribution from thermal broadening.

\begin{figure*}
\includegraphics[angle=-90,scale=.64]{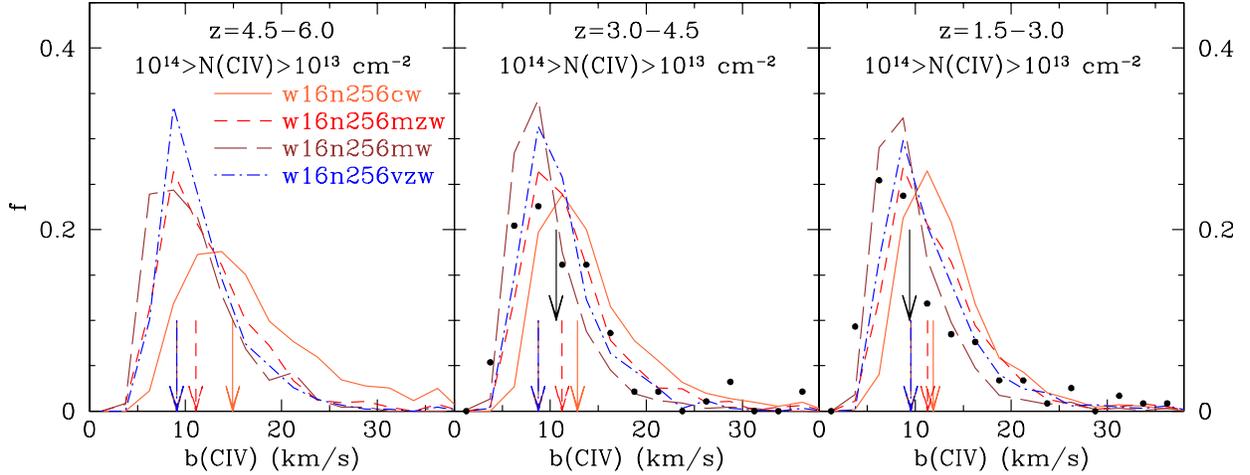}
\caption[]{Line width ($b$-parameter) distributions for
$10^{13}<N$(\hbox{C\,{\sc iv}})$<10^{14}$ cm$^{-2}$ systems for
our various outflow simulations compared to observations by BSR03
(black dots).  The arrows at the bottom show to the median values of
the distribution for each model, and the top arrow is the median of the
BSR03 data.  The data agree with the mzw model at intermediate redshift,
but suggest more compact, cooler \hbox{C\,{\sc iv}} absorbers at low
redshift such as those in the vzw or mw model.}
\label{fig:bparam}
\end{figure*}

Figure~\ref{fig:bparam} shows histograms of $b$-parameters (i.e. Voigt
profile line widths) of components with $10^{13}<N$(\hbox{C\,{\sc
iv}})$<10^{14}$ cm$^{-2}$ in four of our outflow models for our three redshift
bins, compared with BSR03 observations (filled circles in the lower
two redshift bins).  This column density range is optimal because
observations both yield a statistically significant number of lines
and generally approach 100\% completeness.  Arrows at the bottom of
the plot show the median values.

The trends in this plot follow those in Figure~\ref{fig:globalIGM}, middle
right panel (which showed the mean temperature evolution of the IGM).
The cw model has significantly larger $b$-parameters owing to substantial
shock heat input from outflows, and appears to be clearly discrepant when
compared to observations.  Indeed, the cw model actually has more lines
($\sim66\%$) at $z>4.5$ that arise from collisional ionized \hbox{C\,{\sc
iv}} ($T\ga 10^5$ K, $b_{thermal}>11.8\kms$) than at $z<4.5$ where
\hbox{C\,{\sc iv}} is photoionized in cooler gas; this partly explains
the $b_{med}$ evolution downward.  Observations from BSR03 (solid points)
suggest the mzw model is the best fit at intermediate redshift, while
the vzw model better approximates the low-$z$ absorbers.
The momentum-driven wind models (mw, mzw, vzw) are remarkable in that
$b_{med}$ does not vary by more than 1 $\kms$ from $z=6\rightarrow
1.5$, especially considering that one of the main points of this paper
is that \hbox{C\,{\sc iv}} absorbers show significant physical
evolution.  

\begin{figure*}
\includegraphics[angle=-90,scale=.64]{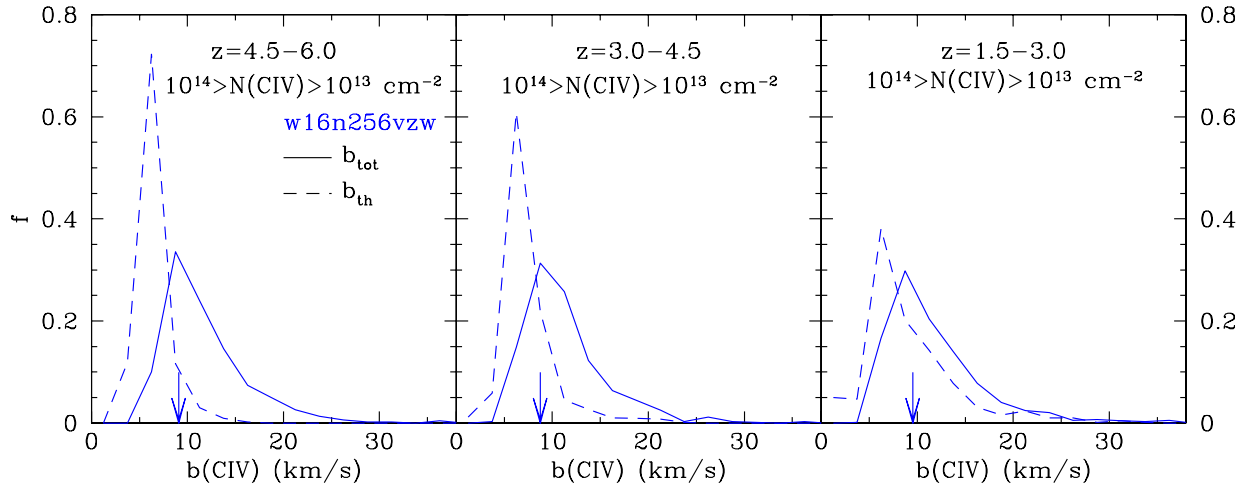}
\caption[]{Histograms of total linewidth (solid lines) for the vzw
model and thermal linewidth ($b_{th}$) derived from the temperature of
the \hbox{C\,{\sc iv}} absorbing gas (dashed lines).  Thermal
broadening comprises the majority of the total linewidth, especially
at low redshift, showing that the $b$-parameter distribution can
provide a sensitive test of feedback energy input into the IGM.}
\label{fig:btempcomp}
\end{figure*}

As an informative exercise, we subdivide the width of the vzw model
line into a thermal component, $b_{th}$, that depends only on
temperature, and a spatial component, $b_{sp}$, due to Hubble
expansion across the physical extent of the absorber.  We ignore
turbulent broadening since the IGM at these overdensities is very
quiescent \citep{rau05}.  Specifically, we assign $b_{tot} =
\sqrt{b_{th}^2 + b_{sp}^2}$, where $b_{th}=3.7 \sqrt{T/10^4}$ for
carbon, and $b_{sp} = H(z)\;w$(\hbox{C\,{\sc iv}}) where
$w$(\hbox{C\,{\sc iv}}) is the spatial extent of the \hbox{C\,{\sc
iv}}.  Figure~\ref{fig:btempcomp} shows the total (solid) and thermal
(dashed) $b$-parameter histograms for the vzw model, with $b_{th}$
calculated from the temperatures at the line centers.  The thermal
widths are the most significant component of the total line width,
especially at low redshift, demonstrating (as expected) that at the
moderate overdensities of \hbox{C\,{\sc iv}} absorbers the Hubble
broadening is sub-dominant.  This implies that $b$-parameters can
effectively trace the IGM temperature in \hbox{C\,{\sc iv}} absorbing
gas.

In summary, the line width distributions and the low-column density end of
the CDD together put significant constraints on the wind speeds of early
outflows:  They must be strong enough to pollute the diffuse IGM at early
times, but weak enough so as not to overheat the IGM.  Some heating is
accommodated, and in fact during the Age of \hbox{C\,{\sc iv}} there is a
hint that without wind heating, line widths are too narrow.  More careful
comparisons will be necessary to place more detailed constraints, but
the fact that our locally-calibrated momentum-driven wind scenarios
naturally fall within these relatively tight constraints represents a
significant success for these models.

\subsection{Pixel Optical Depth Statistics} \label{sec: pod}

The pixel optical depth (POD) method provides a powerful way to measure
the metallicity in the diffuse IGM across a wide range of densities.
The idea is to directly utilize optical depth information in the
spectrum rather than fit lines, so that even weak absorption that
would not yield a statistically significant line identification can
be quantified.  \citet{son98} pioneered this approach, which was also
used by \citet{dav98} to quantify \hbox{O\,{\sc vi}} absorption in the
crowded high-redshift Ly-$\alpha$ forest.  More recently, \citet{agu05}
and \citet{son05} have expanded on POD analyses to precisely quantify
\hbox{C\,{\sc iv}} absorption in the IGM.  In this section we compare
POD analyses of simulated spectra with such observations.

In POD analyses, a histograms of optical depth ratios of two ions is
constructed.  This is done by identifying significant absorption in one
ion, and examining the location in the spectrum where the other ion should
be, and inferring the optical depth ratio of the latter to the former.
In our case, we will consider \hbox{C\,{\sc iv}}/\hbox{H\,{\sc i}}
and \hbox{C\,{\sc iii}}/\hbox{C\,{\sc iv}} ratios, in order to compare
with data from \citet{agu05}.  Since \hbox{C\,{\sc iii}} has a rest
wavelength of 977\AA, we must also correct for coincident absorption
from \hbox{H\,{\sc i}} Lyman series lines; we apply the method detailed
by \citet{agu02} to do so.  Specifically, the lowest order unsaturated
Lyman-series line is used to determine the \hbox{H\,{\sc i}} optical
depth, which is then subtracted from all Lyman lines leaving the
absorption component of any metal lines.  Higher-order Ly$\alpha$ lines
with their lower oscillator strengths allow one to probe much higher
\hbox{H\,{\sc i}}, up to $\tau$(\hbox{H\,{\sc i}})$>1000$ as opposed to
$\tau$(\hbox{H\,{\sc i}})$\sim3$ using Ly$\alpha$ alone.  The POD method
is surprisingly effective at extracting metal lines in the Lyman forest,
although the effectiveness declines at $z>4$ due to the thick Ly$\alpha$
forest.  While the full POD histograms are rich in information, for
brevity here we focus on the median of the optical depth ratios as a
function of \hbox{H\,{\sc i}} or \hbox{C\,{\sc iv}} optical depth.

We apply the POD method to the 30 lines-of-sight extracted from each
simulation, where we include all 25 ions and continuum-fit the spectra
blueward of Ly$\alpha$ as described in \S\ref{sec: ionize}.  We compare
our results with POD statistics from \citet{agu05} for a sample of 6
quasars with $<z>\approx 3.1$.

\begin{figure*}
\includegraphics[angle=-90,scale=.68]{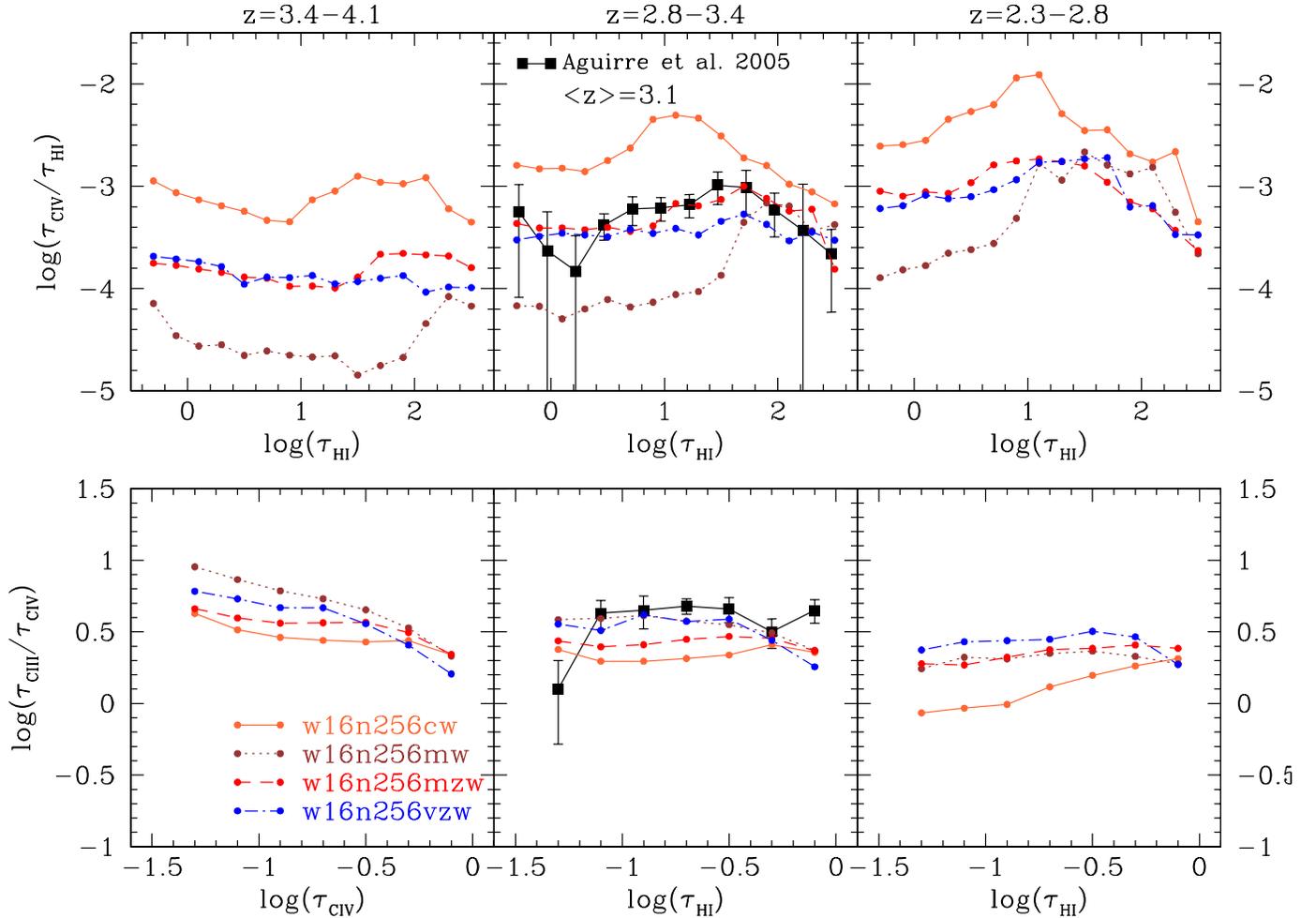}
\caption[]{Median pixel optical depth ratios for our outflow models in
three redshift bins, compared to observations by \citet{agu05} (solid
lines).  The mzw model provides the best fit to the observed
\hbox{C\,{\sc iv}}/\hbox{H\,{\sc i}} ratio where \hbox{H\,{\sc i}} is
a proxy for overdensity, indicating that the metals are properly
spatially distributed in this model.  The vzw model does nearly as
well, while the cw model injects far too many metals and the mw model
is only able to enrich the most overdense regions.  The vzw and mw
models adequately fit the \hbox{C\,{\sc iii}}/\hbox{C\,{\sc iv}} ratio
while the cw and mzw models underestimate this ratio due to excessive
heating of the IGM.  The mismatch at $\tau$(\hbox{C\,{\sc iii}})$\sim
-0.1$ is likely a result of our simulated spectra having lower $S/N$
than the observations resulting in worse Lyman series subtraction at
high optical depths in the \hbox{C\,{\sc iii}} forest.  The
\hbox{C\,{\sc iv}}/\hbox{H\,{\sc i}} ratio grows toward lower redshift
due to increasing metal enrichment, while the \hbox{C\,{\sc
iii}}/\hbox{C\,{\sc iv}} decreases due to heating of the IGM and
decreasing physical densities. }
\label{fig:pod}
\end{figure*}

\subsubsection{\hbox{C\,{\sc iv}}/\hbox{H\,{\sc i}} POD Statistics}

Figure~\ref{fig:pod} (top panels) shows the median \hbox{C\,{\sc
iv}}/\hbox{H\,{\sc i}} ratio as a function of \hbox{H\,{\sc i}}
optical depth for a variety of redshift ranges, for the cw, mw, mzw,
and vzw models (left to right).  This POD statistic shows even more
dramatic differences between the various feedback models than the CDDs.
For comparison, we plot the data from \citet{agu05} in the $z=2.8-3.4$
panels.

The overall trend is that the median POD ratio rises with time.  This
is primarily because the \hbox{H\,{\sc i}} optical depth at a given
overdensity is dropping more rapidly with time than the \hbox{C\,{\sc
iv}} optical depth because as the universe becomes less dense
\hbox{H\,{\sc i}} becomes more ionized, while \hbox{C\,{\sc iv}}
ionizing to \hbox{C\,{\sc v}} is offset somewhat by \hbox{C\,{\sc
iii}} ionizing up to \hbox{C\,{\sc iv}}.  However, a secondary
non-trivial effect is that the IGM is becoming increasingly enriched.
While the median POD ratio is relatively flat with \hbox{H\,{\sc
i}} optical depth, the peak does shift to lower \hbox{H\,{\sc i}}
optical depth with time, owing to winds pushing metals further out
into the IGM.

Comparing to observations, it is clear that the cw model overpredicts the
median POD ratio at $z\approx 3.1$, consistent with over-enrichment of the
IGM as has been seen from the $\Omega$(\hbox{C\,{\sc iv}}) and CDD comparisons.
What is new here is that the shape of the curve is also discrepant, in
that the peak value is produced around $\log\tau$(\hbox{H\,{\sc i}})$\sim 1$,
while observations prefer a peak at $\log\tau$(\hbox{H\,{\sc i}})$\sim 1.5$.
This is indicating that metals are too widely distributed in the cw model,
in addition to being overproduced.

The mw model represents the other extreme, where it fits only the high
density regions ($\log\tau$(\hbox{H\,{\sc i}})$\ga 2$) but does not
distribute the metals widely enough to match at $\log\tau$(\hbox{H\,{\sc
i}})$\la 1.5$.  Clearly, the winds are too weak in this model to enrich
the diffuse IGM.

In contrast, the mzw does and excellent job fitting the full range of
\hbox{H\,{\sc i}} optical depths, indicating the winds are both producing
the correct amount of \hbox{C\,{\sc iv}} as well as distributing it
correctly versus the underlying \hbox{H\,{\sc i}} absorption.  The vzw
model is nearly as good, although it appears to produce slightly too
little \hbox{C\,{\sc iv}} absorption overall.  Without doing a more
careful comparison, we can only say that both these models are in
reasonable agreement with observations.

\citet{agu05} applied the POD method on simulations including a 10
$\hmpc$, $2\times216^3$ particle \gad~simulation from SH03 (the Q4
simulation) with the cw feedback model, but without metal-line cooling.
They added an ad hoc cooling method wherein SPH particles with cooling
times less than Hubble times instantaneously reach $T\sim10^4$ K;
resulting in a \hbox{C\,{\sc iv}}/\hbox{H\,{\sc i}} ratio increasing by
about $\sim\times 10$ with cooling.  Using a test simulation, an 8
$\hmpc$ box vzw simulation with $2\times 128^3$ particles with
and without metal-line cooling, we find the metal-line cooling increases
\hbox{C\,{\sc iv}} absorption only by between 60-90\%, likely because in
the denser regions the IGM is continuously reheated by feedback, while
in the underdense regions the cooling time even with metals approaches
the Hubble time.  Hence the additional effects of metal line cooling
cannot reconcile the cw model with observations.

\subsubsection{\hbox{C\,{\sc iii}}/\hbox{C\,{\sc iv}} Statistics}

The \hbox{C\,{\sc iii}}/\hbox{C\,{\sc iv}} POD relation is used by S03
to provide a constraint on the temperature of the absorbing
\hbox{C\,{\sc iv}}.  This is a difficult measurement to obtain since
\hbox{C\,{\sc iii}} (977\AA) is buried in the Ly$\beta$ forest, but
their method applied identically to observations and simulations shows
that the ratio can still be instructive.  They conclude that the high
ratios ($\tau$(\hbox{C\,{\sc iii}})$/\tau$(\hbox{C\,{\sc iv}})$>3$ at
$z>3$) rule out large fractions of \hbox{C\,{\sc iv}}-absorbing gas at
$T>10^5$K where collisional ionization dominates and the ratio drops
precipitously (refer to their Figure 7 to see how this ratio depends
on density and temperature using the QG background).  \citet{agu05}
find that their \gad\ simulation remained much too hot to reproduce the
observed ratio, and ratios were 0.65 dex too low even when including
metal cooling, only a modest improvement.

We find that our cw model indeed produces too low a median POD ratio.
This indicates that the IGM is too hot in this model compared
with observations, as we also saw in the $b$-parameter comparison.
In contrast, our momentum-wind models fare considerably better.  The mzw
model still appears to heat the IGM too much, but the vzw and mw models
(with relatively low wind speeds that do not add much heat to the IGM)
fare quite well.

S03 points out that the observed \hbox{C\,{\sc iii}}/\hbox{C\,{\sc
iv}} ratios are inconsistent with the absorbing gas being at $T\ga
10^5$K.  However, even the vzw and mw models have a substantial
fraction of their \hbox{C\,{\sc iv}}-absorbing gas collisionally
ionized, about 30\% at $z\sim 3$.  Despite this, these models match
the POD ratios quite well, showing that some collisionally-ionized
\hbox{C\,{\sc iv}} is still accommodated by observations.

S03 further observes a gradual decline in $\tau$(\hbox{C\,{\sc
iii}})$/\tau$(\hbox{C\,{\sc iv}}) with time.  All wind models
produce a gradual decline qualitatively consistent with these data.
In photo-ionized gas this decline is mainly driven by Hubble expansion
increasing the IGM ionization level, and secondarily by the increasing
temperature of the IGM.

In summary, pixel optical depth techniques provide a complementary
technique to Voigt profile fitting for examining IGM \hbox{C\,{\sc
iv}} absorption.  The overall results from both of these techniques
support the same basic conclusions: Our cw model overproduces metals,
distributes them too widely, and makes the IGM too hot, while the mw model
underproduces metals, does not distribute them widely enough, and keeps
the IGM (possibly) too cool.  In between lie the mzw and vzw models that
seem to span the allowed range of metal and energy input.  More generally,
these observations already provide a non-trivial bracketing of the allowed
energy and metal input into the IGM as a function of redshift.  More careful
comparisons with observations, particularly pushing to higher and lower
redshifts, should allow us to provide strong constraints on the properties
of galactic outflows across cosmic time.

\section{Numerical Resolution}\label{sec: resolution}

Numerical simulations by necessity have a finite dynamic range, and
even advanced ones such as ours cannot fully resolve the relevant scales
from star formation to the Hubble volume.  Hence we must ensure that our
results are not sensitive to issues of numerical resolution.  For this
purpose we have run with 8, 16 and 32 $\hmpc$ box lengths, as described
in \S\ref{sec: sims}, which span a factor of 4 in spatial resolution and a
factor of 64 in mass resolution .  The results presented so far have been
derived from the $16\hmpc$ volumes for each outflow model, because as we
will show in this section, their results show good convergence relative
to the higher-resolution $8\hmpc$ runs (at least for the momentum-driven
wind models), but the $32\hmpc$ runs do not show good convergence.
We present simple arguments that explain this behavior in terms of the
Jeans length in the moderately-overdense IGM giving rise to \hbox{C\,{\sc
iv}} absorption.  Finally, we show that the constant wind model shows poor
convergence at all resolutions.  We focus on the convergence properties of
the observable quantities that we have compared to in previous sections.

\begin{figure}
\includegraphics[angle=-90,scale=.32]{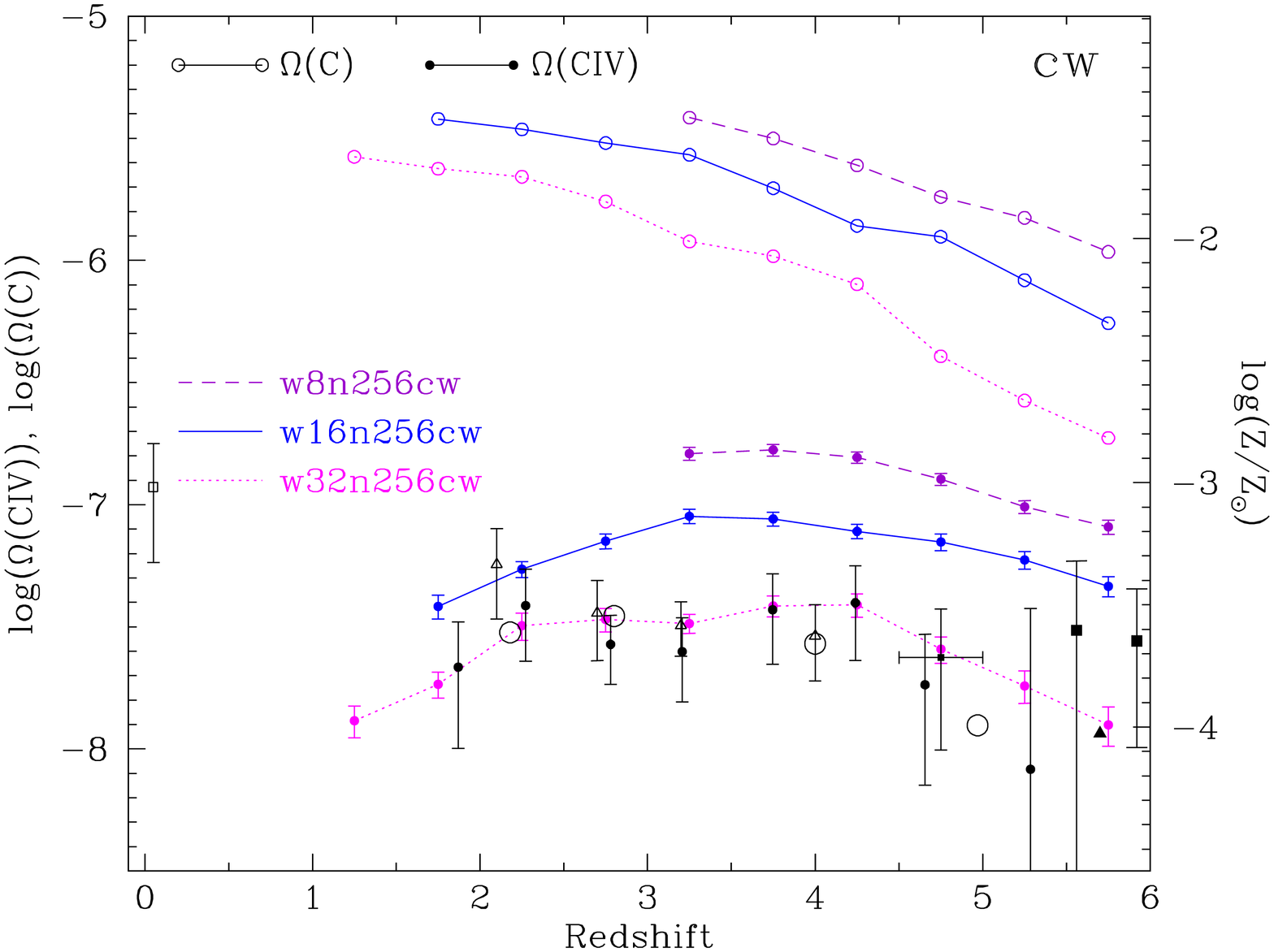} 
\includegraphics[angle=-90,scale=.32]{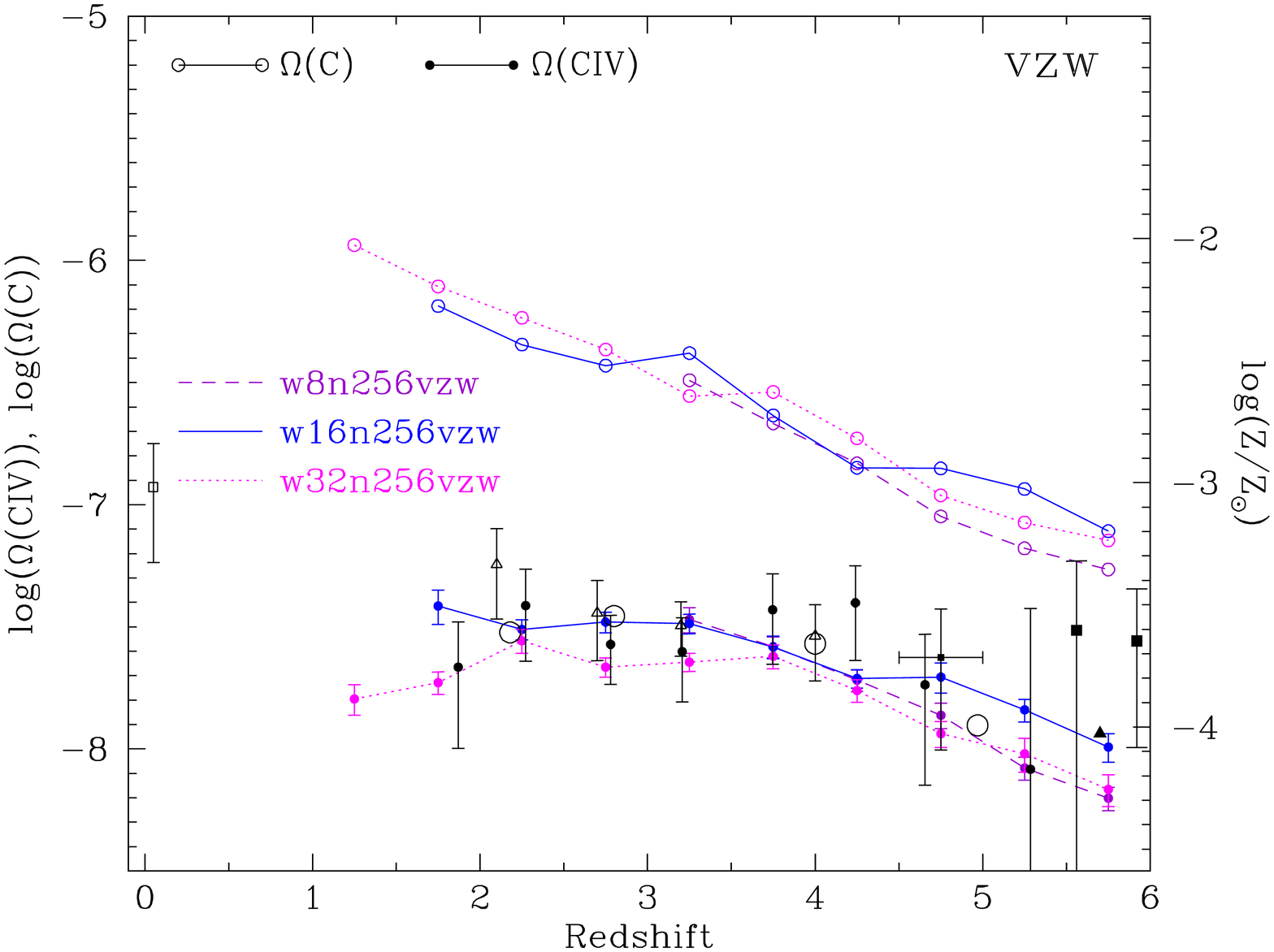} 
\caption[]{The numerical resolution dependence of $\Omega$(C) (as
measured along our lines of sight) and $\Omega$(\hbox{C\,{\sc iv}}) in
the cw and vzw models in the 8, 16, and 32 $\hmpc$ boxes with $256^3$
gas and dark particles.  The same data points described in
Figure~\ref{fig:omegac4} are displayed.  $\Omega$(C) and
$\Omega$(\hbox{C\,{\sc iv}}) converge quite well in momentum-driven
wind models, unlike for the constant wind case.  The
$\Omega$(\hbox{C\,{\sc iv}}) measurement suggests slightly larger
ionization corrections at lower redshift in larger boxes.  }
\label{fig:omega_res}
\end{figure}

In Figure~\ref{fig:omega_res} we show $\Omega$(\hbox{C\,{\sc iv}}) and
$\Omega$(C) (or more accurately the carbon in the IGM) for the cw
(top) and vzw (bottom) models.  The cw model shows poor convergence,
with more metals and \hbox{C\,{\sc iv}} absorption being produced in
higher resolution simulations.  This is in stark contrast with the
global star formation rate in these models, which shows excellent
convergence at late times as demonstrated by SH03 (higher-resolution
simulations do resolve earlier star formation).  In the cw model, all
galaxies roughly enrich the same volume independent of galaxy mass
since they eject winds at the same speed.  Hence higher resolution
simulations that resolve star formation in smaller halos at earlier
times distribute metals much more widely.  To obtain a convergent IGM
enrichment in a constant wind model requires resolving every galaxy
that has ever driven a wind, which is exceptionally challenging
computationally.  Since the disagreement with observations is even
greater in higher resolution runs, this model is unlikely to be
successful in explaining IGM enrichment.

The vzw model, on the other hand, shows much better resolution
convergence over most of the redshift range in both
$\Omega$(\hbox{C\,{\sc iv}}) and $\Omega$(C), with a minor exception
at $z\ga 5$ where the stochastic nature of biased early star formation
may be producing differences between the realizations.

The impact of a finite volume is also seen in the vzw model at $z\la
3$.  Here, the $32\hmpc$ volume produces a larger global \hbox{C\,{\sc
iv}} ionization correction, owing to larger and hotter structure being
formed in the larger volume.  The differences are small, but
illustrate that large volumes are needed as well as high resolution in
order to properly model the IGM.  Hence it is desirable to use the
largest possible volume that is resolution-converged.

\begin{figure*}
\includegraphics[angle=-90,scale=.64]{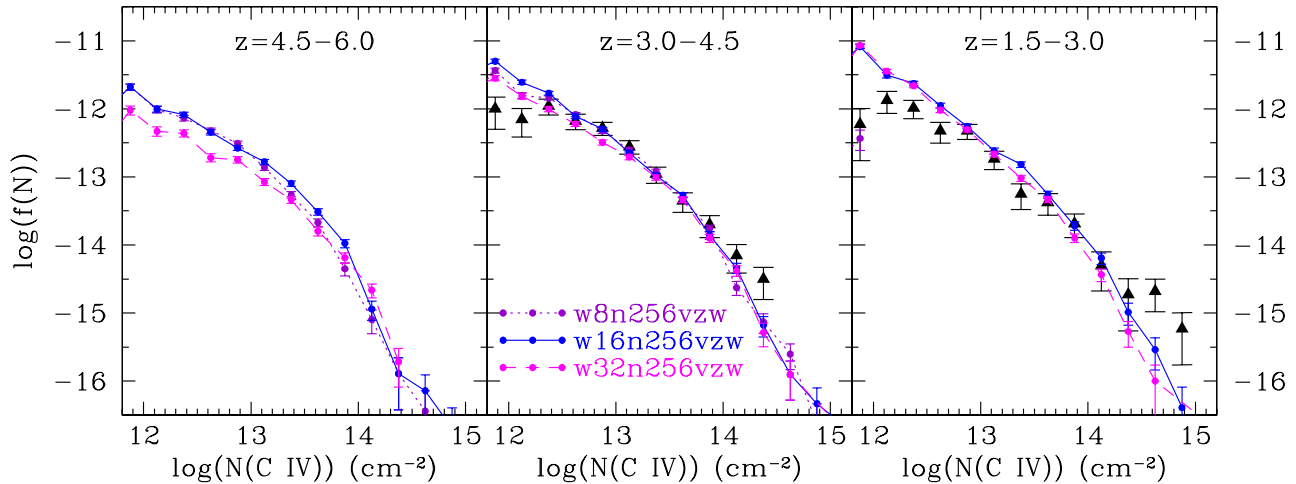} 
\caption[]{Column density distributions of \hbox{C\,{\sc iv}} for the
w8n256vzw, w16n256vzw, and w32n256vzw simulations, compared to the
BSR03 points (black triangles).  The 8 and 16~$\hmpc$ simulations
appear resolution converged, while the 32 $\hmpc$ box underestimates
weak absorbers at high-$z$ and possibly stronger absorbers at low-$z$.
}
\label{fig:c4cdd_res}
\end{figure*}

Despite these minor issues, our overall conclusions should be robust
as long as the simulations are capable of accurately resolving the
properties of \hbox{C\,{\sc iv}} absorbers.  As
Figure~\ref{fig:c4cdd_res} shows, this is a good assumption for the
CDD, with the 8 and $16\hmpc$ vzw volumes showing good convergence
over most of the redshift range.  The other momentum-driven winds
follow similar trends, while cw is not converged (as with
$\Omega$(\hbox{C\,{\sc iv}}); not shown).  The deviations in the vzw
case are primarily associated with the $32\hmpc$ volume.  It
underestimates weak absorbers at high-$z$ because it cannot fully
resolve all metal production, and it underestimates strong absorbers
at low-$z$ because strong \hbox{C\,{\sc iv}} systems arise in small
galactic halos that it cannot resolve.  Hence our 16~$\hmpc$ volume is
optimal.  Unfortunately, we cannot directly determine if our
16~$\hmpc$ volume is robust at $z<3$, since we have only evolved our
8~$\hmpc$ volume to $z=3$ due to CPU time considerations.  Some
effects of multi-phase collapse in high density regions may therefore
persist in this volume, which may explain the deficit present in all
models compared to data at the largest column densities.

\begin{figure*}
\includegraphics[angle=-90,scale=.64]{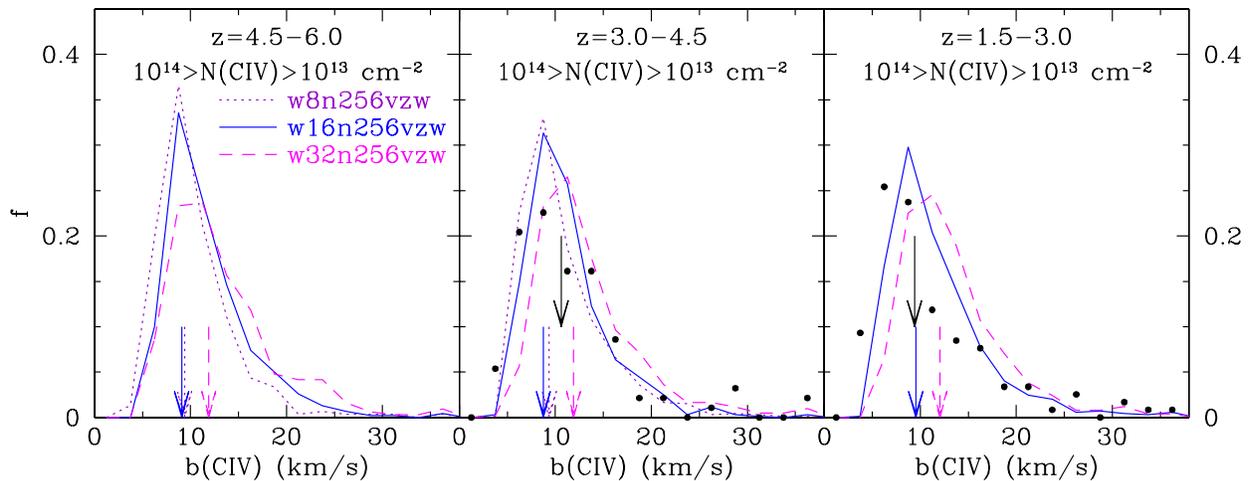}
\caption[]{The $b$-parameters distributions in our three volumes for the
vzw model.  This demonstrates the w8n256vzw and w16n256vzw absorbers
are resolution-converged and fit the observations by BSR03, while the
w32n256vzw absorbers are too wide.  Using a Jeans length argument (see
text), we show that the 32 $\hmpc$ often has too poor resolution to
properly resolve diffuse \hbox{C\,{\sc iv}} absorbers.  }
\label{fig:bparam_res}
\end{figure*}

The linewidths provide an even more stringent test of resolution,
because they are keenly sensitive to small fluctuations in diffuse
gas.  Figure~\ref{fig:bparam_res} shows the $b$-parameter distribution
for the vzw model in three redshift intervals.  While the 8 and
$16\hmpc$ agree very well, the w32n256vzw model clearly shows wider
$b$-parameters at all redshifts.  If we split the width of the lines
up into their thermal and spatial component as we did in \S\ref{sec:
bparam}, we find that the increase is mostly in $b_{sp}$.  At $z>4.5$,
we find $w$(\hbox{C\,{\sc iv}})$ = 12.7$, 11.5, and 16.5 kpc for the
8, 16, and 32 $\hmpc$ boxes respectively.  Hence the line widths, as
with the CDDs, are well converged up to $16\hmpc$ but falter at lower
resolution.

Can we understand why the 16~$\hmpc$ volumes are converged but
32~$\hmpc$ ones are not?  To do so, we estimate the minimum spatial
resolution needed to properly model IGM fluctuations at any given
density.  We begin with the minimum SPH resolution, which is
approximately $0.5~l_{box}/n_{part}\times(\rho/\bar{\rho})^{-1/3}$, or
$0.5~\bar{d}_{mis} (\rho/\bar{\rho})^{-1/3}$ where $\bar{d}_{mis}$ is
the mean interparticle spacing (i.e. a single resolution element
generally requires two SPH particles in any given direction).  The
physical minimum optically-thin absorber size depends on the Jeans
length, the limit below which the thermal kinetic energy prevents the
IGM from collapsing.  The Jeans length is $r_{j} =
975~\sqrt{T_4/[(\rho/\bar{\rho})(1+z)^3]}$~kpc at a given density
$\rho$, where $T_4=T/10^4$.  Equating the minimum SPH resolution to
$r_j$, we derive the following expression for the maximum
interparticle spacing allowable in a simulation to resolve an
absorber,

\begin{equation}
d_{res} = 170~(\rho/\bar\rho)^{-1/6} (4/(1+z))^{-3/2} \sqrt{T_4}~\hkpc.
\end{equation}

Because of the inverse density dependence in $d_{res}$, SPH simulations
actually tend to resolve the low-density IGM fluctuations slightly better than
the high-density ones, at a given temperature.  However, particularly
at high redshifts, low-density fluctuations tend to be colder (see
Figure~\ref{fig:avec4lines}), making them the dominant resolution
constraint.  The typical value of 170~$\hkpc$ above is uncomfortably
close to our 32~$\hmpc$ volume's $\bar{d}_{mis} = 125 \hkpc$, hence
this volume is starting to fail to resolve some absorbers, resulting in
poor convergence.

At low-$z$, \citet{sim06} observed strong metal-line absorbers to have
$T\sim10^4$~K and $\rho/\bar\rho\sim100$ between $z=2.7\rightarrow
1.8$.  These are similar to the physical parameters of our strongest
\hbox{C\,{\sc iv}} systems at these redshifts.  At $z=2$, we obtain
$\bar{d}_{res} = 51 \hkpc$ for these parameters, which would not quite
be resolved even in our 16~$\hmpc$ runs ($\bar{d}_{mis} = 63 \hkpc$).
Hence the deficit in strong systems in our models may be due to lack
of resolution.

To summarize, the resolution convergence for the vzw model is quite
good up to our $16\hmpc$ volume, which is precisely why we chose to
quote predictions for this volume throughout the paper.  Our $32\hmpc$
shows poorer convergence, but it also suggests that at lower redshifts
($z\la 3$) the small $16\hmpc$ volume may slightly impact strong
absorber statistics.  We have checked that our results for the vzw
case are generally applicable to all the momentum-driven wind models.
In contrast, the cw model shows poor resolution convergence, and the
disagreements with observations are significantly exacerbated as one
moves to higher resolution.

\section{Conclusions} \label{sec: sum}

We have explored various models of galactic outflows incorporated into
cosmological hydrodynamic simulations of structure formation, and compared
them with observations of IGM enrichment at $z\ga 1.5$.  With no outflows,
we show \citep[in agreement with previous studies, e.g.][]{agu01} that
a smattering of metals are distributed out to moderate overdensities
through dynamical interactions, but cannot come close to enriching the
diffuse IGM to observed values.  Hence high-velocity outflows are required
to enrich the IGM to observed values.  In order to allow flexibility and
perform controlled tests of various scenarios, we implement these outflows
``by hand", tying the wind velocities ($v_w$) and mass loading factors
($\eta$) to galaxy properties through various parameterizations.

In this paper we show tests of five outflow models.  Most of these models,
and all successful ones, are based on momentum-driven wind scenarios that
are favored by observations of local starburst galaxies.  We consider
a constant-wind (cw) model that produces strong, early outflows; a weak
momentum-wind model (mw) that yields little IGM heating; an ``observed"
wind model (zw) obtained by taking the observed scalings of local outflows
at face value (not a momentum-driven scenario); a strong momentum-wind
model (mzw) calibrated by local starburst observations of \citet{mar05};
and a variable momentum wind model (vzw), where we have included a
spread in wind speeds in accord with observations of \citet{rup05}.
Our models are chosen to span an interesting range in outflow strengths,
while broadly reproducing the global star formation rate density evolution
down to $z=1.5$.

Our first main result is that the outflow models favored by local
starburst observations, when applied to galaxies forming hierarchically
across cosmic time, are able to reproduce a wide range of \hbox{C\,{\sc
iv}} absorption line observations during the epoch of peak cosmic star
formation ($z\sim 1.5-5.5$).  We further demonstrate that this agreement
is non-trivial, requiring particular levels of metal and energy input
into the IGM as a function of time.  In particular, winds should have
high enough speeds to enrich the diffuse IGM at early times, but low
enough speeds not to overheat the IGM.  Winds should also have mass
loading factors high enough to suppress early star formation, but low
enough to produce enough early metals.  Generically, momentum-driven wind
models produce the desired scalings, because they heavily suppress early
star formation but are able to distribute many of the metals formed into
the IGM due to the high early mass loading factors in small galaxies.
Yet, their low wind speeds from small galaxies do not overheat the IGM.
It is therefore quite remarkable and non-trivial that our stronger
momentum-driven wind scenarios (mzw and vzw), based on physical models
calibrated by local observations of outflows, naturally reproduce a
wide variety of observations.

Our second key result is that the lack of evolution observed in the
properties of \hbox{C\,{\sc iv}} absorption such as the global mass
density in \hbox{C\,{\sc iv}} systems ($\Omega$(\hbox{C\,{\sc iv}})
from $z\sim 5.5\rightarrow 1.5$ is not necessarily indicative of a
lack of continual enrichment, as has been argued previously
\citep[e.g.][]{son01,sca02}.  In all our models, most of the metals
are ejected into the IGM during the heydey of star formation at $z\la
6$, but our successful models also inject sufficient heat into the IGM
to generate a drop in the \hbox{C\,{\sc iv}} ionization fraction that
roughly balances the increase in metallicity.  The success of the mzw
and vzw models in matching $\Omega$(\hbox{C\,{\sc iv}}) evolution
renders unnecessary a putative population of early star formation that
enriches a large fraction of the IGM at $z\ga 6$.

Comparing various models to $\Omega$(\hbox{C\,{\sc iv}}) data showed that
\hbox{C\,{\sc iv}} is overproduced in the cw and zw models, primarily
because early star formation is insufficiently suppressed and copious
metals are produced.  While the large heat input from the cw model
partially counteracts metal overproduction by lowering the \hbox{C\,{\sc
iv}} ionization fraction, it is insufficient to bring this model into
agreement with data.  Note that a comparison to observed $z\sim 6$ galaxy
luminosity functions by \citet{dav06} also indicated that star formation
is not suppressed sufficiently early on in the cw model.  One could
envision lowering the wind speed in the cw model to reduce the heat input
and lower $\Omega$(\hbox{C\,{\sc iv}}), but this would exacerbate the
overcooling problem at high redshifts.  The highest-$z$ observations of
$\Omega$(\hbox{C\,{\sc iv}}) tend to favor models that suppress early star
formation, which occurs naturally in the momentum-wind driven scenarios
(mzw, vzw) owing to high mass loading factors in small galaxies.

The column density distributions of \hbox{C\,{\sc iv}} absorbers present a
test of the spatial distribution of metals.  Models with overly weak winds
such as mw cannot pollute the low overdensity ($\rho/\bar\rho\sim$~few)
IGM to reproduce the population of weak absorbers seen at $z\sim 3-4$.
Once winds are strong enough (vzw, mzw, even cw and zw), the shape and
(lack of) evolution of the CDD is roughly as observed.

Observations of the \hbox{C\,{\sc iv}} linewidth ($b$-parameter)
distribution shows little evolution, which again is generally reproduced
in all our models.  The linewidths provide constraints on the temperature
of absorbing gas, which seem to accommodate some but not too much feedback
energy input, as provided by the vzw, mzw, and mw models.

Pixel optical depth (POD) ratios provide a different way of characterizing
\hbox{C\,{\sc iv}} absorption, yet yield broadly similar conclusions.
The \hbox{C\,{\sc iv}}/\hbox{H\,{\sc i}} POD ratio tests the amount and
spatial distribution of the metals, because \hbox{H\,{\sc i}} absorption
is correlated with underlying gas density.  Comparisons with our models
show that cw distributes metals too widely, mw not widely enough, and
mzw and vzw lying in the intermediate range that agrees with data.
The \hbox{C\,{\sc iii}}/\hbox{C\,{\sc iv}} POD ratio is a sensitive
measure of temperature, and shows that the cw and possibly even mzw heat
the IGM too much, while the vzw and mw models match well.

In short, vzw and to a slightly lesser extent mzw are the only models
that match all observations of \hbox{C\,{\sc iv}} absorption we have
considered, while reproducing a promising match to the global observed
star formation history.  These observations provide sensitive tests of
the metallicity and temperature structure of the IGM, making the
agreement of these models an impressive achievement.  The fact that
these models are consistent with local observations of outflows
provides an interesting connection between starbursts-driven outflows
at all epochs.  

Our simulations also provide insights into the physical effects of
outflows on the IGM.  Both the vzw and mzw models show relatively
low volume filling factor of $Z>10^{-3}Z_\odot$ gas, increasing from
$\sim 2\rightarrow 10$\% from $z=6\rightarrow 2$.  Hence widespread
enrichment of the IGM is not required to match observations, and in
fact is disfavored as models that distribute metals more widely tend to
overproduce \hbox{C\,{\sc iv}} absorption.  These results broadly agree
with measurements of the quiescence of the IGM at high redshift that
suggest outflows cannot have impacted a significant volume \citep{rau05}.
The metallicity-overdensity relation in the IGM is not a power-law as
is sometimes assumed, and instead shows relatively flat enrichment
at high overdensities, dropping off steeply at lower overdensities.
An interesting prediction of our models is that a significant
fraction of \hbox{C\,{\sc iv}} lines are collisionally ionized.
However, this does not violate temperature constraints on the IGM as seen
from $b$-parameter distributions or \hbox{C\,{\sc iii}}/\hbox{C\,{\sc
iv}} POD ratios.  The median \hbox{C\,{\sc iv}} absorber traces higher
overdensity and hotter gas at lower redshifts, and the optical depth is
only weakly correlated with underlying gas density, unlike the tight
relationship inferred for Ly$\alpha$ absorbers \citep[e.g.][]{dav99}.
This primarily owes to the large range of ionization fractions
in \hbox{C\,{\sc iv}} systems, from cooler photoionized absorbers to
collisionally ionized systems reaching $T\sim 3\times 10^5$K.  The idea
that \hbox{C\,{\sc iv}} absorption arises almost exclusively in diffuse,
photoionized gas following a tight density-temperature relation similar
to \hbox{H\,{\sc i}} absorbers is not favored by our models.

While we present five wind models in this paper, we have actually
simulated over a dozen wind models, and decided to focus on these five
because they are physically and/or observationally motivated, and best
illustrate the relationship between outflow properties and IGM enrichment.
Nevertheless, the range of outflow models investigated in this work is
far from exhaustive, as one could envision more complex recipes that
might still be consistent with observations.  Hence our favored models
should be regarded as plausible, but not uniquely successful.

Further uncertainties in the exact model that best describes galactic
outflows comes from a range of systematic uncertainties we have not
extensively considered in this work.  One uncertainty is the impact
of Helium reionization (investigated briefly in \S\ref{sec: omega}),
which can impact the comparison of \hbox{C\,{\sc iv}} absorption at
early epochs.  Unfortunately including this effect requires fairly
precise knowledge of the topology and epoch of Helium reionization,
which is poorly constrained at present.  A second uncertainty is that
our simulations were run with a cosmology that is slightly different
than that now favored by the 3rd-year WMAP release.  In particular, the
new values of $\sigma_8$ (now 0.75 as opposed to 0.9 as we used) and
$\Omega_m$ (0.24 vs. 0.30) will significantly lower the halo collapse
fraction at high redshift \citep{spe06}.  This means that the mass
loading factors will need to be reduced relative to equation~\ref{eqn:
massload} to match observations of high-redshift star formation rates.
The net effect should be small: Given that observations dictate a
certain global star formation rate, any wind model that matches it will
produce a similar amount of metals.  A third uncertainty involves the
possible contribution of Population III stars to early IGM enrichment.
Our models, while not requiring it, cannot rule out some contribution
from early massive star formation that might have higher yields and
larger supernova energy input.  More generally we have not considered
the possibility of IMF variations, although recent work hints at this
\citep{far06}.  A fourth uncertainty arises from possible contributions
from AGN feedback in unbinding enriched gas from galaxies, as bright
quasars have been observed at $z>6$ \citep{fan06}.  A fifth uncertainty
is the effect of enriched stellar winds tecycling metals into the ISM,
mostly from asymptotic giant branch stars, which will increase the
net carbon yield from stars.  Despite these caveats, the fact that a
plausible feedback model exists connecting local starbursts and high-$z$
IGM enrichment represents significant progress towards quantifying the
impact of outflows on galaxies and the IGM across cosmic time.

Further constraints on outflows can be provided by examining IGM metal
evolution to $z\sim 0$, studying early IGM metal enrichment extending
into the reionization epoch, matching the mass-metallicity evolution
in galaxies, and reproducing the abundances patterns and gradients in
galaxy clusters, among other things.  We are currently investigating
all these aspects, especially the impact of various outflow models on
galaxy populations \citep[see][]{dav06}.  On the theoretical side, it is
of course preferable to generate and drive winds self-consistently out
of galaxies rather than having to include them ``by hand", and progress
is being made towards that end \citep[e.g.][]{sca06}.  Observationally,
we stress the need to obtain $z>4.5$ observations of \hbox{C\,{\sc iv}}
capable of resolving thermal linewidths in order to constrain the physical
nature of early feedback and metallicity enrichment.  In short, there
remains much work to be done in order to obtain a fully self-consistent
model of metal evolution and distribution in the universe.  Our work
represents a small but important step forward in this area.

\section*{Acknowledgments}  \label{sec: ack}

We thank A. Aguirre, K. Finlator, L. Hernquist, N. Katz, M. Pettini,
E. Ryan-Weber, J. Schaye, and R. Simcoe for their helpful input.  We thank
V. Springel and L. Hernquist for the use of \gad~before public release,
and E. Scannapieco for providing us with his database of \hbox{C\,{\sc
iv}} lines.  The simulations were run on the Xeon Linux Supercluster
at the National Center for Supercomputing Applications, and on our
department's 100-processor Beowulf system at the University of Arizona.
Support for this work, part of the Spitzer Space Telescope Theoretical
Research Program, was provided by NASA through a contract issued by the
Jet Propulsion Laboratory, California Institute of Technology under a
contract with NASA.  Support for this work was also provided by NASA
through grant number HST-AR-10647.01 from the SPACE TELESCOPE SCIENCE
INSTITUTE, which is operated by AURA, Inc. under NASA contract NAS5-26555.

\label{lastpage}


\begin{thebibliography}{}
\bibitem[\protect\citeauthoryear{Adelberger et al.} {2003}]{ade03}Adelberger, K. L., Steidel, C. C., Shapley, A. E., \& Pettini, M. 2003, ApJ, 584, 45
\bibitem[\protect\citeauthoryear{Adelberger et al.} {2005}]{ade05} Adelberger K. L., Shapley A. E., Steidel C. C., Pettini M., Erb D. K., \& Reddy N. A. 2005, ApJ, 629, 636
\bibitem[\protect\citeauthoryear{Aguirre et al.} {2001}]{agu01} Aguirre, A., Hernquist, L., Schaye, J.,  Katz, N., Weinberg, D. H., \& Gardner, J. 2001, ApJ, 561, 521
\bibitem[\protect\citeauthoryear{Aguirre et al.} {2002}]{agu02} Aguirre, A., Schaye, J., \& Theuns, T. 2002, ApJ , 576, 1
\bibitem[\protect\citeauthoryear{Aguirre et al.} {2005}]{agu05} Aguirre, A., Schaye, J. Hernquist, L., Kay, S., Springel, V., \& Theuns, T. 2005, ApJ, 620, L13
\bibitem[\protect\citeauthoryear{Aracil et al.} {2004}]{ara04} Aracil, B., Petitjean, P., Picho, C., \& Bergeron, J. 2004, A\&A, 419, 811
\bibitem[\protect\citeauthoryear{Balogh et al.} {2001}]{bal01} Balogh, M. L., Pearce, F. R., Bower, R. G., \& Kay, S. T. 2001, MNRAS, 326, 1228
\bibitem[\protect\citeauthoryear{Becker et al.} {2006}]{bec06} Becker, G. D., Sargent, W. L. W., Rauch, M., \& Simcoe, R. A. 2006, ApJ, 640, 69
\bibitem[\protect\citeauthoryear{Bertone \& White} {2006}]{ber06} Bertone, S. \& White, S. D. M. 2006,  MNRAS, 367, 247
\bibitem[\protect\citeauthoryear{Bokensberg, Sargent, \& Rauch} {2003}]{bok03} Boksenberg, A., Sargent, W. L. W., \& Rauch, M. 2003, ASP Conference Proceedings, Vol. 297, 447, eds. Edited E. Perez, R.M.G. Delgado, \& G. Tenorio-Tagle (BSR03)
\bibitem[\protect\citeauthoryear{Bolton et al.} (2006)]{bol06}  Bolton, J. S., Haehnelt, M. G., Viel, M., \& Carswell, R. F. 2006, MNRAS, 366, 1378
\bibitem[\protect\citeauthoryear{Carswell et al.} {1987}]{car87} Carswell, R. F., Webb, J. K., Baldwin, J. A., \& Atwood, B. 1987, ApJ, 319, 709
\bibitem[\protect\citeauthoryear{Cen et al.} {2005}]{cen05} Cen, R., Nagamine, K., \& Ostriker, J. P. 2005, ApJ, 635, 86
\bibitem[\protect\citeauthoryear{Cole et al.} {2001}]{col01} Cole, S., et al. 2001, MNRAS, 326, 255
\bibitem[\protect\citeauthoryear{Croft et al.} {1998}]{cro98} Croft, R. A. C., Weinberg, D. H., Katz, N., \& Herquist, L. 1998, ApJ, 495, 44
\bibitem[\protect\citeauthoryear{Dav\'e et al.} {1997}]{dav97} Dav\'e, R., Hernquist, L., Weinberg, D. H., \& Katz, N. 1997, ApJ, 477, 21
\bibitem[\protect\citeauthoryear{Dav\'e et al.} {1998}]{dav98} Dav\'e, R., Hellsten, U., Hernquist, L., Katz, N., \& Weinberg, D. H. 1998, ApJ, 509, 661
\bibitem[\protect\citeauthoryear{Dav\'e et al.} {1999}]{dav99} Dav\'e, R., Hernquist, L., Katz, N., \& Weinberg, D. H. 1999, ApJ, 511, 521
\bibitem[\protect\citeauthoryear{Dav\'e et al.} {2001}]{dav01} Dav\'e, R. et al. 2001, ApJ, 552, 473
\bibitem[\protect\citeauthoryear{Dav\'e \& Tripp} {2001}]{dav01b} Dav\'e, R. \& Tripp, T. M. 2001, ApJ, 553, 528
\bibitem[\protect\citeauthoryear{Dav\'e et al.} {2006}]{dav06} Dav\'e, R., Finlator, K., \& Oppenheimer, B. D. 2006, MNRAS, in press, astro-ph/0511532
\bibitem[\protect\citeauthoryear{Dekel \& Silk} {1986}]{dek86} Dekel, A. \& Silk, J. 1986, ApJ, 303, 39
\bibitem[\protect\citeauthoryear{Eisenstein \& Hu} {1999}]{eis99} Eisenstein, D. J. \& Hu, W. 1999, ApJ, 511, 5
\bibitem[\protect\citeauthoryear{Ellison et al.} {1999}]{ell99} Ellison, S. L., Pettini, M., Lewis, G. F., Songaila, A., Cowie, L. L. 1999, Ap\&SS, 269, 201
\bibitem[\protect\citeauthoryear{Ellison et al.} {2000}]{ell00} Ellison, S. L., Songaila, A., Schaye, J., \& Pettini, M. 2000, AJ, 120, 1175
\bibitem[\protect\citeauthoryear{Engelbracht et al.} {2006}]{eng06} Engelbracht, C. W. et al. 2006, ApJ, 642, L127
\bibitem[\protect\citeauthoryear{Erb et al.} {2006}]{erb06} Erb, D. K., Shapley, A. E., Pettini, M., Steidel, C. C., Reddy, N. A, \& Adelberger, K. L. 2006, ApJ, accepted, astro-ph/0602473
\bibitem[\protect\citeauthoryear{Fan et al.} {2006}]{fan06}  Fan, X. et al. 2006, AJ, 132, 117
\bibitem[\protect\citeauthoryear{Fardal et al.} {2006}]{far06}  Fardal, M. A., Katz, N., Weinberg, D. H., \& Dav'e, R. 2006, MNRAS, submitted, astro-ph/0604534
\bibitem[\protect\citeauthoryear{Ferrara et al.} {2000}]{fer00} Ferrara, A., Pettini, M., \& Shchekinov, Y. 2000, MNRAS, 319, 539
\bibitem[\protect\citeauthoryear{Fujita et al.} {2004}]{fuj04} Fujita, A., Mac Low, M.-M., Ferrara, A., Meiksin, A. 2004, ApJ, 613, 159
\bibitem[\protect\citeauthoryear{Frye et al.} {2003}]{fry03} Frye, B. L., Tripp, T. M., Bowen, D. B., Jenkins, E. B., \& Sembach, K. R. 2003, in ``The IGM/Galaxy Connection: The Distribution of Baryons at z=0'', ASSL Conference Proceedings Vol. 281, 231, eds. J.L. Rosenberg \& M.E. Putman
\bibitem[\protect\citeauthoryear{Gnedin \& Ostriker} {1997}]{gne97} Gnedin, N. Y. \& Ostriker, J. P. 1997, ApJ, 486, 581
\bibitem[\protect\citeauthoryear{Haardt \& Madau} {1996}]{haa96} Haardt, F. \& Madau, P. 1996, ApJ, 461, 20
\bibitem[\protect\citeauthoryear{Haardt \& Madau} {2001}]{haa01} Haardt, F. \& Madau, P. 2001, in ``Clusters of galaxies and the high redshift universe observed in X-rays, Recent results of XMM-Newton and Chandra", XXXVIth Rencontres de Moriond , eds. D.M. Neumann \& J.T.T. Van.
\bibitem[\protect\citeauthoryear{Heckman et al.} {2000}]{hec00} Heckman, T. M., Lehnert, M. D., Strickland, D. K., \& Armus, L. 2000, ApJS, 129, 493
\bibitem[\protect\citeauthoryear{Heckman} {2003}]{hec03} Heckman, T. M. 2003, in Avila-Reese V., Firmani C., Frenk C. S., Allen C., eds, Rev. Mex. Astron. Assoc., Vol. 17, Galaxy Evolution: Theory \& Observations, p. 47
\bibitem[\protect\citeauthoryear{Hellsten et al.} {1998}]{hel98} Hellsten, U., Hernquist, L., Katz, N., \& Weinberg, D. H. 1998, ApJ, 499, 172
\bibitem[\protect\citeauthoryear{Hernquist et al.} {1996}]{her96} Hernquist, L., Katz, N., Weinberg, D. H., \& Miralda-Escud\'e, J. 1996, ApJ, 457, :51
\bibitem[\protect\citeauthoryear{Hui \& Gnedin} {1997}]{hui97} Hui, L. \& Gnedin, N. Y. 1997, MNRAS, 292, 27
\bibitem[\protect\citeauthoryear{Hopkins} {2004}]{hop04} Hopkins, A. M. 2004, ApJ, 615, 209
\bibitem[\protect\citeauthoryear{Jenkins et al.} {2001}]{jen01} Jenkins, A., Frenk, C. S., White, S. D. M., Colberg, J. M., Cole, S., Evrard, A. E., Couchman, H. M. P., \& Yoshida, N. 2001, MNRAS, 321, 372
\bibitem[\protect\citeauthoryear{Katz, Weinberg, Hernquist} {1996}]{kat96} Katz, N., Weinberg, D. H., \& Hernquist, L. 1996, ApJS, 105, 19
\bibitem[\protect\citeauthoryear{Kennicutt} {1998}]{ken98} Kennicutt, R. C. 1998, ApJ, 498, 541
\bibitem[\protect\citeauthoryear{Kere\v{s} et al.} {2005}]{ker05} Kere\v{s}, D., Katz, N., Weinberg, D. H., \& Dav\'e, R. 2005, MNRAS, 363, 2
\bibitem[\protect\citeauthoryear{Kirkman et al.} {2005}]{kir05} Kirkman, D., Tytler, D., Suzuki, N., Melis, C., Hollywood, S., James, K., So, G., Lubin, D., Jena, T.,; Norman, M. L., \& Paschos, P. 2005, MNRAS, 360, 1373
\bibitem[\protect\citeauthoryear{Madau et al.} {1996}]{mad96} Madau, P., Ferguson, H. C., Dickinson, M. E., Giavalisco, M., Steidel, C. C., \& Fruchter, A. 1996, MNRAS, 283, 1388
\bibitem[\protect\citeauthoryear{Maller \& Bullock} {2004}]{mal04} Maller, A. H., \& Bullock, J. S. 2004, MNRAS, 355, 694
\bibitem[\protect\citeauthoryear{Martin} {1999}]{mar99} Martin, C. L. 1999, ApJ, 513, 156
\bibitem[\protect\citeauthoryear{Martin} {2005}]{mar05} Martin, C. L. 2005, ApJ, 621, 227
\bibitem[\protect\citeauthoryear{McKee \& Ostriker} {1977}]{mck77} McKee, C. F. \& Ostriker, J. P.\ 1977, ApJ, 218, 148
\bibitem[\protect\citeauthoryear{Mo \& Miralda-Escud\'e} {1996}]{mo96} Mo, H. J. \& Miralda-Escud\'e, J. 1996, ApJ, 469, 589
\bibitem[\protect\citeauthoryear{Murray, Quatert, \& Thompson} {2005}]{mur05} Murray, N., Quatert, E., \& Thompson, T. A. 2005, ApJ, 618, 569
\bibitem[\protect\citeauthoryear{Olbers} {1826}]{olb26} Olbers, H. W. M. 1826
\bibitem[\protect\citeauthoryear{Page et al.} {2006}]{pag06} Page et al. 2006, ApJ, submitted, astro-ph/0603450
\bibitem[\protect\citeauthoryear{Pettini et al.} {2001}]{pet01} Pettini M., Shapley A. E., Steidel C. C., Cuby J.-G., Dickinson M., Moorwood A. F. M., Adelberger K. L., \& Giavalisco M. 2001, ApJ, 554, 981
\bibitem[\protect\citeauthoryear{Pettini et al.} {2003}]{pet03} Pettini, M., Madau, P., Bolte, M., Prochaska, J.X., Ellison, S.L., \& Fan, X. 2003, ApJ, 594, 695
\bibitem[\protect\citeauthoryear{Porciani \& Madau} {2005}]{por05} Porciani, C. \& Madau, P. 2005, ApJ, 625, L43
\bibitem[\protect\citeauthoryear{Prochaska et al.} {2003}]{pro03} Prochaska, J. X., Gawiser, E., Wolfe, A. M., Castro, S., \& Djorgovski, S. G. 2003, ApJ, 595, L9
\bibitem[\protect\citeauthoryear{Press et al.} {1993}]{pre93} Press, W. H., Rybicki, G. B., \& Schneider, D. P. 1993, ApJ, 414, 64
\bibitem[\protect\citeauthoryear{Rauch et al.} {1997}]{rau97} Rauch, M., Haehnelt, M. G., \& Steinmetz, M. 1997, ApJ, 481, 601
\bibitem[\protect\citeauthoryear{Rauch et al.} {2001}]{rau01} Rauch, M., Sargent, W. L. W., \& Barlow, T. A. 2001, ApJ, 554, 823
\bibitem[\protect\citeauthoryear{Rauch et al.} {2005}]{rau05} Rauch, M., Becker, G. D., Viel, M., Sargent, W. L. W., Smette, A., Simcoe, R. A., Barlow, T. A., Haehnelt, M. G. 2005, ApJ, 632, 58
\bibitem[\protect\citeauthoryear{Rupke, Veilleux \& Sanders} {2005}]{rup05} Rupke, D. S., Veilleux, S., \& Sanders, D. B. 2005, ApJS, 160, 115
\bibitem[\protect\citeauthoryear{Ryan-Weber et al.} {2006}]{rya06} Ryan-Weber, E. V., Pettini, M., \& Madau, P. 2006, MNRAS, accepted, astro-ph/0607029
\bibitem[\protect\citeauthoryear{Scannapieco et al.} {2002}]{sca02} Scannapieco, E., Ferrara, A., \& Madau, P. 2002, ApJ, 574, 590
\bibitem[\protect\citeauthoryear{Scannapieco et al.} {2006}]{sca06} Scannapieco, E., Pichon, C., Aracil, B., Petitjean, P., Thacker, R.J., Pogosyan, D., Bergeron, J., \& Couchman, H.M.P. 2006, MNRAS, 365, 615 
\bibitem[\protect\citeauthoryear{Schaerer} {2003}]{sch03b} Schaerer, D. 2003, A\&A, 397, 527
\bibitem[\protect\citeauthoryear{Schaye et al.} {2000}]{sch00} Schaye, J., Theuns, T., Rauch, M., Efstathiou, G., \& Sargent, W. L. W. 2000, MNRAS, 318, 817
\bibitem[\protect\citeauthoryear{Schaye et al.} {2003}]{sch03} Schaye, J., Aguirre, A., Kim, T.-S., Theuns, T., Rauch, M., \& Sargent, W.L.W. 2003, ApJ, 596, 768 (S03)
\bibitem[\protect\citeauthoryear{Shapley et al.} {2003}]{sha03} Shapley, A. E., Steidel, C. C., Pettini, M., \& Adelberger, K. L. 2003, ApJ, 588, 65
\bibitem[\protect\citeauthoryear{Simcoe et al.} {2006}]{sim06} Simcoe, R.A., Sargent, W.L.W., Rauch, M., \& Becker, G. 2006, ApJ, 637, 648
\bibitem[\protect\citeauthoryear{Simcoe} {2006}]{sim06b} Simcoe, R. A. 2006, ApJ, submitted, astro-ph/0605710
\bibitem[\protect\citeauthoryear{Sokasian et al.} {2003}]{sok03} Sokasian, A., Abel, T., \& Hernquist, L. 2003, MNRAS, 340, 473
\bibitem[\protect\citeauthoryear{Songaila \& Cowie} {1996}]{son96} Songaila, A., Cowie, L.L. 1996, AJ, 112, 335
\bibitem[\protect\citeauthoryear{Songaila} {1998}]{son98} Songaila, A. 1998, AJ, 115, 2184
\bibitem[\protect\citeauthoryear{Songaila} {2001}]{son01} Songaila, A. 2001, ApJ, 561, L153
\bibitem[\protect\citeauthoryear{Songaila} {2005}]{son05} Songaila, A. 2005, AJ, 130, 1996
\bibitem[\protect\citeauthoryear{Spergel et al.} {2006}]{spe06} Spergel et al. 2006, ApJ, submitted, astro-ph/0603449
\bibitem[\protect\citeauthoryear{Springel \& Hernquist} {2002}]{spr02} Springel, V. \& Hernquist, L.  2002, MNRAS, 333, 649
\bibitem[\protect\citeauthoryear{Springel \& Hernquist} {2003a}]{spr03a} Springel, V. \& Hernquist, L.  2003, MNRAS, 339, 289
\bibitem[\protect\citeauthoryear{Springel \& Hernquist} {2003b}]{spr03b} Springel, V. \& Hernquist, L.  2003, MNRAS, 339, 312 (SH03)
\bibitem[\protect\citeauthoryear{Springel} {2005}]{spr05} Springel, V. 2005, MNRAS, 364, 1105
\bibitem[\protect\citeauthoryear{Theuns et al.} {2000}]{the00} Theuns, T., Schaye, J., \& Haehnelt, M. G. 2000, MNRAS, 315, 600
\bibitem[\protect\citeauthoryear{Theuns et al.} {2002a}]{the02a} Theuns, T., Bernardi, M., Frieman, J., Hewett, P., Schaye, J., Sheth, R. K., \& Subbarao, M. 2002, ApJ, 574, L111
\bibitem[\protect\citeauthoryear{Theuns et al.} {2002b}]{the02} Theuns, T., Viel, M., Kay, S., Schaye, J., Carswell, R.F., \& Tzanavaris, P. 2002, ApJ, 578, L5
\bibitem[\protect\citeauthoryear{Tremonti et al.} {2004}]{tre04} Tremonti, C. A., et al. 2004, ApJ, 613, 898
\bibitem[\protect\citeauthoryear{Sutherland \& Dopita} {1993}]{sut93} Sutherland, R. S. \& Dopita, M. A. 1993, ApJS, 88, 253
\bibitem[\protect\citeauthoryear{Woosley \& Weaver} {1995}]{woo95} Woosley, S. E. \& Weaver, T. A. 1995, ApJS, 101, 181
\bibitem[\protect\citeauthoryear{Zheng et al.} {2004}]{zhe04} Zheng, W., et al. 2004, ApJ, 605, 631
\end{thebibliography}
\end{document}